\documentclass[11pt, a4paper]{article}
\usepackage{jcappub}
\usepackage{siunitx}

\usepackage[dvipsnames]{xcolor}

\pdfoutput=1

\usepackage[T1]{fontenc}
\usepackage{lmodern}
\usepackage{calc}
\usepackage{graphicx}
\usepackage{booktabs}
\usepackage{textcomp}
\usepackage{xspace}
\usepackage{relsize}
\usepackage{amssymb}
\usepackage{amsmath}
\usepackage{listings}
\usepackage{microtype}
\usepackage{multirow}
\usepackage{tabularx}
\usepackage{array}
\usepackage{placeins}
\usepackage{cuted}
\usepackage{soul} 
\usepackage{fixltx2e}
\usepackage{slashed}
\usepackage{bm}
\usepackage[numbers,sort&compress]{natbib}
\usepackage[labelfont=bf,font=small]{caption}
\usepackage[skip=-2pt]{subcaption}
\usepackage[clockwise,figuresright]{rotating}
\usepackage{tikz}
\usepackage[normalem]{ulem}
\usepackage[utf8]{inputenc}

\usepackage{etoolbox}
\AfterEndEnvironment{strip}{\leavevmode}

\allowdisplaybreaks

\newcolumntype{L}{>{\raggedright\let\newline\\\arraybackslash\hspace{0pt}}X}
\newcolumntype{R}{>{\raggedleft\let\newline\\\arraybackslash\hspace{0pt}}X}
\newcolumntype{C}{>{\centering\let\newline\\\arraybackslash\hspace{0pt}}X}

\newcommand{\gambitinstitute}[1]{\expandafter\csname #1\endcsname\label{#1}}

\newcommand{\aachen}{Institute for Theoretical Particle Physics and Cosmology (TTK), RWTH Aachen University, D-52056 Aachen, Germany}
\newcommand{\queens}{Department of Physics, Engineering Physics and Astronomy, Queen's University, Kingston ON K7L 3N6, Canada}
\newcommand{\mcdonald}{Arthur B. McDonald Canadian Astroparticle Physics Research Institute, Kingston ON K7L 3N6, Canada}
\newcommand{\perimeter}{Perimeter Institute for Theoretical Physics, Waterloo ON N2L 2Y5, Canada}
\newcommand{\imperial}{Department of Physics, Imperial College London, Blackett Laboratory, Prince Consort Road, London SW7 2AZ, UK}
\newcommand{\cambridge}{Cavendish Laboratory, University of Cambridge, JJ Thomson Avenue, Cambridge, CB3 0HE, UK}

\newcommand{\oslo}{Department of Physics, University of Oslo, N-0316 Oslo, Norway}
\newcommand{\adelaide}{Department of Physics, University of Adelaide, Adelaide, SA 5005, Australia}

\newcommand{\monash}{School of Physics and Astronomy, Monash University, Melbourne, VIC 3800, Australia}

\newcommand{\okc}{Oskar Klein Centre for Cosmoparticle Physics, AlbaNova University Centre, SE-10691 Stockholm, Sweden}

\newcommand{\uq}{School of Mathematics and Physics, The University of Queensland, St.\ Lucia, Brisbane, QLD 4072, Australia}
\newcommand{\gottingen}{Institut f\"ur Astrophysik, Georg-August Universit\"at G\"ottingen, Friedrich-Hund-Platz~1, 37077 G\"ottingen, Germany}

\newcommand{\kicc}{Kavli Institute for Cosmology, University of Cambridge, Madingley Road, Cambridge, CB3 0HA, UK}
\newcommand{\caius}{Gonville \& Caius College, Trinity Street, Cambridge, CB2 1TA, UK}

\makeatletter

\newcommand{\preprintnumber}[1]{\gdef\@preprintnumber{\begin{flushright}{#1}\end{flushright}}}

\g@addto@macro\bfseries{\boldmath}
\makeatother

\let\underscore\_
\renewcommand{\_}{\discretionary{\underscore}{}{\underscore}}

\makeatletter
\let\orgdescriptionlabel\descriptionlabel
\renewcommand*{\descriptionlabel}[1]{%
  \let\orglabel\label
  \let\label\@gobble
  \phantomsection
  \protected@edef\@currentlabel{#1}%
  \let\label\orglabel
  \orgdescriptionlabel{#1}%
}
\makeatother

\newcommand\postnewlinemarker{\hbox{\ensuremath{\hookrightarrow}}}
\newcommand\cpp[1]{{\lstinline!#1!}}  
\newcommand\cpppragma[1]{{\CPPcommentstyle#1}}
\newcommand\yaml[1]{{\lstset{style=yaml}\lstinline!#1!\lstset{style=cpp}}}
\newcommand\yamlvalue[1]{{\YAMLvaluestyle\ttfamily#1}}
\newcommand\yamlcomment[1]{{\YAMLcommentstyle\ttfamily#1}}

\newcommand\term[1]{{\lstset{style=terminal}\lstinline!#1!\lstset{style=cpp}}}
\newcommand\termalt[1]{{\lstset{style=terminalalt}\lstinline!#1!\lstset{style=cpp}}}
\newcommand\fortran[1]{{\lstset{style=fortran}\lstinline!#1!\lstset{style=cpp}}}
\newcommand\py[1]{{\lstset{style=python}\lstinline!#1!\lstset{style=cpp}}}
\newcommand\customtilde{{\raisebox{0.2ex}{\scalebox{0.6}{\boldmath$\sim$}}}}
\newcommand\mathematica[1]{{\lstset{style=Mathematica}\lstinline!#1!\lstset{style=cpp}}}
\newcommand\guminline[1]{{{\lstset{style=gum}\lstinline!#1!}}}
\newcommand\textinline[1]{{{\lstset{style=text}\lstinline!#1!}}}

\def\be{\begin{equation}}
\def\ee{\end{equation}}
\def\ba{\begin{eqnarray}}
\def\ea{\end{eqnarray}}
\newcommand{\bea}{\begin{eqnarray}}
\newcommand{\eea}{\end{eqnarray}}

\lstnewenvironment{lstlistingyaml}{\lstset{style=yaml}}{\lstset{style=cpp}}
\lstnewenvironment{lstlistingterm}{\lstset{style=terminal}}{\lstset{style=cpp}}
\lstnewenvironment{lstlistingfortran}{\lstset{style=fortran}}{\lstset{style=cpp}}
\lstnewenvironment{lstcpp}{\lstset{style=cpp}}{\lstset{style=cpp}}
\lstnewenvironment{lstcppalt}{\lstset{style=cppalt}}{\lstset{style=cpp}}
\lstnewenvironment{lstcppnum}{\lstset{style=cppnum}}{\lstset{style=cpp}}
\lstnewenvironment{lstyaml}{\lstset{style=yaml}}{\lstset{style=cpp}}
\lstnewenvironment{lstgum}{\lstset{style=gum}}{\lstset{style=cpp}}
\lstnewenvironment{lstterm}{\lstset{style=terminal}}{\lstset{style=cpp}}
\lstnewenvironment{lsttermalt}{\lstset{style=terminalalt}}{\lstset{style=cpp}}
\lstnewenvironment{lsttext}{\lstset{style=text}}{\lstset{style=cpp}}
\lstnewenvironment{lstfortran}{\lstset{style=fortran}}{\lstset{style=cpp}}
\lstnewenvironment{lstpy}{\lstset{style=python}}{\lstset{style=cpp}}
\lstnewenvironment{lstmathematica}{\lstset{style=mathematica}}{\lstset{style=cpp}}

\newcommand{\tmpname}{}
\newcommand{\tmplistingname}{}
\makeatletter
\newif\ifATOlabelname
\lst@Key{labelname}{Listing}{\def\ATOlabelname{#1}\global\ATOlabelnametrue}
\makeatother
\lstnewenvironment{lstcpplabel}[1][]{
  \lstset{style=cpp,#1} 
  \ifATOlabelname
    \renewcommand{\tmpname}{\lstlistingname}
    \renewcommand{\tmplistingname}{\lstlistlistingname}
    \renewcommand{\lstlistingname}{\ATOlabelname}
    \renewcommand{\lstlistlistingname}{List of \lstlistingname s}
  \fi
}{
  \renewcommand{\lstlistingname}{\tmpname}
  \renewcommand{\lstlistlistingname}{\tmplistingname}
  \lstset{style=cpp}
}
\definecolor{solarized@base03}{HTML}{002B36}
\definecolor{solarized@base02}{HTML}{073642}
\definecolor{solarized@base01}{HTML}{586e75}
\definecolor{solarized@base00}{HTML}{657b83}
\definecolor{solarized@base0}{HTML}{839496}
\definecolor{solarized@base1}{HTML}{93a1a1}
\definecolor{solarized@base2}{HTML}{EEE8D5}
\definecolor{solarized@base3}{HTML}{FDF6E3}
\definecolor{solarized@yellow}{HTML}{B58900}
\definecolor{solarized@orange}{HTML}{CB4B16}
\definecolor{solarized@red}{HTML}{DC322F}
\definecolor{solarized@magenta}{HTML}{D33682}
\definecolor{solarized@violet}{HTML}{6C71C4}
\definecolor{solarized@blue}{HTML}{268BD2}
\definecolor{solarized@cyan}{HTML}{2AA198}
\definecolor{solarized@green}{HTML}{859900}
\definecolor{darkred}{HTML}{550003}
\definecolor{darkgreen}{HTML}{00AA00}
\definecolor{orchid}{HTML}{AF06F5}

\newcommand\YAMLstringstyle{\footnotesize\color{solarized@green}\mdseries}
\newcommand\YAMLkeystyle{\footnotesize\color{solarized@blue}\ttfamily}
\newcommand\YAMLvaluestyle{\footnotesize\color{blue}\mdseries}
\newcommand\ProcessThreeDashes{\llap{\color{cyan}\mdseries-{-}-}}

\newcommand\CPPcommentstyle{\color{solarized@violet}\footnotesize\ttfamily}
\newcommand\CPPdirectivestyle{\color{solarized@magenta}\footnotesize\ttfamily}
\newcommand\termplainstyle{\footnotesize\ttfamily}

\newcommand\YAMLcommentstyle{\color{solarized@orange}\ttfamily}

\newcommand\processLongMacroDelimiter
{%
\CPPdirectivestyle%
\#define%
}

\lstdefinestyle{cpp}
{
  language=C++,
  basicstyle=\footnotesize\ttfamily,
  basewidth={0.53em,0.44em}, 
  numbers=none,
  tabsize=2,
  breaklines=true,
  escapeinside={@}{@},
  showstringspaces=false,
  numberstyle=\tiny\color{solarized@base01},
  keywordstyle=\color{solarized@orange},
  stringstyle=\color{solarized@red}\ttfamily,
  identifierstyle=\color{solarized@blue},
  commentstyle=\CPPcommentstyle,
  directivestyle=\CPPdirectivestyle,
  emphstyle=\color{solarized@green},
  frame=single,
  rulecolor=\color{solarized@base2},
  rulesepcolor=\color{solarized@base2},
  literate={~} {\customtilde}1,
  moredelim=*[directive]\ \ \#,
  moredelim=*[directive]\ \ \ \ \#
}

\lstdefinestyle{cppalt}
{
  language=C++,
  basicstyle=\footnotesize\ttfamily,
  basewidth={0.53em,0.44em}, 
  numbers=none,
  tabsize=2,
  breaklines=true,
  escapeinside={*@}{@*},
  showstringspaces=false,
  numberstyle=\tiny\color{solarized@base01},
  keywordstyle=\color{solarized@orange},
  stringstyle=\color{solarized@red}\ttfamily,
  identifierstyle=\color{solarized@blue},
  commentstyle=\CPPcommentstyle,
  directivestyle=\CPPdirectivestyle,
  emphstyle=\color{solarized@green},
  frame=single,
  rulecolor=\color{solarized@base2},
  rulesepcolor=\color{solarized@base2},
  literate={~}{\customtilde}1,
  moredelim=**[is][\processLongMacroDelimiter]{BeginLongMacro}{EndLongMacro} 
}

\lstdefinestyle{cppnum}
{
  language=C++,
  basicstyle=\footnotesize\ttfamily,
  basewidth={0.53em,0.44em}, 
  numbers=none,
  tabsize=2,
  breaklines=true,
  escapeinside={@}{@},
  numberstyle=\tiny\color{solarized@base01},
  showstringspaces=false,
  keywordstyle=\color{solarized@orange},
  stringstyle=\color{solarized@red}\ttfamily,
  identifierstyle=\color{solarized@blue},
  commentstyle=\CPPcommentstyle,
  directivestyle=\CPPdirectivestyle,
  emphstyle=\color{solarized@green},
  frame=single,
  rulecolor=\color{solarized@base2},
  rulesepcolor=\color{solarized@base2},
  literate={~} {\customtilde}1,
  moredelim=*[directive]\ \ \#,
  moredelim=*[directive]\ \ \ \ \#
}

\lstdefinestyle{python}
{
  language=Python,
  basicstyle=\footnotesize\ttfamily,
  basewidth={0.53em,0.44em},
  numbers=none,
  tabsize=2,
  breaklines=true,
  escapeinside={@}{@},
  showstringspaces=false,
  numberstyle=\tiny\color{solarized@base01},
  keywordstyle=\color{blue},
  stringstyle=\color{orange}\ttfamily,
  identifierstyle=\color{darkred},
  commentstyle=\color{purple},
  emphstyle=\color{green},
  frame=single,
  rulecolor=\color{solarized@base2},
  rulesepcolor=\color{solarized@base2},
  literate = {~}{\customtilde}1
             {\ as\ }{{\color{blue}\ as\ \color{black}}}3
             {.set}{{\color{black}.}{\color{darkred}set}}4
}

\lstdefinestyle{fortran}
{
  language=Fortran,
  basicstyle=\footnotesize\ttfamily,
  basewidth={0.53em,0.44em},
  numbers=none,
  tabsize=2,
  breaklines=true,
  escapeinside={@}{@},
  showstringspaces=false,
  numberstyle=\tiny\color{solarized@base01},
  keywordstyle=\color{blue},
  stringstyle=\color{orange}\ttfamily,
  identifierstyle=\color{Periwinkle},
  commentstyle=\color{purple},
  emphstyle=\color{green},
  morekeywords={and, or, true, false},
  frame=single,
  rulecolor=\color{solarized@base2},
  rulesepcolor=\color{solarized@base2},
  literate={~}{\customtilde}1
}

\lstdefinestyle{terminal}
{
  language=bash,
  basicstyle=\termplainstyle,
  numbers=none,
  tabsize=2,
  breaklines=true,
  escapeinside={@}{@},
  frame=single,
  showstringspaces=false,
  numberstyle=\tiny\color{solarized@base01},
  keywordstyle=\color{solarized@orange},
  stringstyle=\color{solarized@red}\ttfamily,
  identifierstyle=\color{black},
  commentstyle=\color{solarized@violet},
  emphstyle=\color{solarized@green},
  frame=single,
  rulecolor=\color{solarized@base2},
  rulesepcolor=\color{solarized@base2},
  morekeywords={gambit, cmake, make, mkdir, gum, python, wget, tar, cp, pippi, mpirun},
  deletekeywords={test},
  literate = {/gambit}{{/}{\color{black}}gambit}6
             {gambit/}{{\color{black}}gambit{/}}6
             {gum/}{{\color{black}}gum{/}}4
             {/include}{{/}{\color{black}}include}8
             {cmake/}{{\color{black}}cmake/}6
             {.cmake}{{.}{\color{black}}cmake}6
             {.gum}{{.}{\color{black}}gum}6
             {.tar}{{.}{\color{black}}tar}4
             {source/}{{\color{black}}source{/}}7
             { type}{{\color{black}}{}type}5
             {~}{\customtilde}1
             {math}{{\color{solarized@orange}}math}4
}

\lstdefinestyle{terminalalt}
{
  language=bash,
  basicstyle=\footnotesize\ttfamily,
  numbers=none,
  tabsize=2,
  breaklines=true,
  escapeinside={*@}{@*},
  frame=single,
  showstringspaces=false,
  numberstyle=\tiny\color{solarized@base01},
  keywordstyle=\color{solarized@orange},
  stringstyle=\color{solarized@red}\ttfamily,
  identifierstyle=\color{black},
  commentstyle=\color{solarized@violet},
  emphstyle=\color{solarized@green},
  frame=single,
  rulecolor=\color{solarized@base2},
  rulesepcolor=\color{solarized@base2},
  morekeywords={gambit, cmake, make, mkdir},
  deletekeywords={test},
  literate = {\ gambit}{{\ }{\color{black}}gambit}7
             {/gambit}{{/}{\color{black}}gambit}6
             {gambit/}{{\color{black}}gambit{/}}6
             {/include}{{/}{\color{black}}include}8
             {cmake/}{{\color{black}}cmake/}6
             {.cmake}{{.}{\color{black}}cmake}6
             {~}{\customtilde}1
}

\lstdefinestyle{text}
{
  language={},
  basicstyle=\footnotesize\ttfamily,
  identifierstyle=\color{black},
  numbers=none,
  tabsize=2,
  breaklines=true,
  escapeinside={*@}{@*},
  showstringspaces=false,
  frame=single,
  rulecolor=\color{solarized@base2},
  rulesepcolor=\color{solarized@base2},
  literate={~}{\customtilde}1
}

\lstdefinestyle{yaml}
{
  language=bash,
  escapeinside={@}{@},
  keywords={true,false,null},
  otherkeywords={},
  keywordstyle=\color{solarized@base0}\bfseries,
  basicstyle=\footnotesize\color{black}\ttfamily,
  identifierstyle=\YAMLkeystyle,
  sensitive=false,
  commentstyle=\YAMLcommentstyle,
  morecomment=[l]{\#},
  morecomment=[s]{/*}{*/},
  stringstyle=\YAMLstringstyle\ttfamily,
  moredelim=**[s][\YAMLkeystyle]{,}{:},   
  moredelim=**[l][\YAMLvaluestyle]{:},    
  morestring=[b]',
  morestring=[b]",
  literate =    {---}{{\ProcessThreeDashes}}3
                {>}{{\textcolor{solarized@red}\textgreater}}1
                {gtr}{\textgreater}1
                {grt}{\textgreater}1
                {|}{{\textcolor{solarized@red}\textbar}}1
                {\ -\ }{{\mdseries\color{black}\ -\ \negmedspace}}3
                {\}}{{{\color{black} \}}}}1
                {\{}{{{\color{black} \{}}}1
                {[}{{{\color{black} [}}}1
                {]}{{{\color{black} ]}}}1
                {~}{\customtilde}1,
  breakindent=0pt,
  breakatwhitespace,
  columns=fullflexible
}

\lstdefinestyle{gum}
{
  language=bash,
  escapeinside={@}{@},
  keywords={true,false,null,all},
  otherkeywords={},
  keywordstyle=\color{solarized@base02}\bfseries,
  basicstyle=\footnotesize\color{black}\ttfamily,
  identifierstyle=\color{solarized@magenta},
  sensitive=false,
  commentstyle=\color{solarized@cyan}\ttfamily,
  morecomment=[l]{\#},
  morecomment=[s]{/*}{*/},
  stringstyle=\footnotesize\color{solarized@base01}\mdseries\ttfamily,
  moredelim=**[l][\footnotesize\color{solarized@base02}\mdseries]{:},    
  morestring=[b]',
  morestring=[b]",
  literate =    {---}{{\ProcessThreeDashes}}3
                {grt}{{\textcolor{solarized@magenta}\textgreater}}1
                {gtr}{{\textcolor{solarized@base02}\textgreater}}1
                {/>}{{\textcolor{solarized@magenta}\textgreater}}1
                {/<}{{\textcolor{solarized@magenta}\textless}}1
                {lss}{{\textcolor{solarized@base02}\textless}}1
                {pls}{{\textcolor{solarized@magenta}+}}1
                {mns}{{\textcolor{solarized@magenta}-}}1
                {|}{{\textcolor{solarized@base02}\textbar}}1
                {\ -\ }{{\mdseries\color{solarized@base02}\ -\ \negmedspace}}3
                {\}}{{{\color{solarized@base02} \}}}}1
                {\{}{{{\color{solarized@base02} \{}}}1
                {[}{{{\color{solarized@base02} [}}}1
                {]}{{{\color{solarized@base02} ]}}}1
                {~}{\customtilde}1,
  breakindent=0pt,
  breakatwhitespace,
  columns=fullflexible
}

\lstdefinestyle{mathematica}
{
  language={Mathematica},
  basicstyle=\footnotesize\ttfamily,
  basewidth={0.53em,0.44em},
  numbers=none,
  tabsize=2,
  breaklines=true,
  postbreak=,
  escapeinside={@}{@},
  numberstyle=\tiny\color{black},
  showstringspaces=false,
  numberstyle=\tiny\color{solarized@base01},
  keywordstyle=\color{solarized@orange},
  stringstyle=\color{solarized@red}\ttfamily,
  identifierstyle=\color{solarized@orange}\ttfamily,
  commentstyle=\color{solarized@gray}\ttfamily,
  directivestyle=\color{solarized@orange}\ttfamily,
  emphstyle=\color{solarized@green},
  frame=single,
  rulecolor=\color{solarized@base2},
  rulesepcolor=\color{solarized@base2},
  literate={~} {\customtilde}1,
  moredelim=*[directive]\ \ \#,
  moredelim=*[directive]\ \ \ \ \#,
  mathescape=false
}

\newcommand{\doublecross}[2]{\hyperref[#2]{\textbf{#1}}}
\newcommand{\doublecrosssf}[2]{\hyperref[#2]{\textbf{\textsf{#1}}}}

\newcommand{\startglossary}{\section{Glossary}\label{glossary}Here we explain some terms that have specific technical definitions in \GB.\begin{description}}
\newcommand{\finishglossary}{\end{description}}

\newcommand{\bcode}{\begin{lstlisting}}
\newcommand{\ecode}{\end{lstlisting}}
\newcommand{\metavarf}[1]{\textit{\color{darkgreen}\footnotesize\textrm{#1}}}

\newcommand{\metavar}{\metavarf}

\newcommand{\ie}{i.e.\ }
\newcommand{\eg}{e.g.\ }

\newcommand{\gambit}{\textsf{GAMBIT}\xspace}

\newcommand{\darkbit}{\textsf{DarkBit}\xspace}

\newcommand{\scannerbit}{\textsf{ScannerBit}\xspace}

\newcommand{\neutrinobit}{\textsf{NeutrinoBit}\xspace}

\newcommand{\GB}{\gambit}

\newcommand{\mpi}{\textsf{MPI}\xspace}

\newcommand\Mathematica{\textsf{Mathematica}\xspace}

\newcommand\pippi{\textsf{pippi}\xspace}
\newcommand\MultiNest{\textsf{MultiNest}\xspace}
\newcommand\multinest{\MultiNest}
\newcommand\Polychord{\textsf{PolyChord}\xspace}
\newcommand\polychord{\Polychord}

\newcommand\diver{\textsf{Diver}\xspace}

\newcommand\xx{\raisebox{0.2ex}{\smaller ++}\xspace}
\newcommand\Cpp{\textsf{C\xx}\xspace}

\newcommand\plainC{\textsf{C}\xspace}
\newcommand\Python{\textsf{Python}\xspace}
\newcommand\python{\Python}
\newcommand\Cython{\textsf{Cython}\xspace}
\newcommand\cython{\Cython}
\newcommand\Fortran{\textsf{Fortran}\xspace}
\newcommand\YAML{\textsf{YAML}\xspace}

\newcommand\cmake{\textsf{CMake}\xspace}
\newcommand\gpp{\textsf{g++}\xspace}
\newcommand\gfortran{\textsf{gfortran}\xspace}
\newcommand\icpc{\textsf{icpc}\xspace}
\newcommand\ifort{\textsf{ifort}\xspace}

\newcommand\beq{\begin{equation}}
\newcommand\eeq{\end{equation}}

\usepackage{tikz}
\usetikzlibrary{trees}
\usetikzlibrary{shapes}
\usetikzlibrary{backgrounds}
\usetikzlibrary{calc}
\usetikzlibrary{positioning}
\usepackage{adjustbox}

\definecolor{PatrickBlue}{HTML}{00549F}
\definecolor{PBCompInflation}{HTML}{9F9A00}
\definecolor{PBCompNeutrinos}{HTML}{9F0055}
\definecolor{PBCompNuisance}{HTML}{9F4A00}
\definecolor{PBCompBlue}{HTML}{00059F}
\definecolor{PBCompTurquoise}{HTML}{009F9A}
\definecolor{PBCompNeutrons}{HTML}{559F00}

\makeatletter
\patchcmd{\ttlh@hang}{\parindent\z@}{\parindent\z@\leavevmode}{}{}
\patchcmd{\ttlh@hang}{\noindent}{}{}{}
\makeatother

\newcommand{\mplred}{\overline{m}_\text{P}}

\usepackage{upgreek}

\newcolumntype{P}{>{\raggedright\arraybackslash}X}

\newcommand{\mev}{\,\textrm{MeV}}

\def\calP{{\cal P}}

\def\|{{ \, || \,}}

\def\npiv{N_\star}

\def\dd{{\textnormal{d}}}

\newcommand{\cosmobit}{\textsf{CosmoBit}\xspace}

\newcommand{\exoclass}{\textsf{ExoCLASS}\xspace}
\newcommand{\darkages}{\textsf{DarkAges}\xspace}
\newcommand{\plc}{\textsf{plc}\xspace}
\newcommand{\class}{\textsf{CLASS}\xspace}
\newcommand{\camb}{\textsf{CAMB}\xspace}
\newcommand{\classy}{\textsf{classy}\xspace}
\newcommand{\alterbbn}{\textsf{AlterBBN}\xspace}
\newcommand{\multimodecode}{\textsf{MultiModeCode}\xspace}
\newcommand{\montepython}{\textsf{MontePython}\xspace}
\newcommand{\montepythonlike}{\textsf{MontePythonLike}\xspace}
\newcommand{\cosmomc}{\textsf{CosmoMC}\xspace}
\newcommand{\cosmosis}{\textsf{Cosmosis}\xspace}
\newcommand{\cobaya}{\textsf{Cobaya}\xspace}
\newcommand{\mastercode}{\textsf{MasterCode}\xspace}
\newcommand{\hepfit}{\textsf{HEPFit}\xspace}
\newcommand{\pybind}{\textsf{pybind11}\xspace}

\newcommand{\classinput}{\cpp{input\_dict}\xspace}

\newcommand{\mplike}{\py{Likelihood}\xspace}
\newcommand{\mpdata}{\py{Data}\xspace}

\newcommand{\lcdm}{{\Lambda\mathrm{CDM}}}
\newcommand{\Tcmb}{T_{\rm CMB}}
\newcommand{\Tgamma}{T_{\gamma}}
\newcommand{\Tnu}{T_{\nu}}

\newcommand{\omegacdm}{\omega_{\rm cdm}}
\newcommand{\omegar}{\omega_{\rm r}}
\newcommand{\omegab}{\omega_{\rm b}}
\newcommand{\omegalambda}{\omega_{\Lambda}}
\newcommand{\taur}{\tau_{\rm reio}}
\newcommand{\yhe}{Y_{\rm p}}

\newcommand{\Neff}{N_{\rm eff}}

\usepackage{pifont}
\newcommand{\cmark}{\ding{51}}%
\newcommand{\xmark}{\ding{55}}%

\renewcommand{\arraystretch}{1.2}
\newcommand\purl[1]{\protect\url{#1}} 

\usepackage{mathtools}
\usepackage{siunitx}
\usepackage{afterpage}
\usepackage{colortbl}
\usepackage{bm}
\usepackage{arydshln}
\usepackage{wasysym}

\renewcommand{\vec}{\bm}

\newcolumntype{L}[1]{>{\raggedright\arraybackslash}p{#1}}

\title{CosmoBit: A GAMBIT module for computing cosmological observables and likelihoods}

\author{The GAMBIT Cosmology Workgroup:}
\author[1,2,3]{Janina J. Renk,}
\author[4]{Patrick St\"ocker,}
\author[1,2]{Sanjay Bloor,}
\author[1]{Selim Hotinli,}
\author[5]{Csaba Bal{\'a}zs,}
\author[6]{Torsten Bringmann,}
\author[5]{Tom\'as E. Gonzalo,}
\author[7,8,9]{Will Handley,}
\author[10]{Sebastian Hoof,}
\author[2]{Cullan Howlett,}
\author[4]{Felix Kahlhoefer,}
\author[1,2]{Pat Scott,}
\author[11,12,13]{Aaron C. Vincent}
\author[14]{and Martin White}

\emailAdd{janina.renk@fysik.su.se}
\emailAdd{stoecker@physik.rwth-aachen.de}
\emailAdd{sanjay.bloor12@imperial.ac.uk}

\affiliation[1]{\imperial}
\affiliation[2]{\uq}
\affiliation[3]{\okc}
\affiliation[4]{\aachen}
\affiliation[5]{\monash}
\affiliation[6]{\oslo}
\affiliation[7]{\cambridge}
\affiliation[8]{\kicc}
\affiliation[9]{\caius}
\affiliation[10]{\gottingen}
\affiliation[11]{\queens}
\affiliation[12]{\mcdonald}
\affiliation[13]{\perimeter}
\affiliation[14]{\adelaide}

\date{Received: date / Accepted: date}

\abstract
{We introduce \cosmobit, a module within the open-source \GB software framework for exploring connections between cosmology and particle physics with joint global fits. \cosmobit provides a flexible framework for studying various scenarios beyond $\Lambda$CDM, such as models of inflation, modifications of the effective number of relativistic degrees of freedom, exotic energy injection from annihilating or decaying dark matter, and variations of the properties of elementary particles such as neutrino masses and the lifetime of the neutron. Many observables and likelihoods in \cosmobit are computed via interfaces to \alterbbn, \class, \darkages, \montepython, \multimodecode, and \plc. This makes it possible to apply a wide range of constraints from large-scale structure, Type Ia supernovae, Big Bang Nucleosynthesis and the cosmic microwave background. Parameter scans can be performed using the many different statistical sampling algorithms available within the \GB framework, and results can be combined with calculations from other \GB modules focused on particle physics and dark matter.  We include extensive validation plots and a first application to scenarios with non-standard relativistic degrees of freedom and neutrino temperature, showing that the corresponding constraint on the sum of neutrino masses is much weaker than in the standard scenario.
}

\begin{document}

\hfill{\small TTK-20-27}

\hfill{\small gambit-code-2020}

\vspace*{-2\baselineskip}

\maketitle


\section{Introduction}
\label{sec:intro}

The practice of carrying out combined likelihood analyses using multiple datasets in order to compare models and estimate their parameters is well established, both in cosmology and particle physics.  Packages such as \cosmomc \cite{CosmoMC,Lewis:2013hha}, \montepython \cite{Audren:2012wb,brinckmann2018montepython}, \cobaya \cite{Torrado:2020dgo} and \cosmosis \cite{Zuntz:2014csq} provide the ability to use combinations of observables from e.g.\ large-scale structure (LSS), Type Ia supernovae (SNe Ia), Big Bang Nucleosynthesis (BBN) and the cosmic microwave background (CMB) to constrain a number of different cosmological models.  These range from the canonical Lambda-Cold Dark Matter ($\Lambda$CDM) cosmology, to theories of inflation, dark energy, additional neutrinos, exotic energy injection, and many others.

On the particle physics side, `global fitting' frameworks such as \gambit \cite{gambit,grev}, \mastercode \cite{MasterCodemSUGRA} and \hepfit \cite{hepfit} allow phenomenologists to carry out combined fits to data from the Large Hadron Collider (LHC), direct and indirect dark matter searches, flavour physics, precision Standard Model (SM) measurements, and neutrino physics.  These have been applied to a number of different theories for physics beyond the SM (BSM), ranging from supersymmetry \cite{CMSSM,MSSM,EWMSSM,Mastercode17,Costa:2017gup} to extended Higgs sectors \cite{SSDM,SSDM2,HP,Chowdhury17,Chiang:2018cgb}, axions \cite{Axions,XENON1T}, effective theories of flavour \cite{Bhom:2020lmk} and additional neutrinos \cite{RHN}.  Many of these BSM scenarios have additional cosmological implications not accounted for in the particle physics fits.  Examples are inflationary implications of different axion theories, impacts on BBN of decaying heavy relics, and effects of neutrinos on the power spectrum of cosmological perturbations at small scales.  Likewise, most beyond-$\Lambda$CDM theories are effective theories describing the impacts of new states on cosmology, with completions in terms of concrete BSM models leading to a host of potential signals in traditional particle physics experiments.

To date, no tool has been developed for combining the wealth of cosmological and particle physics datasets available, in order to simultaneously constrain theories from both sides. In this paper, as the first development in this direction, we present \cosmobit, which provides observable and likelihood calculations for BBN, the CMB, LSS, SNe Ia and other cosmological probes within the \gambit framework. First released in 2017, \gambit is an open-source global fitting framework that is particularly well suited for the purpose of combining information from different branches of physics, because it has been designed from the beginning to be easily extendable to new physics models and new experimental datasets. In particular, it has a modular design that allows many calculations to be reused with minimal effort if the physics model of interest is changed. \gambit can produce results in both the Bayesian and frequentist statistical frameworks using a host of advanced statistical sampling algorithms accessible via a dedicated sampling module called \scannerbit \cite{ScannerBit}, with massive, multi-level parallelisation ensuring computational efficiency. When combined with existing \gambit modules specialised for collider \cite{ColliderBit}, flavour \cite{FlavBit}, dark matter \cite{DarkBit}, neutrino \cite{RHN} and precision \cite{SDPBit} physics, \cosmobit allows one to use \gambit to perform joint cosmological and particle global fits for the first time.

\gambit, The Global and Modular BSM Inference Tool, is a highly flexible and modular framework for performing global analyses of physical theories. For this purpose all calculations of observables and likelihoods are separated into self-contained sub-calculations (called \textbf{module functions}), each of which provides a specific result (its \textbf{capability}) based on a set of input information (its \textbf{dependencies}). \gambit then identifies the functions required for a specific analysis at runtime and dynamically connects them in the most efficient way possible (\textbf{dependency resolution}). The result is a directed acyclic graph connecting the outputs and inputs of different functions (the \textbf{dependency tree}), which is automatically adapted if new observables or likelihoods are added. \gambit may also make use of functions provided by external packages (\textbf{backends}) written in \plainC, \Cpp, \Python 2, \Python 3, \Mathematica and all variants of \Fortran. A function provided to \gambit by a \textbf{backend} is called a \textbf{backend function}.

To run \gambit, the user provides an input file in \YAML format, selecting the \textbf{model} to analyse, choosing a sampling algorithm (\textbf{scanner}) to employ, and providing a list of all \textbf{likelihoods} and \textbf{observables} to be calculated. In this context a \textbf{model} refers to a collection of parameters, which can be interpreted by suitable \textbf{module functions} and constitute the starting points of the \textbf{dependency tree}. Chosen \textbf{likelihoods} and \textbf{observables} correspond to the \textbf{capabilities} of the \textbf{module functions} that should constitute the end points of the \textbf{dependency tree}. The aim of a global analysis with \gambit is then to identify the regions in parameter space that best explain the observed data and draw inferences on the underlying model. Depending on the precise question of interest, \gambit can either calculate the frequentist profile likelihood or the Bayesian posterior. This choice has strong implications for how to best sample the model parameters, and \gambit provides many different sampling strategies and statistical tools for this purpose.

In this paper, we provide a detailed description of the \cosmobit module, as well as a number of examples of global fits that can be performed with it, addressing the following cosmologically-interesting questions. What is the impact of uncertainties in the neutron lifetime on the cosmological parameters inferred from BBN? How do cosmological constraints on the sum of neutrino masses change when including realistic neutrino oscillation likelihoods? Which models of inflation are consistent with or even preferred by data? What are the cosmological constraints on dark matter models with new relativistic degrees of freedom (dark radiation)? How do decaying or annihilating relics change the reionisation history and the neutrino temperature, and what are the resulting constraints? Of course, the use of \cosmobit is not limited to these topics, and future releases will continue to expand the scope of the code.

We begin in Sec.\ \ref{sec:exec-sum} by giving an executive summary of the main features of \cosmobit. We provide background and definitions for the standard $\Lambda$CDM cosmology in Sec.\ \ref{sec:lcdm}, describe the BBN, CMB and late-time observables and likelihoods offered by \cosmobit in Sec.\ \ref{sec:lcdm-obs}, and describe support for extensions to $\Lambda$CDM in Sec.\ \ref{sec:beyond_lcdm}.  We give an example of the use of \cosmobit in Sec.\ \ref{sec:constraints}, focused on the determination of neutrino masses in the presence of additional ultra-relativistic species and a non-standard neutrino temperature, and conclude in Sec.\ \ref{sec:conclusion}.  We also provide extensive appendices, consisting of a quickstart guide (Appendix \ref{app:quickstart}), precise internal definitions of the cosmological models implemented in \gambit (Appendix \ref{app:models}), description of interfaces to external cosmology codes relevant for \cosmobit (Appendix \ref{app:backends}), and a list of all module functions provided by \cosmobit (Appendix \ref{app:capabilities}).

\cosmobit is open source and part of the \gambit \textsf{1.5} release, available from \href{http://gambit.hepforge.org}{gambit.hepforge.org} under the terms of the standard 3-clause BSD license.\footnote{\href{http://opensource.org/licenses/BSD-3-Clause}{http://opensource.org/licenses/BSD-3-Clause}.} Likelihood and posterior samples, as well as all input files required to reproduce all validation and example plots from this paper, can be downloaded from \textsf{Zenodo} \cite{CosmoBit_module_zenodo}.

%
\section{Executive Summary} \label{sec:exec-sum}
%

\begin{figure}
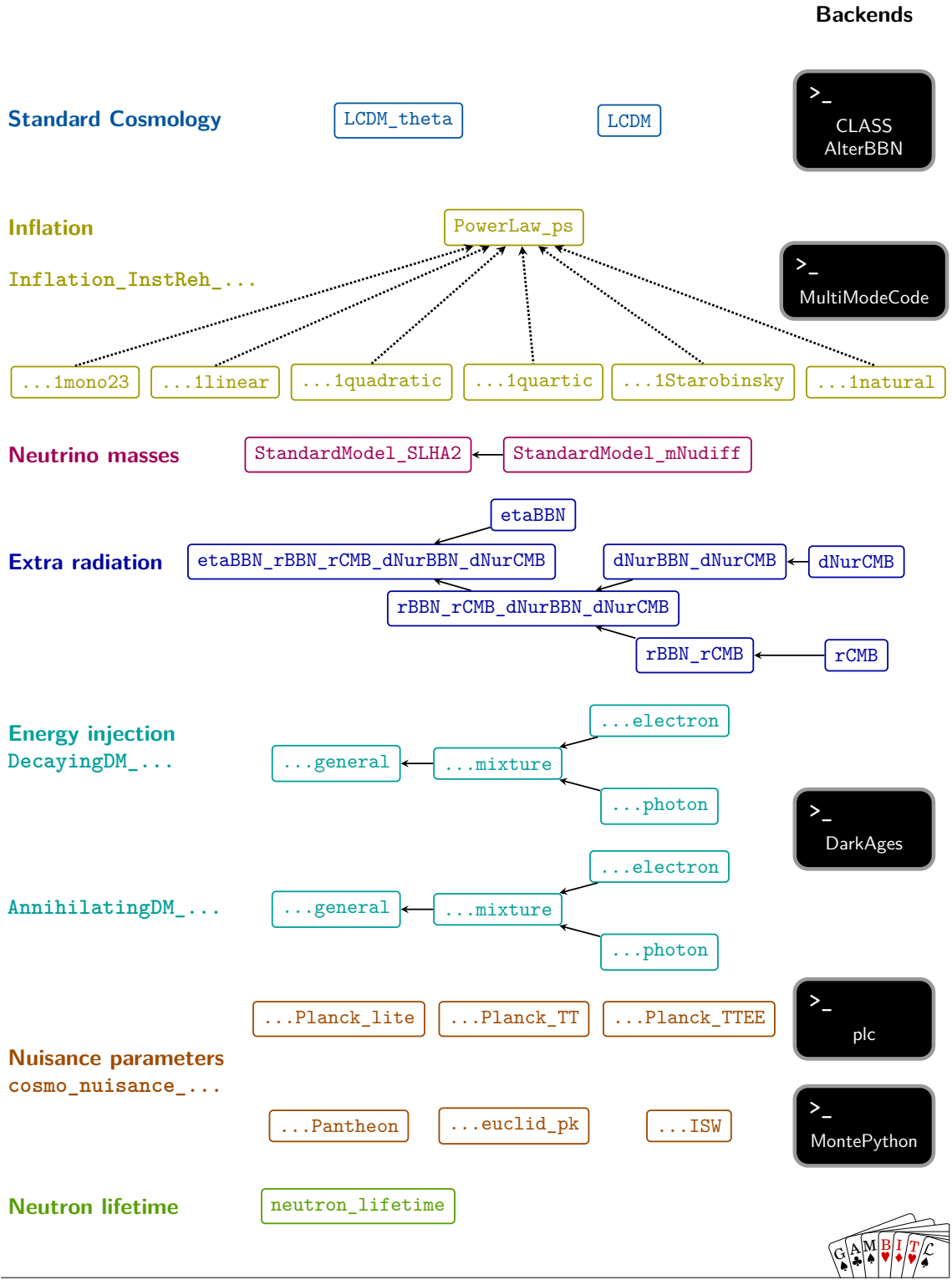

  \centering
   \vspace{-2em}
   \include{include/model_tree_cosmobit}
   \vspace{-1em}
   \caption{Overview of the \textbf{models} implemented in the first release of \cosmobit.  The models are grouped into different categories indicated by different colours. Solid arrows indicate child-to-parent model relations, where a parent model is a more general version that can be unambiguously defined from the parameters of a child model.  Dashed arrows indicate ``friend model'' translations \cite{gambit}, via which parameters in one model family can be interpreted in terms of the parameters of another. A cosmological model, an inflationary model and a neutrino model are required for all \cosmobit scans; other models are optional. With the exception of nuisance parameter models, the user can only choose one model from each group. The black boxes on the right indicate \textbf{backends} (external libraries) required in order to make full use of the respective models.}\label{fig:model_tree}
\end{figure}

In the first release of \cosmobit, we provide a number of \textbf{models} relevant for studying standard $\Lambda$CDM cosmology and several of its extensions:
\begin{itemize}
  \item \textsf{Cosmological models} defining the $\Lambda$CDM cosmology (see Sec.\ \ref{sec:lcdm} and Appendix~\ref{app:models-lcdm});
  \item \textsf{Inflation models} providing the shape of the primordial scalar and tensor power spectra (see Sec.\ \ref{sec:inflation} and Appendix~\ref{app:models-inflation});
  \item \textsf{Neutrino mass models} providing a realistic treatment of massive neutrinos (see Sec.\ \ref{sec:Nu-masses} and Appendix~\ref{app:models-NuMass});
  \item \textsf{Non-standard radiation models} providing modifications of the effective number of relativistic degrees of freedom (see Sec.\ \ref{sec:non-std-Neff} and Appendix~\ref{app:models-dNur});
  \item \textsf{Exotic energy injection models} providing contributions from annihilating or decaying dark matter (see Sec.\ \ref{sec:energy-injection} and Appendix~\ref{app:models-energyinjection});
  \item A \textsf{neutron lifetime} model providing the option to vary the neutron lifetime and explore its effects on cosmology (see Appendix~\ref{app:models-neutron});
  \item \textsf{Cosmological nuisance parameters} needed for likelihoods with nuisance parameters (see Appendix~\ref{app:models-nuisance}).
\end{itemize}
An overview of these model groups is given in Fig.~\ref{fig:model_tree}.

\afterpage{\clearpage}

The \textbf{likelihoods} available in \cosmobit to constrain these models come from BBN (Sec.\ \ref{sec:BBN}), the CMB (Sec.\ \ref{sec:CMB}) and late-time cosmology (Sec.\ \ref{sec:late_time}).  The latter includes not just BAO and SNe~Ia but also weak lensing and galaxy clustering. \cosmobit can employ the help of two external libraries when computing likelihoods:
\begin{itemize}
  \item \plc \cite{Aghanim:2015xee,Aghanim:2019ame}: the official Planck likelihood code, which computes likelihoods for CMB temperature, polarisation, and lensing spectra (Sec.\ \ref{sec:CMB}, Appendix~\ref{app:plc-interface});
  \item \montepython \cite{Audren:2012wb,brinckmann2018montepython}: a package for cosmological parameter inference providing an extensive database of cosmological likelihoods (Sec.\ \ref{sec:late_time}, Appendix~\ref{app:mp-interface}).

  \end{itemize}
To calculate these likelihoods, \cosmobit requires theoretical predictions for the relevant \textbf{observables}. We employ the following public external codes as \textbf{backends} for these calculations:
\begin{itemize}
  \item \alterbbn \cite{Arbey:2011nf,Arbey:2018zfh}: computes the abundances of light elements formed during BBN for different non-standard physics scenarios (Sec.\ \ref{sec:BBN}, Appendix~\ref{app:alterbbn-interface});
  \item \class \cite{Blas:2011rf}: the cosmic linear anisotropy solving system. \class solves the Boltzmann equations in order to determine the evolution of the linear power spectrum, primarily during the recombination epoch (Appendix~\ref{app:class-interface});
  \item \darkages \cite{Stocker:2018avm}: a tool for computing the efficiency of energy injection by exotic particle decays or annihilations. The computed efficiencies are used by the \class extension \exoclass to determine the impact of exotic electromagnetic energy injection on the CMB (Appendix~\ref{app:darkages-interface});
  \item \multimodecode \cite{Mortonson:2010er,Price:2014xpa}: a tool for computing primordial power spectra for different (multifield) inflation models (Appendix~\ref{app:multimode-interface}).
  \end{itemize}

Some backends can help to decrease the runtime of a fit if a scanner supporting the use of \emph{fast-slow} parameters is employed.  This functionality allows different sets of parameters to be changed with a different frequency, such that `fast' (typically nuisance) parameters -- used for calculations that evaluate quickly -- are changed far more frequently than `slow' parameters necessary for calculations requiring a longer runtime.  If from one point in parameter space to the next, no parameter changing the outcome of a calculation is modified, the calculation is skipped; instead, the previously-computed results are reported. \alterbbn, \class, \exoclass, \darkages and \montepython currently support this feature. At the moment the only available fast-slow scanner in \gambit is \polychord. However, a native support of the fast-slow functionality is under development and will be released in a future version of \gambit, allowing this scanning technique to be used for all likelihood calculations.

An overview of the different backends and their purposes is given in Table~\ref{tab:CB-backend-ov}.
Note, however, that the modularity of \gambit allows the user to replace any of these backends by a different external library of their choice. For example, instead of using \class for the computation of cosmological observables, one could implement an interface to \textsf{CAMB}~\cite{Lewis:1999bs} or \textsf{PyCosmo}~\cite{Tarsitano:2020ddh}. Likewise, the calculation of the primordial element abundances could easily be provided by e.g.~\textsf{PArthENoPE}~\cite{Pisanti:2007hk,Consiglio:2017pot} or \textsf{PRIMAT}~\cite{Pitrou:2018cgg} instead of \alterbbn.
%

\begin{table}[tb!]
{ \centering
 \small
   \makebox[\linewidth]{
  \begin{tabular}{l p{6cm} c p{3.5cm}}
    \toprule
    Backend        & Capabilities \& purpose                           &  Fast-slow?  &  Model group(s)                                     \\
    \midrule
    \alterbbn      &  calculate light element abundances               & \cmark       &  \textsf{\color{PatrickBlue}Standard Cosmology}     \\
                   &   $\rightarrow$ helium abundance for \class run   &              &  \textsf{\color{PBCompNeutrinos}Neutrino Masses}    \\
                   &   $\rightarrow$ BBN Likelihood                    &              &  \textsf{\color{PBCompBlue}Extra radiation}         \\
    \midrule
    \class         &  calculate cosmological observables               & \cmark       &  \textsf{\color{PatrickBlue}Standard Cosmology}     \\
                   &   $\rightarrow$ Planck likelihoods                &              &  \textsf{\color{PBCompNeutrinos}Neutrino Masses}    \\
                   &   $\rightarrow$ \montepython likelihoods          &              &  \textsf{\color{PBCompBlue}Extra radiation}         \\\noalign{\vskip 2pt}
    \exoclass      &  as for \class, but can also be used with
                      annihilating and decaying dark matter models              & \cmark       & +\textsf{\color{PBCompTurquoise}Energy Injection}   \\
    \midrule
    \darkages      &  calculate spectrum and efficiency of energy
                      injection into CMB                               & \cmark       &  \textsf{\color{PBCompTurquoise}Energy Injection}   \\
                   &   $\rightarrow$ passed to \exoclass               &              &                                                     \\
    \midrule
    \montepython   &  calculate cosmological likelihoods               & \cmark       &  \textsf{\color{PBCompNuisance}Nuisance parameters} \\
    \midrule
    \multimodecode &  calculate primordial power  spectrum             & \xmark       &  \textsf{\color{PBCompInflation}Inflation}          \\
                   &  $\rightarrow$ passed to \class                   &              &                                                     \\
    \midrule
    \plc           &  calculate Planck likelihoods                     & \xmark       &  \textsf{\color{PatrickBlue}Standard Cosmology}     \\
                   &                                                   &              &  \textsf{\color{PBCompNuisance}Nuisance parameters} \\
    \bottomrule
  \end{tabular}
  }
  \caption{Overview of the \textbf{backends} (external codes) called by \cosmobit.  For each backend, the role in calculations within \gambit is given. The table also indicates which backends can exploit the use of \emph{fast-slow} parameters by skipping their respective calculation if no input parameter has changed from one point to the next in parameter space.  The last column provides a list of model groups from which a backend is able to make use of model parameters.
  \label{tab:CB-backend-ov}}
  }
\end{table}

\section{The cosmological standard scenario}
\label{sec:lcdm}

The current concordance model for cosmology is the flat $\Lambda$CDM universe, typically defined
in terms of 6 independent physical parameters.
Here we will mostly follow the same convention as common in the literature
and define them
as the density of baryons ($\omega_{\rm b}$) and cold dark matter ($\omega_{\rm cdm}$),
the present Hubble expansion rate ($H_0$),
the optical depth
at reionisation ($\tau_{\rm reio}$), and the amplitude ($A_s$) and spectral index ($n_s$)
of the primordial scalar perturbations.  These parameters are shown in the ``Standard Cosmology'' section of Table \ref{tab:cosmo_params}.  Note that in some analyses (e.g.\ by Planck), $H_0$ is swapped for the angular acoustic scale $\theta_*$.
We assume a flat Universe (curvature $\Omega_k=0$) as the combination of CMB and BAO data leads to $\Omega_k=0.0007 \pm 0.0019$, consistent with zero~\cite{Aghanim:2018eyx}.

In this Section, for convenience we briefly recap the definitions of
these parameters and the main features of $\Lambda$CDM cosmology.
We focus on what is currently implemented in \cosmobit; for more thorough reviews of
standard cosmology, see e.g. Refs.~\cite{Lahav:2019bbc,PDG20}.

\subsection{Background evolution} 
\label{sec:lcdm_bg}

The assumptions of homogeneity and isotropy on large scales, i.e. the cosmological principle,
uniquely determines that the geometry of the Universe is described by the
Friedmann-Lema\^itre-Robertson-Walker (FLRW) metric~\cite{1922ZPhy...10..377F, 1933ASSB...53...51L, Robertson:1935zz, 1933ZA......7..153R, PDG20}.
With this ansatz, Einstein's field equations yield the two Friedmann equations
governing the evolution of the scale factor $a(t)$ (equivalently the redshift $z\equiv a_0/a-1$, where $a_0 \equiv a(t=0) = 1$ is the scale factor today).
The first of these states that the Hubble expansion rate,
$H\equiv \dot a/a$ (where $\dot{a} \equiv \mathrm{d}a/\mathrm{d}t$), is proportional to the square root of the total energy density; it can be written as
\be \label{eq:FriedmannEq}
\sum_i \rho_i/\rho_c = 1+ k/\dot a^2 = 1\,,
\ee
where the last equality follows from assuming a flat universe ($k=0$). Here
$\rho_{\rm c}\equiv 3\, \mplred^2 H^2$
denotes the critical density required for exact flatness,
with the reduced Planck mass $\mplred=(8\pi G)^{-1/2}$,
and $\rho_i$ is the energy density of cosmological
component $i$, including that of the cosmological constant $\Lambda$.
The second Friedmann equation describes cosmic acceleration, stating that $\ddot a$ is proportional
to the trace of the stress-energy tensor, $T^\mu_{~\mu}=\sum_i (\rho_i+3p_i)$,
where $p_i$ is the pressure of component $i$.

The stress-energy tensors of non-interacting components are conserved separately,
\mbox{$\nabla_\nu T^{\mu\nu}=0$}.
For a constant equation of state, $w_i\equiv p_i/\rho_i$, the respective energy density then
dilutes as
  $\rho_i\propto a^{-3(w_i+1)}$ \label{eq:friedmann}
with the expansion of the Universe. This implies thermalisation; in standard Big Bang cosmology
the early Universe was dominated by radiation
($w_r=1/3$), followed by an era of matter domination ($w_m=0$), before entering the regime where
the cosmological constant $\Lambda$ dominates ($w_\Lambda=-1$).
Today, only a small fraction of the total energy content comes from radiation, i.e. relativistic particles.
The major contributions come from dark energy and two forms of non-relativistic matter, namely dark matter and baryons.
The contributions of the matter components are often parametrised in terms  of
$\omega_i\equiv \Omega_i \, h^2$, with $\Omega_i \equiv\left( \rho_i/\rho_c\right)_0$
and $h$ given by the present Hubble expansion rate as
$H_0\equiv 100\, h\, {\rm km}\,{\rm s}^{-1}\,{\rm Mpc}^{-1}$.
From the values of the energy densities in matter and radiation today, one can infer the contribution from the cosmological constant.  Using Eq.~(\ref{eq:friedmann}) and assuming the Universe is flat leads to $\omegalambda= (h^2-\omegab-\omegacdm-\omegar)\simeq (h^2-\omegab-\omegacdm)$.

The radiation density at late times receives contributions from the CMB and any other relativistic species. These additional relativistic species are commonly parametrised in terms of an effective number $N_{\rm eff}$ of fermionic degrees of freedom with temperature $\Tnu=(4/11)^{1/3}\,\Tgamma$:
\be \label{eq:rhor-late}
 \rho_r
 =\left[1+ \frac78 \left(\frac{4}{11}\right)^\frac{4}{3}N_{\rm eff}\right]\frac{\pi^2}{15}T_\gamma^4\,,
\ee
where $\Tgamma$ is the photon temperature (for $z=0$ equal to the measured CMB temperature $T_{\gamma 0} =\Tcmb$).
The standard choice for $\Lambda$CDM is $N_{\rm eff}=3.045$ [\citenum{deSalas:2016ztq}; see also \citenum{Akita:2020szl,Froustey:2020mcq}], slightly higher than the three generations of neutrinos in the SM. The deviation from 3 follows from a precise calculation of the neutrino temperature after decoupling, where corrections arise from the fact that neutrinos do not decouple instantaneously and therefore receive some energy from annihilating electron-positron pairs.

In the literature, it is common to fix $\Tcmb$ to the value determined from measurements of the CMB monopole by the COBE/FIRAS mission~\cite{Mather:1998gm} and the recalibration using WMAP velocity data to $\Tcmb = 2.72548 \pm 0.00057$\,K~\cite{Fixsen2009}. In principle, however, $\Tcmb$ is a free parameter of $\lcdm$. Given the constraining power of current datasets, neglecting the errors on $\Tcmb$ is justified; however, with upcoming, more precise data this simplification can lead to biases in the other cosmological parameters~\cite{Yoo:2019dyl}. In addition, recent observed discrepancies between early and late-time measurements of the Hubble constant can be fully recast into a tension between the photon temperature today and at recombination, see Ref.~\cite{Ivanov:2020mfr} (see Sec.~\ref{sec:tensions} for a discussion on the Hubble tension).
For these reasons, we include $\Tcmb$ as a model parameter of $\lcdm$ in \cosmobit. This allows the user full flexibility in the treatment:  $\Tcmb$ can either be varied as model parameter and constrained by the data or fixed to a given value.

During most of the (standard) cosmological evolution, entropy is largely conserved.
This allows the scale factor to be related to the photon temperature as
\be
 T_\gamma \propto \left(g_{\rm eff}^s\right)^{-1/3} a^{-1},
\ee
where $g_{\rm eff}^s(T)$ is the effective number of active entropic relativistic degrees of freedom at temperature $T$,
\be
  g_{\rm eff}^s(T) \approx \sum_{B} g_B \left(\frac{T_B}{\Tgamma}\right)^4 + \frac{7}{8}\sum_{F} g_F \left(\frac{T_F}{\Tgamma}\right)^4
\ee
where the subscripts $B$ and $F$ denote boson and fermion degrees of freedom respectively.
This relation also implies that the Big Bang was hot, as it describes the initial condition of
the Universe as a plasma at high temperature. As the plasma cooled with the expansion,
light elements started to form at temperatures around $T_\gamma\sim1$\,MeV during the epoch of Big Bang Nucleosynthesis (BBN).
Measurements of primordial abundances of these light elements in today's Universe provide the current earliest direct cosmological probe (see Section \ref{sec:BBN}).

Much later, at temperatures $T_\gamma\sim1$\,eV (corresponding to a redshift of $z_{\rm rec}\sim1100$),
protons and electrons ``re''-combined to form neutral hydrogen. At that point, the photons were thus no
longer tightly coupled to a charged plasma, and freely propagated until the Universe reionised due
to the first starlight around $z_{\rm reio}\sim10$. These freely propagating photons form the present observable CMB. Observables related to the CMB photons allow precision inference of cosmological
parameters (see Section \ref{sec:CMB}). The most important parameter related to astrophysical processes -- but still included in the list of $\Lambda$CDM parameters -- is the effective optical depth at reionisation,
$\tau_{\rm reio}$. It describes the partial reabsorption of CMB photons by the plasma in the late Universe.

\subsection{Perturbations} \label{sec:perturbations}

Even the very early Universe was  not {\it perfectly} isotropic and homogeneous. Instead, tiny primordial perturbations imprinted on the
FLRW metric (and hence the stress-energy tensor) grew under the influence of gravity
at all scales smaller than the horizon.
These perturbations manifested themselves  in the CMB anisotropy spectrum
and, later, in all cosmological structures visible today.

The power spectrum of these primordial perturbations is usually treated as an initial condition
that constitutes part of the $\Lambda$CDM model. In particular, an excellent fit to the data
is obtained by simply describing the spectrum of scalar perturbations in terms of only two
free phenomenological parameters, $A_s$ and $n_s$. These parameters quantify the amplitude and spectral index of a scale-free power spectrum, given by
\be
\label{Psdef}
\calP_\zeta(k) \equiv\frac{k^3}{2\pi^2}\langle|\zeta_k|^2\rangle \simeq A_s\left(\frac{k}{k_\star}\right)^{n_s-1}\,,
\ee
where by default we choose a comoving pivot scale of $k_\star=0.05\,\rm{Mpc}^{-1}$.  A value of $n_s=1$ corresponds to a fully scale-invariant spectrum; the best-fit value in $\lcdm$ is only slightly smaller than this.
Here, we express scalar perturbations in terms of the (gauge-invariant) curvature perturbation on uniform-density hypersurfaces
$\zeta\equiv -\Psi-H\,\delta\rho/\dot{\bar\rho}$~\cite{Bardeen:1983qw}, where $\Psi$  is a scalar metric perturbation related to the intrinsic Ricci scalar curvature of constant time hypersurfaces as $R^{(3)} = 4\nabla^2 \Psi/a^2$. On super-Hubble scales, $k\ll aH$, $\zeta$ has a single constant non-vanishing adiabatic solution~\citep{Weinberg:2003sw} which satisfies Eq.~(\ref{Psdef}).\footnote{Another gauge-invariant measure of scalar perturbations (often adopted  in the study of inflation, for example) is the comoving curvature perturbation $\mathcal{R}=\Psi-H/(\bar{\rho}+\bar{p})\delta q$, where $\delta q$ is a perturbation to the 3-momentum field. Note that $\mathcal{R}$ is equal to $\zeta$ up to a term scaling with $[k/(aH)]^2$ which vanishes on super-Hubble scales (for a review, see e.g.\ Ref.\ \cite{Baumann:2009ds}).

Note that if present, multiple dynamical degrees of freedom can also contribute non-vanishing isocurvature fluctuations, which can evolve on super-Hubble scales. The \multimodecode package, to which \cosmobit is interfaced, provides some capacity for evaluating these effects during inflation, and will be discussed in detail in the following sections.}

The most widely-accepted cosmological scenario for generating such primordial perturbations
is inflation, an early epoch of accelerated expansion of space where sub-Hubble quantum fluctuations are
driven beyond the Hubble radius (and in the progress become classical perturbations, amplified to 
macroscopic scales).
As the Universe continues to expand after inflation ends, these perturbations then progressively `re-enter the 
horizon' (in the sense that they become sub-Hubble, $k\gtrsim aH$).
In particular, the simplest realisation of slow-roll inflation is capable of producing, to leading
order, perturbations of the type just discussed (see Section \ref{sec:inflation} for
possible extensions).

Well within the Hubble scale, it is customary to instead express the scalar perturbations in terms
of the density contrast, $\delta\equiv \delta \rho/{\bar\rho}$. As the gauge-dependence of $\delta$ is negligible on sub-Hubble scales (for standard gauge choices), its  interpretation is more intuitive than that of $\zeta$.
The difference between $\delta$ and $\zeta$ is parametrised by the transfer function $\mathcal{T}$, which explicitly factors out the time-dependence of perturbations under the effect of gravity as the Universe expands. It is defined as
\be
\label{eq:Pm}
 \left\langle|\delta_i(t,k)|^2\right\rangle \equiv \mathcal{T}^2_i(t,k)\,\calP_\zeta(k)\,.
\ee
Different components of the Universe $i$ generally evolve differently as they become smaller than the Hubble scale.
In the linear regime, $\delta_i\ll 1$, the transfer function is obtained by solving the equations
of motion resulting from the linearised field equations.
In the following, we briefly recap how such small perturbations in the main $\lcdm$ components evolve
during the epochs of radiation, matter and cosmological constant domination.

During radiation domination, adiabatic perturbations have a relative normalisation between radiation and matter of 
$\delta_r/\delta_m=4/3$ when $k\lesssim aH$.  Any density contrast in the radiation fluid
then quickly decays. Baryons are tightly coupled to the photons
in this regime, $\delta_b\simeq\delta_\gamma$, but perturbations in a non-interacting CDM component
will grow logarithmically (and in fact somewhat faster closer to matter-radiation equality, as governed by
the M\'esz\'aros equation~\cite{Meszaros:1974tb}).

During matter domination, linear CDM perturbations then start to grow at the same rate as the universe, $\delta_{\rm cdm}(t)\propto a(t)$, while radiation and baryon perturbations oscillate and quickly decay once they become sub-Hubble.
Pressure terms in the photon-baryon fluid lead to characteristic oscillations visible in the CMB (Section \ref{sec:CMB}). Soon after recombination, the baryons are released from the photons and start to fall into the gravitational potentials built up by the CDM perturbations, until $\delta_b\simeq \delta_{\rm cdm}$.
In the linear regime, the CDM and baryonic density contrasts continue to grow at the same pace until the onset of
$\Lambda$ domination, which stalls this growth.

Once a perturbation at a given scale $k$ starts to approach the non-linear regime, $\delta_i^k\sim1$,
the linear equations of motion can no longer adequately describe the non-linear evolution, which eventually
leads to the gravitational collapse and virialisation of (proto-)halos.
In $\lcdm$, structure formation proceeds hierarchically,
with the smallest halos forming first and later merging to form larger halos.

At the scale of galaxy clusters, the matter power spectrum
defined in Eq.~(\ref{eq:Pm}) is still only in the mildly non-linear regime today.
On small non-linear scales, the matter power spectrum carries valuable information for cosmology. For example, it is sensitive to the mass and number of neutrinos, and to the effects of gravity on galactic scales.  However, the modeling of the matter power spectrum on these small scales is subject to uncertainties related to baryonic feedback in the process of galaxy formation. Accurate modelling of these non-linearities is therefore important when using small-scale observables to test cosmological models (see e.g.~Ref.~\cite{Chisari:2019tus}).
Observables on large, linear scales are less affected by these modeling uncertainties and, hence, are commonly used to constrain cosmological parameters.

Such observables include the scale of baryon acoustic oscillations (BAOs)-- 
fluctuations in the matter density, inherited from acoustic oscillations in the photon baryon fluid, and revealed by galaxy surveys -- and cluster counts (see Section \ref{sec:late_time}).

\begin{table}
{\small
\begin{tabular}{L{2.5cm} p{7.8cm}p{1.5cm}p{1.8cm}}
  \toprule
  Likelihood & Brief description & No. of nuisances & External\,\,\,\, library \\
  \midrule
  \textbf{BBN}      & Light element abundance measurements. Contains $\prescript{4}{}{\rm He}$ \cite{PDG20} and deuterium \cite{Cooke:2017cwo} by default; can also include $\prescript{3}{}{\rm He}$ and $\prescript{7}{}{\rm Li}$. See Sec.\ \ref{app:alterbbn-interface} for details. & 0 & \alterbbn \\
  \midrule
  \textbf{CMB}      & CMB Likelihoods from \emph{Planck}, both 2018~\cite{Aghanim:2019ame} and 2015~\cite{Aghanim:2015xee}. See Sec.\ \ref{app:plc-interface} for details. &  &   \\
  \emph{Low-$\ell$}       & \emph{Planck low-$\ell$} likelihoods, TT, EE, TTEE. & 1 &    \plc\\
    \emph{High-$\ell$}      & \emph{Planck high-$\ell$} likelihoods, TT, (TTTEEE),  & 16 (34) &    \plc\\
                            & or `lite' version. & 1 (`lite') &  \plc \\
    \emph{Planck} lensing   & \emph{Planck} lensing likelihoods. & 1  &  \plc \\
    BK14       & Likelihoods from BICEP2/KECK array (2014) \cite{Array:2015xqh}. & 10 &    \montepython \\
    POLARBEAR & Likelihood from measurement of CMB $B$-Modes with POLARBEAR (2014) \cite{Ade:2014afa}. & 0 & \montepython \\
  \midrule
  \textbf{SNe~Ia}     & Type Ia supernovae. &  &    \\
  JLA       & Joint light-curve analysis (JLA) between SDSS-II and SNLS supernovae \cite{Betoule:2014frx}.  & 4 & \montepython \\
  Pantheon & Likelihood from combined Pantheon sample~\cite{Scolnic:2017caz}. & 1 & \montepython  \\
  \midrule
    \textbf{Galaxy Clustering}    & Surveys analysing the clustering of galaxies.  &  &    \\
   BAO & Correlated distance measurements using BAO from the combined SDSS BOSS-DR12 sample~\cite{Alam:2016hwk}, and eBOSS-DR14 LRGs~\cite{Ata:2017dya} and QSOs~\cite{Bautista_2018} (as described in~\cite{CosmoBit_numass}). Also 6dF~\cite{2011MNRAS.416.3017B}, SDSS-MGS~\cite{2015MNRAS.449..835R}, WiggleZ \cite{Kazin:2014qga}, DES-Y1~\cite{Abbott_2018} and earlier data releases. & 0 & \montepython \\
    Growth rate of structure & Growth rate of large scale structure measurements from BOSS-DR12 (correlated with BAO measurements)~\cite{Alam:2016hwk}. & 0 & \montepython \\
   Power spectra & Binned measurements of the galaxy power spectra from SDSS-DR4~\cite{Tegmark:2006az} and DR7~\cite{Reid:2009xm}, and WiggleZ~\cite{Parkinson:2012vd} & 0 & \montepython \\

  \midrule
    \textbf{Weak lensing}   & Weak lensing likelihoods. & & \\
    KiDS & Likelihood from KiDS tomographic weak lensing survey, from Ref.~\cite{Kohlinger:2017sxk}. &  &  \montepython  \\
    CFTHLenS & Likelihood from CFHTLenS \cite{Heymans:2013fya}. & 1 & \montepython \\
    \midrule
    \textbf{Misc.} & & & \\
    HST & Hubble space telescope likelihood on the Hubble parameter \cite{Riess:2016jrr}. & 0 & \montepython \\
    ISW & Likelihood based on tomographic analysis of the integrated Sachs-Wolfe effect \cite{Stolzner:2017ged}. & 16 & \montepython \\
    Time delay & Measurements of Quasar lensing time-delays~\cite{Suyu_2013}. & 0 & \montepython \\

    \bottomrule
\end{tabular}
}
\caption{
The most relevant likelihoods available in \cosmobit. For details on adding likelihoods to the \YAML file, we refer the reader to Appendix~\ref{app:backends}.
\label{tab:likelihoods}
}
\end{table}
\afterpage{\clearpage}

%
\section{Observational probes and likelihoods} \label{sec:lcdm-obs}
%

In this section, we review the most important observational probes from the early Universe and late-time
cosmology. Further, we present the likelihoods currently available for use in \cosmobit. These measurements cover a range of different cosmological epochs and will be discussed in chronological order. Table~\ref{tab:likelihoods} includes an overview of a selection of likelihoods provided in the first release of \cosmobit.

In the following we will present the results from a variety of scans carried out with \cosmobit. These serve both as an illustration of the many different applications of \cosmobit and as validation of our implementation of the different models and backend interfaces. Unless explicitly stated otherwise, we present results in the Bayesian framework, i.e.\ in terms of posterior probabilities and credible regions. To obtain these results we use the \multinest scanning algorithm~\cite{Feroz:2007kg,Feroz:2008,1306.2144}. Our priors for the $\Lambda$CDM parameters are given in Table~\ref{tab:lcdm_priors}. We emphasize that since these parameters are tightly constrained by data, our results are largely insensitive to these priors.

\begin{table}[t]
\centering
\begin{tabular}{l l l}
\toprule
Param. & Range & Prior \\
\midrule
$H_0 $                            & [50, 80]\,km\,s$^{-1}$\,Mpc$^{-1}$ & flat \\
$\omega_\mathrm{b} $                    & [0.020, 0.024] & flat \\
$\omega_\mathrm{cdm} $                  & [0.10, 0.15] & flat \\
$\mathrm{ln}\left(10^{10}\,A_s\right)$  & $[2.5,\ 3.5]$ & flat \\
$n_s $                                  & [0.90, 1.10] & flat \\
$\tau_\text{reio} $                                 & [0.004, 0.20] & flat \\
\bottomrule
\end{tabular}
\caption{Priors adopted for 6 free parameters of $\Lambda$CDM.}
\label{tab:lcdm_priors}
\end{table}

\subsection{Light elements from Big Bang Nucleosynthesis} \label{sec:BBN}

Big Bang Nucleosynthesis (BBN; see e.g.\ \cite{Iocco:2008va,Coc:2017pxv} for reviews) refers to the formation of light elements such as deuterium, helium, lithium and beryllium in the early Universe, starting about $3$ minutes after the Big Bang. This corresponds to temperatures of $T \sim 100\,\mathrm{keV}$. The resulting primordial abundances of the different elements can be predicted using well-understood SM physics.  Abundance measurements of these elements in environments such as pristine gas clouds and very metal-poor stars allow us to estimate the primordial values. By comparing the predictions and observations, BBN thus provides the observational probe of cosmology that reaches the furthest back in time of all those at our current disposal.

The predicted element abundances are, for the most part, in excellent agreement with the primordial abundances inferred from observations, validating the standard picture of a hot Big Bang cosmology consistent with SM particle physics. In turn, every non-standard modification to this scenario is strongly constrained by data.

The formation of light elements is driven by a coupled system of nuclear reactions. This reaction chain starts with the formation of deuterium through the combination of a proton and a neutron, i.e.
\be
\label{eq:D-formation}
n + p \to D + \gamma \, .
\ee
After that, the formation of $ ^3{\rm He} $ and $ ^4{\rm He} $ follow through the reactions
\be
\label{eq:He-formation}
D + D \to n + \prescript{3}{}{\rm He} \quad\text{and}\quad D + \prescript{3}{}{\rm He} \to \prescript{4}{}{\rm He} + p \,.
\ee
After the synthesis of stable helium ($ ^4{\rm He} $), a more complex reaction network is needed to achieve the formation of beryllium, lithium and other subsequent light elements.

Two important quantities dictate the importance of a specific nuclear reaction at a given time: the expansion rate of the Universe, and the reaction rate $ \Gamma \sim \langle\sigma v\rangle n $, where $n$ is the number density of the reagents and $\langle\sigma v\rangle$ the thermally-averaged, velocity-weighted cross-section for the reaction. The expansion rate is given by $ H \sim 0.33 \sqrt{g_{\rm eff}} \, T^2/\mplred$, where $ g_{\rm eff}$ denotes the effective number of relativistic degrees of freedom. At high temperatures, the reaction rate is typically larger than the expansion rate and the process in question is in thermal and chemical equilibrium.

In this equilibrium state, the formation of stable elements through a complex reaction network is not efficient, as any element produced in a specific reaction will be dissociated by the inverse reaction. However, once the reaction rate decreases below the expansion rate of the Universe, the reaction will drop out of chemical equilibrium and favour one specific direction until the number densities of the elements in the reaction eventually freeze out.

\begin{figure}
  \begin{center}
    \includegraphics[width=0.72\textwidth]{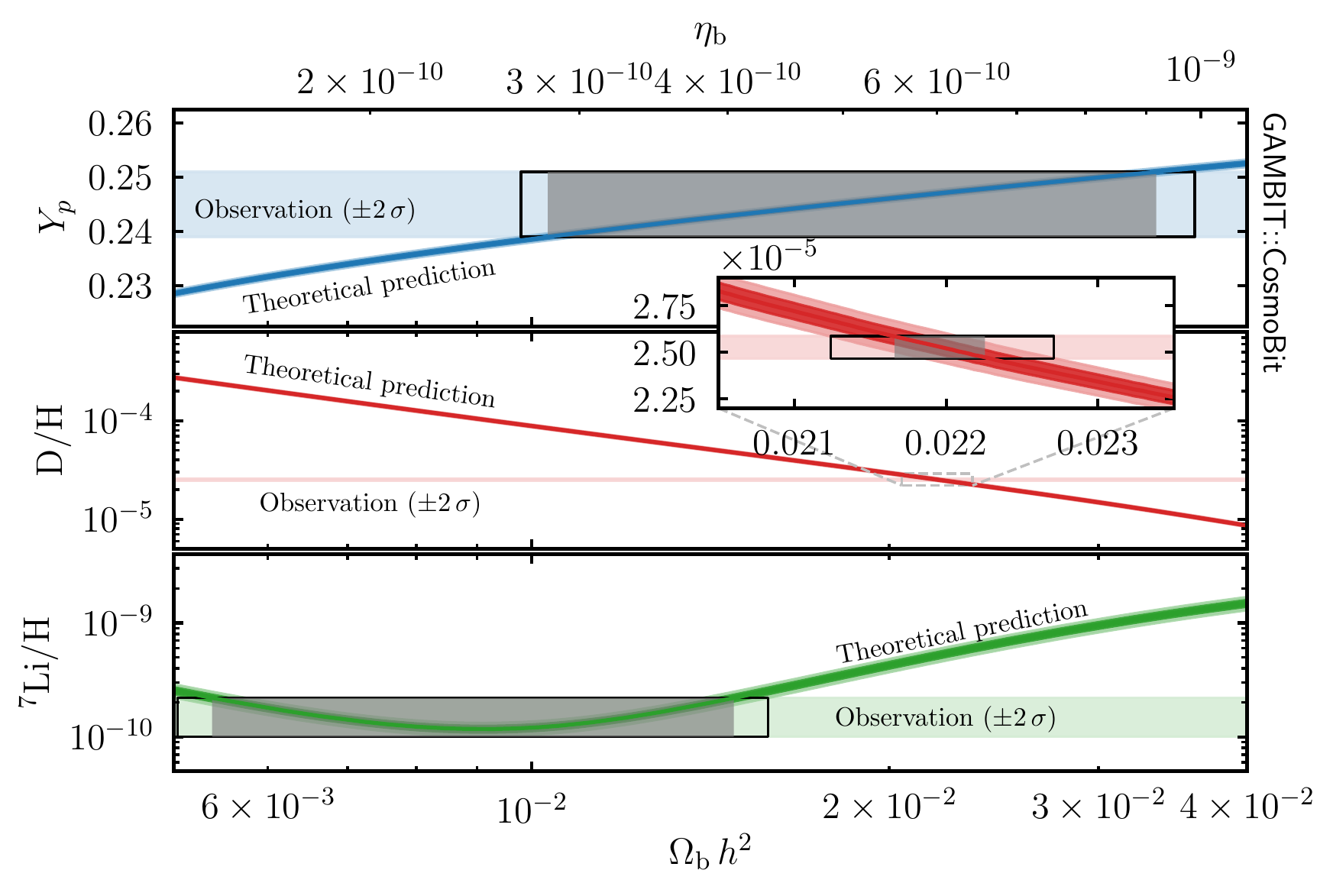}\\
    \includegraphics[width=0.48\textwidth]{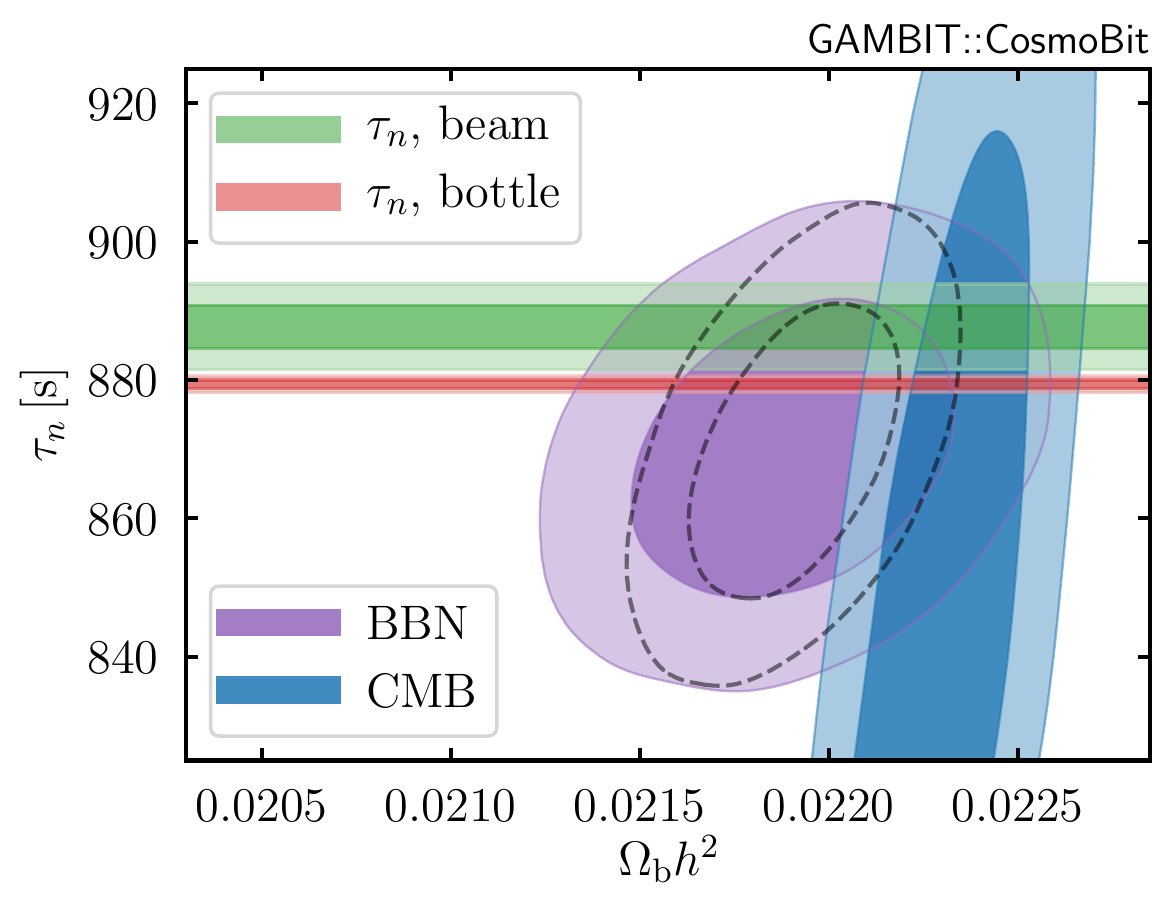}
    \caption{\label{fig:BBN_plots} Comparison of cosmological and astrophysical probes of BBN. {\it Top}: Curved bands indicate the predicted primordial abundances and theoretical uncertainties of helium 4, deuterium and lithium 7 as a function of $ \Omega_{\rm b}h^2 $ (equivalently, $ \eta_\mathrm{b}$) through linear progation of the uncertainties in the reaction rates.  Inner dark shading indicates the $ 1\sigma $ range, and lighter shading indicates the $ 2\sigma $ range. Horizontal bands show the 95\% confidence ranges of the observed abundances~\cite{PDG20,Cooke:2017cwo}. Solid black outlines roughly indicate the inferred uncertainty range on $ \Omega_{\rm b}h^2 $ when theoretical uncertainties on the prediction are accounted for, whereas grey regions show the inferred uncertainty range when theoretical uncertainties on the predictions are neglected. {\it Bottom}: Constraint on $ \Omega_{\rm b}h^2 $ and the neutron lifetime $ \tau_N $ from BBN and CMB data. Horizontal bands refer to the laboratory measurements of the neutron lifetime from ultra-cold neutron (`bottle') experiments \cite{PDG20} and in-flight decays (`beam' experiments) \cite{Yue:2013qrc}. The dashed contour indicates the BBN bound if theoretical uncertainties are neglected.}
  \end{center}
\end{figure}

For this reason, the formation of light elements is very sensitive to modifications of the physics at temperature of around $ T \sim 0.1 \text{--} 1\,\mev $. In particular, any modification to the energy content of the Universe will change its expansion rate and thus affect the final abundances of the light elements. Such modifications include extra relativistic species as well as scenarios of early dark energy. Reaction rates can also be altered by modifying initial conditions such as the baryon-to-photon ratio $\eta_\mathrm{b} = n_b / n_\gamma$ or the (normally absent) neutrino chemical potential, both of which affect the initial number densities of neutrons and protons. We show in the left panel of Fig.~\ref{fig:BBN_plots} how the predicted abundances of various elements depend on $\Omega_\mathrm{b} h^2$ (or equivalently $\eta_\mathrm{b}$) as well as the constraints on these parameters obtained from measurements.

In addition to cosmological uncertainties, the reaction rates are affected by uncertainties in nuclear cross-sections and the lifetime of the neutron. These uncertainties lead to sizeable theoretical uncertainties in the predicted elemental yields, which need to be taken into account when defining the likelihood. As the elements are synthesised through a complex reaction network, these theoretical uncertainties will, in general, be correlated with each other. The bottom panel of Fig.~\ref{fig:BBN_plots} illustrates the effect of varying the neutron lifetime $\tau_n$ on the value of $\Omega_\mathrm{b}$ inferred from BBN and Planck. We include the laboratory measurements and their uncertainties on the neutron lifetime $ \tau_n  $ both from ultra-cold neutron (`bottle') experiments ($\tau_{n\,,\text{bottle}} = 879.4 \pm 0.6\,\mathrm{(stat)} \,\mathrm{s}$) \cite{PDG20} and in-flight decays (`beam' experiments) ($\tau_{n\,,\text{beam}} = 887.7 \pm 1.2\,\mathrm{(stat)} \pm 1.9\,\mathrm{(syst)} \,\mathrm{s}$) \cite{Yue:2013qrc} as horizontal bands. On the one hand, this plot can be interpreted as an independent measurement of the neutron lifetime from cosmology, confirming the value measured in the laboratory. On the other hand, it illustrates the non-negligible impact of the experimental uncertainty in the neutron lifetime, which may lead to a systematic shift in the inferred central values of cosmological parameters. This effect is particularly interesting given the tension between the value of $\tau_n$ inferred from beam experiments and bottle experiments, which may be resolved by Physics beyond the Standard Model~\cite{Fornal:2018eol,Cline:2018ami}.

Finally, we note that the primordial elemental abundances also impact the initial conditions for subsequent observational probes. A notable example is the primordial helium fraction $ \yhe \equiv \rho_{^4{\rm He}} / \rho_b $, which is an important initial condition for the physics of recombination.

\subsubsection*{Likelihood}

To obtain theoretical predictions of the abundances of the light elements, the first release of \cosmobit provides an interface to \alterbbn \cite{Arbey:2011nf,Arbey:2018zfh}.  \alterbbn enables the user to study the effect of various non-standard cosmological scenarios on the abundances of light elements. It offers sophisticated methods to include correlations and uncertainties in the nuclear reaction rates in the abundance and corresponding likelihood calculations. For a detailed description of the interface to \alterbbn we refer the reader to Appendix \ref{app:alterbbn-interface}.

In practice, the covariance matrix describing the correlated theoretical uncertainties on each of the primordial abundances is computationally too expensive to compute for every parameter point in a global fit. We have therefore implemented an alternative fast method to include the correlations in the likelihood calculation. It turns out that the relative uncertainties as well as the correlation coefficients for the elements are almost constant when the relevant inputs are varied. It is therefore a reasonable approximation to fix the correlation coefficients and the relative uncertainties for all elements included in the likelihood and construct the covariance matrix by only calculating the central values with \alterbbn.
For the fast method, we assume a relative theoretical uncertainty in the helium abundance $ \yhe $ and deuterium abundance $ D/H $ of
\be
\frac{\sigma_{\yhe}}{\yhe} = 1.348\times10^{-3} \quad\text{and}\quad \, \frac{\sigma_{D/H}}{D/H}  = 1.596\times10^{-2}\,.
\ee
Additionally, we assume a correlation between $ \yhe $ and $ D/H $ of $ 1.524\,\% $. We base these values on the results we obtain from the built-in uncertainty estimates of \alterbbn for various scenarios, including estimates for the correlations between the abundances, and choose the respective maximal absolute values of these scenarios to remain conservative with our uncertainty estimates. We have explicitly verified that this treatment is not only a good approximation for $\lcdm$ cosmologies but also holds for the various extensions of $\lcdm$ included in \cosmobit and relevant for BBN%
\footnote{To this end, we only consider non-standard values of $ \eta_\text{BBN} $ and $\Delta N_\text{ur,BBN}$ (see appendix \ref{app:models-dNur} for a details on these parameters) when determining the appropriate estimates for the relative uncertainties and their correlations, but we do not consider any variation of the neutron lifetime as this is implicitly included in the calculation of \alterbbn. Furthermore, we only consider scenarios which lie within $ \Delta{\rm log} \mathcal{L} \leq 10 $ from the best fit value to avoid large biases from scenarios, disfavoured by data. Further details can be found in Appendix \ref{app:alterbbn-err}.}.

Given a vector $\vec t$ of $n$ theoretically computed abundances, and the corresponding $n\times n$ covariance matrix $C_{\rm theo}$, the likelihood can be calculated by comparing to the observed abundances $\vec d$:
\begin{equation}
-2{\rm log} \mathcal{L} = (\vec t - \vec d)  \,C^{-1}_{\rm tot} \, (\vec t - \vec d)^\mathrm{T} \, + (2 \pi )^{n} \,  {\rm det} \, C_{\rm tot} \,,
\end{equation}
where the total covariance matrix contains the contributions from both theoretical uncertainties and measurement uncertainties, $C_{\rm tot} =  C_{\rm obs} + C_{\rm theo} $. As shown in Fig.~\ref{fig:BBN_plots} the theoretical uncertainties are non-negligible and substantially increase the allowed ranges of parameters inferred from BBN.

By default, we include the helium and deuterium abundances from Refs.~\cite{PDG20} and \cite{Cooke:2017cwo} respectively,
\be
\yhe = 0.245 \pm 0.003 \, , \qquad D/H = (2.527 \pm 0.030) \times 10^{-5} \; .
\ee
However, it is straightforward for the user to use different measurements or to include a different set of abundances in the likelihood calculation (see Appendix \ref{app:alterbbn-interface}). Of particular interest for future studies may be the abundance of $^7{\rm Li}$, which is in slight tension with the predictions of standard cosmology but also suffers from sizeable theoretical uncertainties.

\subsection{Cosmic Microwave Background} \label{sec:CMB}

After the formation of light elements, baryons and photons in the primordial plasma are coupled via Thomson scattering. 
The radiation pressure of photons causes the photon-baryon plasma to undergo acoustic oscillations. These oscillations are modulated by the effects of the plasma collapsing towards overdense regions through gravitational attraction, and by the expansion of the Universe.
These so-called acoustic oscillations propagate through the plasma at the speed of sound. About $380\,000$ years after the Big Bang, the plasma cools down such that electrons and protons combine to form neutral hydrogen. During this epoch of recombination, photons decouple from baryons and travel almost entirely freely through the Universe. These relic photons are still observable today and form the Cosmic Microwave Background (CMB) radiation. They have an almost perfect blackbody spectrum with a temperature of $T_{\rm CMB} = 2.7255$~K \cite{Fixsen2009} with anisotropies at the level of one part in $10^5$. These anisotropies carry imprints from the acoustic oscillations, primordial density fluctuations, the evolution of the gravitational potentials and today's structure of the Universe through lensing effects.

The CMB is completely described by its temperature anisotropy, $\Theta(\hat{\textbf{n}})$ and polarisation $P(\hat{\textbf{n}})$. The temperature anisotropy $\Theta(\hat{\textbf{n}})$ is defined as
\begin{equation}
  \Theta(\hat{\textbf{n}}) \equiv  \frac{T(\hat{\textbf{n}}) - T_{\rm CMB}}{T_{\rm CMB}} = \sum_{\ell = 0}^{\infty} \sum_{m = -\ell}^{\ell} a_{\ell m} Y_{\ell m}(\hat{\textbf{n}}) \,,
\end{equation}
where $T(\hat{\textbf{n}})$ is the value of the temperature field for a given position $\hat{\textbf{n}}$ on the sky. In the second equality, $\Theta(\hat{\textbf{n}})$ is decomposed in terms of the spherical harmonics $Y_{\ell m}(\hat{\textbf{n}})$ for a given integer multipole $\ell \geq 0$ and angular index $m = -\ell, \ldots, +\ell$, with the coefficients $a_{\ell m}$ giving the weight of each harmonic. The monopole temperature $T_{\rm CMB}$ corresponds to $\ell = 0$, and the dipole, from the Doppler shift caused by our motion relative to the CMB, to $\ell = 1$.

For stochastic quantum fluctuations in the early Universe, the resulting temperature fluctuations are Gaussianly distributed, with zero mean. The cosmologically-interesting information is, therefore, contained within the $2$-point correlations of the multipole moments $a_{\ell m}$, i.e.\ the angular power spectrum of perturbations, $C_\ell$. As the angular power spectrum is rotationally invariant, it depends only on $\ell$, not $m$. The theoretical expectation for the temperature power spectrum $C_\ell^{TT}$ can be written in terms of the temperature auto-correlation
\begin{equation}
  C_\ell^{TT} = \frac{1}{2\ell+1} \sum_{m} \langle {a_{\ell m} a^*_{\ell m}} \rangle\,,
\end{equation}
where $\langle\,\rangle$ denotes an ensemble average over many sky realisations. In reality we have only a single Universe to observe and for any given value of $\ell$, there are $2\ell+1$ measurements with which to constrain a given $C_\ell$, corresponding to different values of $m$. Thus, observations are fundamentally limited to measuring $2\ell+1$ independent modes. This gives rise to an unavoidable cosmic variance on large scales (small $\ell$), where very few realisations of the stochastic random fields are available, and the observed power spectrum may differ from its expectation value.

In addition to the temperature anisotropies in the CMB, polarisation of CMB photons provides additional information about the early Universe. As recombination does not occur instantaneously, the surface of last scattering has some finite thickness. CMB photons scatter off free electrons via Thomson scattering within the surface of last scattering. If the background radiation were isotropic, then the net result of the scatterings would introduce zero polarisation. However, any quadrupolar anisotropies in the spectrum of incoming photons lead to a linear polarisation of the resulting CMB radiation field. The polarisation can be separated into $E$ and $B$ modes, corresponding to `electric' (gradient) and `magnetic' (curl) components of the polarisation field.
Similarly to temperature perturbations, $E$ and $B$ modes can be expressed in terms of (spin-weighted) spherical harmonics and organised into power spectra.
The main origin of primordial \emph{E} modes are scalar perturbations, i.e.~perturbations in the density field; whereas at linear order, \emph{B} modes are sourced by tensor perturbations, i.e.~gravitational waves (such as from inflation). Hence, the polarisation power spectra of the CMB, $C_\ell^{EE}$ and $C_\ell^{BB}$, provide a probe for perturbations in the early Universe.  Cross-correlation spectra such as $C_\ell^{TE}$ provide an additional independent constraint, and are useful in calibrating foregrounds and instrumental effects.

In addition to the polarisation of the CMB, the second important consequence of non-instantaneous recombination is the damping
of power of temperature anisotropies on small scales. During the epoch of recombination, CMB photons will scatter multiple times with free electrons in a random walk process. The average distance a photon travels is the mean free path $\lambda_d$,
\begin{equation}
  \lambda_d \simeq \frac{1}{\sqrt{n_e \sigma_T H}} \,,
\end{equation}
where $\sigma_T$ is the Thomson scattering cross-section and $n_e$ is the number density of electrons. Through $n_e$, the damping scale depends both on the baryon density of the Universe and on the primordial helium fraction $\yhe$. Through the Hubble rate $H$, the damping scale is sensitive to cosmological parameters related to the expansion history of the Universe prior to recombination, including the effective number of free-streaming relativistic degrees of freedom, $N_{\rm eff}$.

The acoustic peaks are another important feature of the CMB power spectrum. In the early Universe baryon acoustic oscillations (BAOs) are sourced by the baryon-photon fluid, which oscillates in time and space with a period determined by its sound speed $c_s$.
The comoving distance travelled by a sound wave of the baryon-photon fluid by time $\eta$ is called the sound horizon, and is defined as
\be
r_s(\eta)=\int_{0}^{\eta}{\rm d} \eta' c_s(\eta')\,,
\ee
which is measured to be approximately $r_s\simeq150$\,Mpc in comoving length. The sound horizon sets the frequency of the harmonic series of modes. Modes caught at extrema of their oscillations source peaks in the CMB power spectrum. Peak positions are captured by the $\theta_*$ parameter, which depends on the angular diameter distance at which we observe the fluctuations, $D_A$, and the sound horizon, as $\theta_*=r_s/D_A$, allowing for the position of the first peak in the CMB power spectrum to give an extremely precise measurement of the geometry of the Universe at early times. Furthermore, analogous to mass oscillating on a spring, oscillations of the baryon-photon plasma in and out of the potential wells are sensitive to the ratio of the density of baryons to the overall density of matter. Increasing baryon density increases the rarefraction inside potential wells while correspondingly decreasing the compression away from the potential wells, changing the ratio of the amplitudes of even and odd peaks.

In addition to the intrinsic temperature fluctuations, CMB anisotropies can be sourced by velocity-dependent Doppler effects at the last scattering surface. Additionally, photons are gravitationally redshifted (or blueshifted) due to potentials at last scattering (the Sachs-Wolfe (SW) effect~\cite{Sachs:1967er}) and time-evolving potentials along the line of sight (the integrated SW (ISW) effect). 
The `early' ISW effect takes place just after recombination, that is, during matter domination, but at a time when the radiation density is not yet completely negligible compared to the non-relativistic one. Thus, this effect can be sensitive to new ultra-relativistic species via the matter-radiation ratio.

Another important observable accessible from measurements of the CMB is the gravitational (weak) lensing potential. Similar to variances of temperature anisotropies of intensity and polarisation, cosmological models also make distinct predictions for the lensing potential power-spectrum $C_\ell^{\phi\phi}$.
Lensing of the CMB is local to the observed direction, and takes the form
\be
\tilde{T}(\hat{\boldsymbol{n}})=T(\hat{\boldsymbol{n}}+\boldsymbol{\alpha}(\hat{\boldsymbol{n}}))\,,
\ee
(similarly for the polarisation field), where the lensing deflection angle is approximated as a pure gradient $\boldsymbol{\alpha}(\hat{\boldsymbol{n}})=\boldsymbol{\nabla}\phi$, and $\phi$ is the lensing potential. Lensing of the CMB (along with the ISW effect) provides sensitivity to the matter distribution and the growth of LSS along the line of sight, providing constraining power on cosmological parameters such as the sum of neutrino masses $\sum m_\nu$. 
Non-relativistic neutrinos slow the growth of LSS on small scales because they contribute to the expansion of the Universe but not to the gravitational clustering of non-relativistic matter.

\begin{figure}
  \begin{center}
    \includegraphics[width=0.8\textwidth]{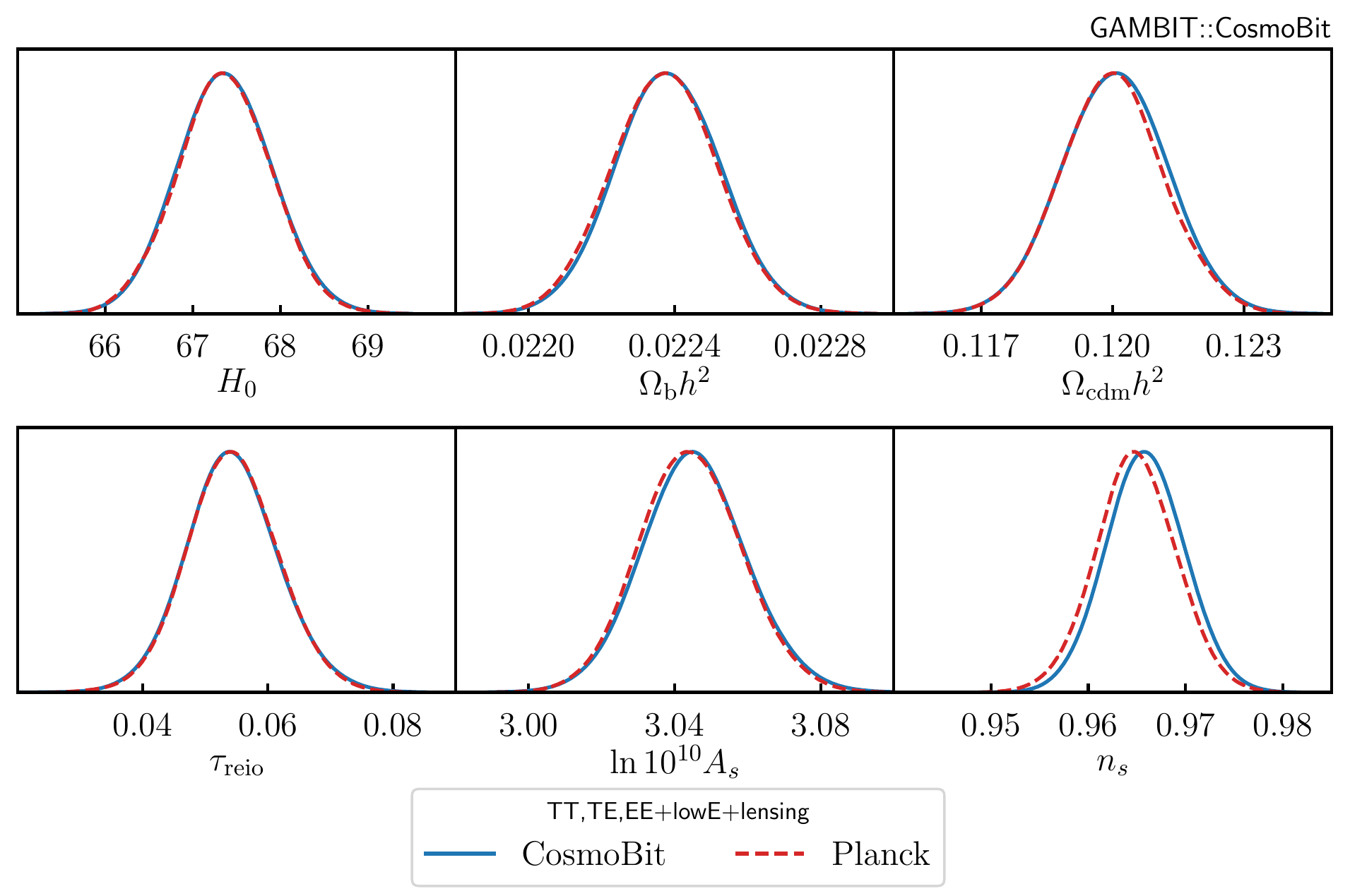}
    \caption{\label{fig:comparison_planck} Comparison of the 1D marginalised posterior probabilities for the $\lcdm$ `baseline' model ($\Sigma m_\nu = 0.06$\,eV) obtained in the Planck 2018 analysis \cite[][Table 1]{Aghanim:2018eyx} with the results obtained using the corresponding \textsf{Plik} likelihoods within \cosmobit. The datasets included in this analysis are CMB TT,TE,EE+lowE+lensing.
}
  \end{center}
\end{figure}

\subsubsection*{Likelihoods}
\label{sec:cmb_likelihoods}

\cosmobit includes an optional Gaussian likelihood for the CMB temperature measured today, with a mean of $\Tcmb = 2.72548$\,K and a standard deviation of $0.00057$\,K~\cite{Mather:1998gm,Fixsen2009}.

For the calculation of likelihoods related to CMB spectra, we use the {\tt Plik} likelihood published within the Planck Likelihood Code (\plc) \cite{Aghanim:2015xee,Aghanim:2019ame}.  This gives access to the full set of Planck 2015 and 2018 likelihoods, i.e.\ the low-$\ell$ (covering multipoles $2\leq\ell\leq29$) and high-$\ell$ ($30\leq\ell\leq2508$) likelihoods for the temperature and polarisation spectra, as well as the lensing likelihood constraining the CMB lensing potential.

The Planck likelihoods contain nuisance parameters to account for uncertainties related to the instrument calibration as well as signal contamination.  There are 34 extra parameters when using high-$\ell$ polarisation likelihoods and 16 when only the temperature anisotropies are considered.  To reduce the number of dimensions in a parameter scan the \plc package also contains a `lite' version of the high-$\ell$ likelihoods with marginalised foregrounds, requiring the use of only one nuisance parameter encoding the absolute calibration, $A_{\rm Planck}$.
The Planck marginalised `lite' likelihood was constructed to reduce the number of nuisance parameters for the likelihood calculation of a standard $\lcdm$ model. It should therefore be used with care when considering non-standard extensions to $\lcdm$ or alternative theories, as it is not necessarily valid for models other than $\lcdm$.

For more details on the likelihood calculation we refer the reader to Refs.\ \cite{Aghanim:2015xee,Aghanim:2019ame}. Documentation of usage of \plc within \gambit is given in Appendix~\ref{app:plc-interface}. Details on the implementation of the nuisance parameter models can be found in Appendix~\ref{app:models-nuisance}.

In Fig.~\ref{fig:comparison_planck}, we show 1D marginalised posteriors from a \cosmobit run for the Planck $\lcdm$ `baseline' model (where $\Sigma m_\nu = 0.06$), using the 2018 TT,TE,EE+lowE+lensing likelihoods in \plc.\footnote{The label `TT,TE,EE' corresponds to a likelihood using the combination of the temperature and E-mode polarisation power spectra, as well as the cross-correlation thereof, for multipoles $\ell > 29$ (\emph{high-l} TTTEEE in \autoref{tab:likelihoods}). The likelihood `lowE' is based on data from the EE power spectrum in the multipole range $2 \leq \ell < 30$. For more details, see Ref.~\cite{Aghanim:2019ame}.} We find excellent agreement with the values obtained by Planck using \camb and \cosmomc~\cite{Aghanim:2018eyx}.
The shift in the $n_s$ posterior is statistically insignificant, and is due to our use of a slightly different helium abundance $\yhe$ to Planck.

\subsection{Late-time cosmology: LSS, BAO and SNe~Ia} \label{sec:late_time}

The utility of the CMB as a tool for inferring the cosmological properties of the Universe from the oldest electromagnetic radiation is complemented by measurements of the late-time expansion and growth history of the Universe. These include measurements of LSS via galaxy surveys, large samples of SNe Ia, measurements of individual strong gravitational lenses, cosmic chronometers, and indirectly, via weak lensing of the CMB. These different probes are sensitive to the late-time Universe in different ways, and so together provide complementary tools via which to extract cosmological parameters.

Compared to the CMB, the main benefit of LSS is the prospect it offers to reduce errors on cosmological measurements, by providing many more Fourier modes in three dimensions, with scaling $N_{\rm modes}\propto k_{\rm max}^3$, where $k_{\rm max}$ is the highest wavenumber accessible to measurements that can be modelled robustly. Measurements of LSS are typically further refined to include the angular or 3-dimensional clustering of galaxies, and weak gravitational lensing.

The first of these, the clustering of galaxies, is sensitive to the distribution of matter in the Universe. However, clustering measurements are contaminated by the fact that galaxies are biased tracers of the underlying matter field, and by the reality that positions in three dimensions can only be obtained from galaxy redshifts, which unavoidably include the effects of the infall towards dense regions of the Universe. On large scales, the transition from radiation to matter domination at the epoch of matter-radiation equality encodes a characteristic peak in the galaxy power spectrum sensitive primarily to $\omega_{m}$. Similarly, as discussed in Sec.~\ref{sec:CMB}, BAO in the photon-baryon fluid of the early Universe leaves residual overdensities in the distribution of galaxies observed today.  This provides a `standard ruler' with which to measure the expansion rate of the Universe. BAO detected in galaxy clustering along the observer's line of sight constrain the Hubble distance $D_{H}(z)=cz/H(z)$ to the redshift of the galaxy sample, relative to size of the sound horizon at the baryon-drag epoch $r_{d}$. BAO transverse to the line of sight measure a ratio containing the angular diameter distance $D_{A}(z)$ (or equivalently the transverse comoving distance $D_{M}(z)=(1+z)D_{A}(z)$), while BAO averaged over all lines of sight measure a ratio containing the volume-averaged distance $D_{V}(z)=D_{A}^{2/3}(z)D_{H}^{1/3}(z)$.  Each of these distance measures can then be used to constrain $\Omega_{m}$, $H_{0}$ and the properties of dark energy~\cite{Meiksin:1998ra,Seo:2005ys,Eisenstein:2006nk}. Lastly, the infall of galaxies towards overdensities introduces a dependence on the growth rate of structure into measurements of galaxy clustering, which allows for testing of different gravitational models~\cite{Linder2007,Song2009,Mueller2018}.

\begin{figure}
  \begin{center}
      \begin{minipage}{.48\textwidth}
      \includegraphics[width=\textwidth]{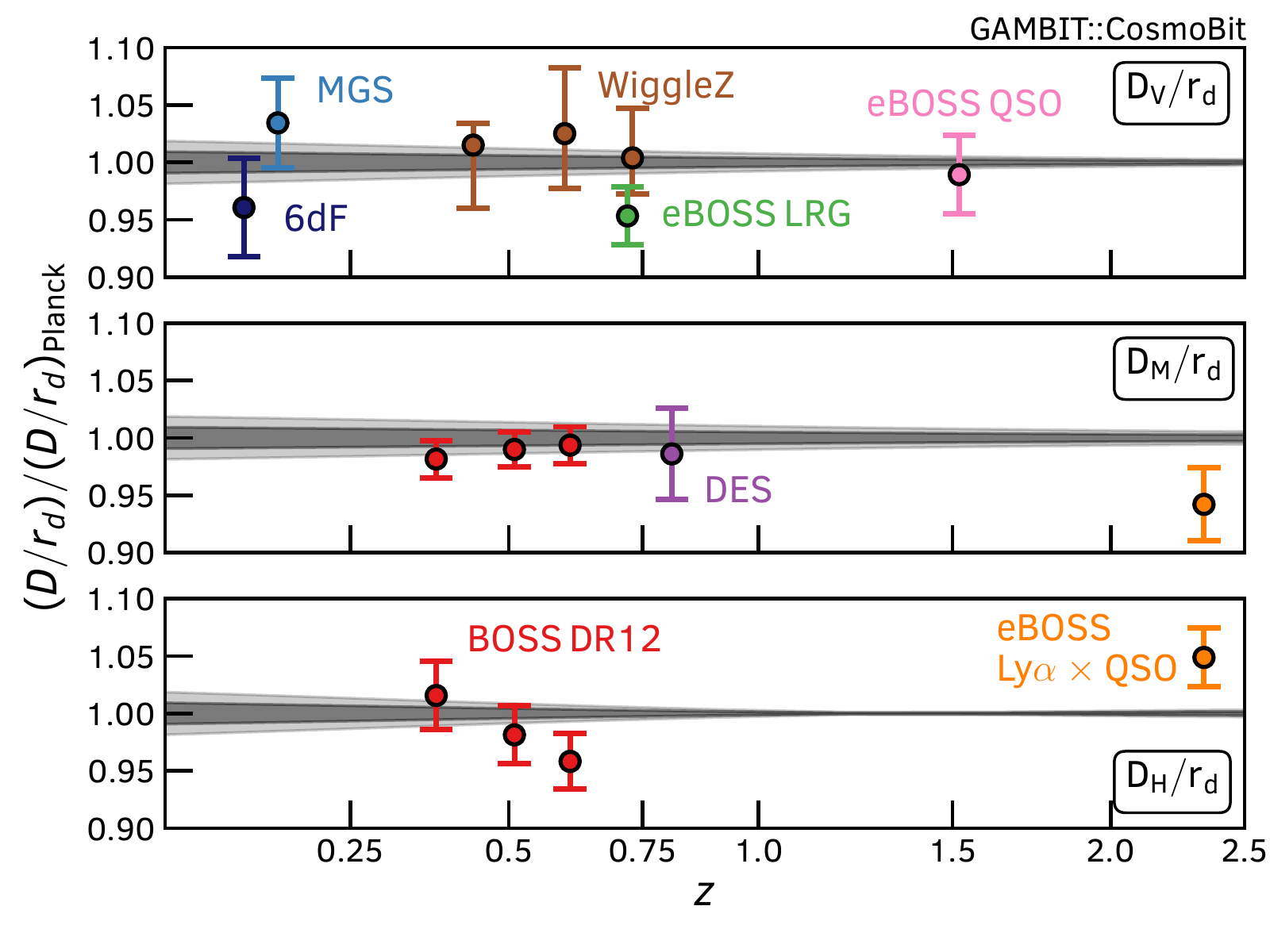}
      \end{minipage}
      \begin{minipage}{.48\textwidth}
      \includegraphics[width=\textwidth]{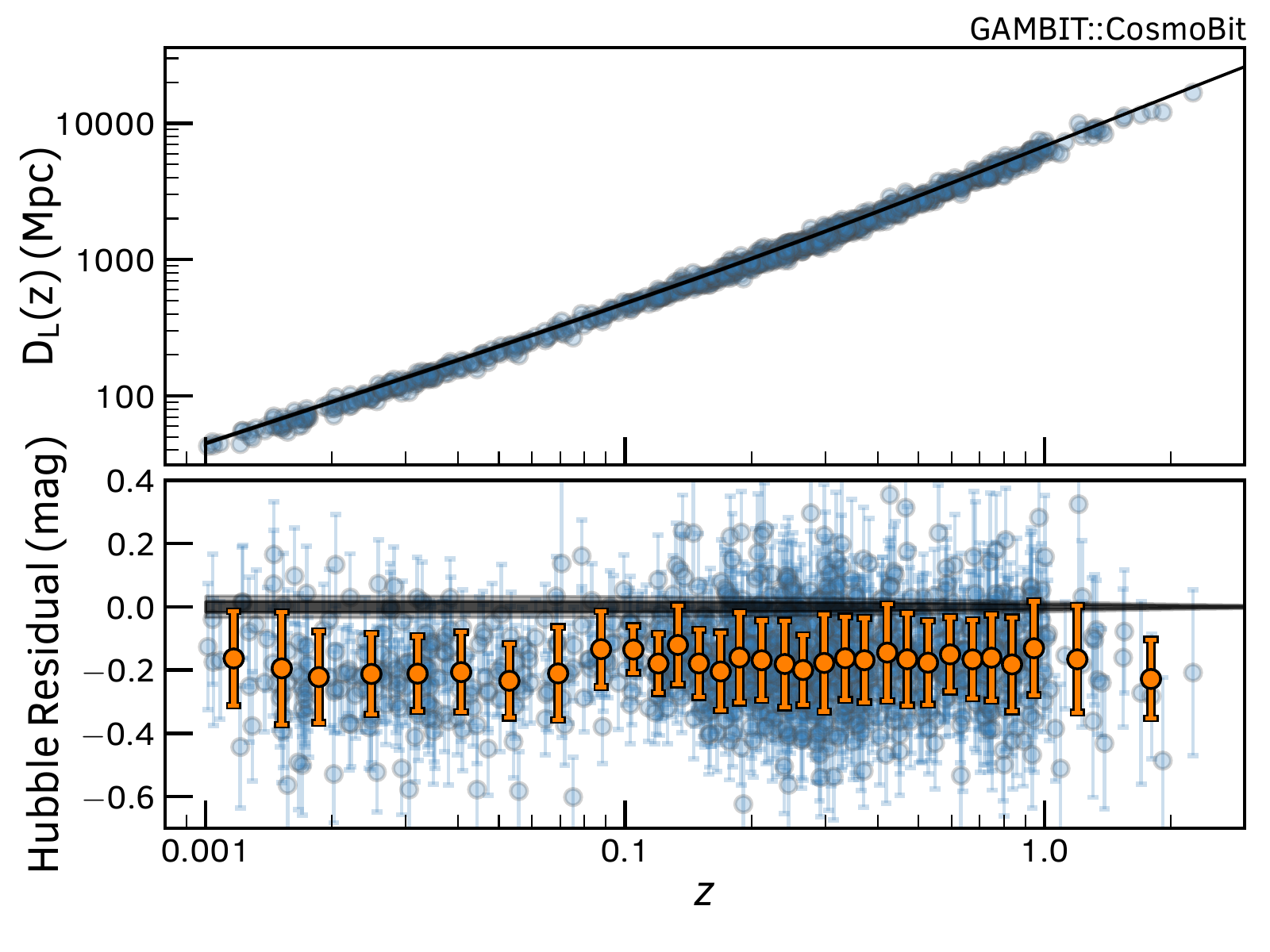}
      \end{minipage}
      \caption{\textit{Left}: A compilation of BAO measurements of the ratio of the transverse comoving ($D_{M}$), Hubble ($D_{H}$) and volume-averaged ($D_{V}$) distances to the sound horizon at the baryon-drag epoch $r_{d}$ as a function of redshift  \cite{2011MNRAS.416.3017B, 2015MNRAS.449..835R,Alam:2016hwk,Kazin:2014qga,Ata:2017dya,Bautista_2018,Abbott_2018,Blomqvist2019}. Gray bands show the 68\% and 95\% credible regions for these distances computed with \cosmobit using the Planck `lite' likelihood and assuming a flat $\lcdm$ model. \textit{Right}: The luminosity distance-redshift relation as measured by the Pantheon supernova sample \cite{Scolnic:2017caz}, with peak supernova magnitude calibrated to match the results from SH0ES \cite{Riess:2016jrr}. The line in the top panel shows the best fit computed with \cosmobit using the Planck `lite' likelihood and assuming a flat $\lcdm$ model. The lower panel shows the difference in the distance moduli (the `Hubble residuals') between the data and this line, which has been converted to a band showing the Planck 68\% and 95\% credible regions. Orange points are averages over the supernovae in bins, chosen to contain roughly constant numbers of SNe Ia per bin, and purely to guide the eye. The obvious discrepancy between the results from Planck and Pantheon in the bottom panel arises from the choice of SH0ES results to calibrate the peak magnitude and highlights the well-known `Hubble tension' (see Sec.~\ref{sec:tensions}).
  \label{fig:bao_sne_data}}
  \end{center}
\end{figure}

The main challenge of clustering measurements is in modelling all these effects: at short distances, high density contrasts and non-linear dynamics often introduce large and poorly-understood theoretical and systematic uncertainties, which can also affect the larger scales via mode coupling. These problems can either be surmounted by careful modelling of systematics and uncertainties (which boosts the number of nuisance parameters~\cite{Baumann:2010tm, Carrasco:2012cv, Senatore:2014eva, Senatore:2014via, Baldauf:2015aha, Carrasco:2013mua, Perko:2016puo, Assassi:2014fva, Lewandowski:2014rca, 2009JCAP:McDonald}), or partially avoided by utilising templates designed to optimally select cosmological signatures robust against complicated non-linearities, such as BAO scale measurements~\cite{2010ApJ...720.1650S,2011ApJ...734...94M,2012MNRAS.427.2132P,Sugiyama:2013gza,Font-Ribera:2013wce,Bautista:2017zgn,Bourboux:2017cbm}. The left panel of Fig.\ \ref{fig:bao_sne_data} shows a compilation of recent measurements of distance scales in the Universe using BAO as a standard ruler, alongside the predictions of a flat $\lcdm$ model computed with \cosmobit, based purely on the constraints imposed by the Planck `lite' likelihood (see Sec.\ \ref{sec:cmb_likelihoods}).

The technique of weak gravitational lensing measures the coherent shape distortion of galaxies in photometric surveys due to the presence of LSS. It mainly probes the integrated shape of the matter power spectrum, without complications due to the fact that the distribution of galaxies is not an unbiased representation of the distribution of matter. This makes it a somewhat `purer' tracer of cosmology than galaxy clustering. However, as the measurements inevitably include non-linear information they too are subject to theoretical uncertainties surrounding the shape of the non-linear matter power spectrum, particularly its suppression/modification due to baryonic effects. The displacement of baryons due to complicated feedback processes affects the gravitational potential within and around dark matter halos, which in turn changes the power spectrum on scales comparable to those of the halos themselves. This propagates through to the modelling of lensing observables, which integrate over these scales \cite{Chisari:2019tus,Semboloni2011,Mead2015}.

Other systematics include those associated with actually observing and recovering accurate shape measurements from photometric catalogues, obtaining photometric redshifts for the lenses and sources for splitting the data into tomographic bins, and modelling of the intrinsic correlations of galaxy shapes beyond that caused by gravitational lensing \cite{Heavens2000,Catelan2001}. Recent years have seen a wealth of new techniques for overcoming these systematics (see \cite{Heymans:2013fya,Troxel2015,Hildebrandt2017,Huang2019} for a few examples), and correspondingly tighter cosmological constraints. Modern lensing surveys primarily report constraints on the parameter $S_{8}=\sigma_{8}\sqrt{\Omega_{m}/0.3}$, where $\sigma^{2}_{8}$ is the variance in the matter field in a sphere of radius $8 h^{-1}\mathrm{Mpc}$, but also contain some information on the dark energy equation of state and curvature of the Universe \cite{Kilbinger2013,Hildebrandt2017,Troxel2018}.

Finally, SNe Ia have been used primarily to constrain the expansion history of the Universe.  They can do so with finer temporal resolution and at lower redshifts than LSS, due to fact that the latter requires averaging over large cosmological volumes to obtain statistically significant measurements. Supernovae at high redshift can constrain the properties of dark energy \cite{Scolnic:2017caz}, while those at low redshift, calibrated using nearby distance anchors, can provide local measurements of $H_{0}$ without the need to assume a cosmological model \cite{Riess:2016jrr, Riess:2019cxk, 2019ApJ...882...34F}. The precision achievable with modern supernova samples is now high enough that systematics such as the effect of contamination and host galaxy properties on supernova magnitudes \cite{2011ApJS..192....1C, Hui2006,Kelly2010}, as well as other observational systematics \cite{Kessler2017, Davis2019}, must be carefully considered and modelled. Nonetheless, SNe Ia have been repeatedly demonstrated to produce robust cosmological constraints~\cite{Betoule:2014frx,Scolnic:2017caz,Scolnic:2017caz}. They have also been anchored with measurements from other probes such as BAO, a technique coined the `inverse distance ladder' \cite{Cuesta2015,Macaulay2019}.

An example Hubble diagram showing measurements from SNe Ia against different cosmological models is shown in the right panel of Fig~\ref{fig:bao_sne_data}. Recently, complementary probes of the expansion rate of the Universe using the time delay between multiply imaged sources behind strong gravitational lenses \cite{2019arXiv190704869W}, and `age-dating' of passive galaxies (typically referred to as `cosmic chronometers') \cite{Stern2010, Moresco2016} have also demonstrated potential to achieve high precision measurements of the Hubble parameter.

\subsubsection*{Likelihoods}

To provide a variety of additional well-tested cosmological likelihoods, \cosmobit includes an interface to \montepython \cite{Audren:2012wb,brinckmann2018montepython}.  \montepython is a tool to perform parameter scans of cosmological models; as a standalone package it allows the user to choose between different sampling algorithms and offers an extended library of cosmological likelihoods. Within \cosmobit we implement an interface to the suite of \montepython likelihood calculations, but not its sampling algorithms. This is because \gambit provides its own advanced and modular sampling algorithms via \scannerbit.
We circumvent any direct call from \montepython to the Boltzmann solver \class needed to obtain theoretical predictions for observables. Instead, \gambit calls \class and passes all information on to \montepython. This is to give \gambit control over all results calculated by \class and to have the option to replace \class by a different Boltzmann solver. For more details on the interface refer to Appendix \ref{app:mp-interface}.

Most likelihoods that are entirely implemented in \montepython can be used out-of-the-box from \cosmobit; we expressly exclude the Planck and WMAP CMB likelihoods, as they require the installation of additional libraries, however the Planck likelihoods are available within \cosmobit directly through the interface to \plc (see Secs.\ \ref{sec:CMB} and \ref{app:plc-interface}). For further exceptions to supported likelihoods, see Appendix \ref{app:mp-interface}.

The interface to \montepython provides access to various probes of the late-time Universe.  Measures of the background evolution can be included through e.g., measurements of recession velocities of SNe~Ia, and the BAO scale. This includes in particular our new state-of-the-art treatment of correlations between different BAO scale measurements, which we use to constrain neutrino masses in a companion paper \cite{CosmoBit_numass}.  The linear matter power spectrum can be probed with weak lensing and galaxy clustering data. The evolution of gravitational potentials can be studied with measurements of the ISW effect through cross-correlations of the CMB temperature spectrum with foreground galaxies. A selection of the most important likelihoods and datasets associated with these tests can be found in Table~\ref{tab:likelihoods}.

Besides testing a model against already available datasets, \montepython also allows the user to forecast the sensitivity of future experiments.  Examples for this are cosmic shear and galaxy clustering likelihood forecasts for Euclid and SKA \cite{2013JCAP...01..026A,2019JCAP...02..047S}.

For an up-to-date compilation of all available likelihoods, we refer the reader to the \montepython website.\footnote{\url{https://brinckmann.github.io/montepython\_public/}} For details on the interface to the likelihoods within \montepython, and how to include them in a \GB analysis, see Sec.\ \ref{app:mp-interface}.

\subsection{Cosmological parameter tensions}
\label{sec:tensions}
At present, a tension exists between early and late-time measurements~\cite{Bernal:2016gxb,Aghanim:2018eyx,Verde:2019ivm,Schoneberg:2019wmt,Addison:2017fdm,Riess:2019cxk}. This is most pronounced in the parameter estimates of the Hubble constant: Planck ~\cite{Aghanim:2018eyx} infer a value of $H_0 = 67.4\pm0.5\:{\rm km}\,{\rm s}^{-1}\,{\rm Mpc}^{-1}$ using CMB data, whilst Riess et al. \cite{Riess:2019cxk} report $74.0\pm1.4\:{\rm km}\,{\rm s}^{-1}\,{\rm Mpc}^{-1}$ from Type Ia supernovae. This tension is generally mirrored in other comparisons of observables from the early and late Universe. More minor tensions in other parameters such as $S_8$~\cite{Kohlinger:2017sxk} arguably exist between datasets, and appear to be growing.

With these more subtle and hidden tensions in mind, interest has been gathering in quantifying tensions~\cite{Charnock:2017,Inman1989,Hobson2002,Battye2015,Seehars2014,Nicola2019,Kunk2006,Karpenka2015,MacCrann2015,Adhikari2019,Douspis,Raveri2018} across the full multidimensional parameter space, rather than inspecting a specific projection of the likelihood/posterior such as $H_0$ or $S_8$. This is entirely within the ``global fit'' framework that \gambit encourages, and we anticipate that \cosmobit will prove very useful to researchers investigating global tensions between cosmological and particle physics datasets.

Whilst it is plausible that some of these discrepancies may be explained by an underestimation of systematic errors, the fact that many independent teams observe consistent effects is suggestive that some of the tension resides in a failing of our concordance model. Over the next few years the hunt is on to locate extensions and modifications to $\Lambda$CDM that are capable of relaxing these tensions in a statistically robust way. We believe that unified frameworks such as \cosmobit will prove essential to researchers aiming to uncover the next standard model of cosmology.

\begin{figure}
  \begin{center}
    \includegraphics[width=0.495\textwidth]{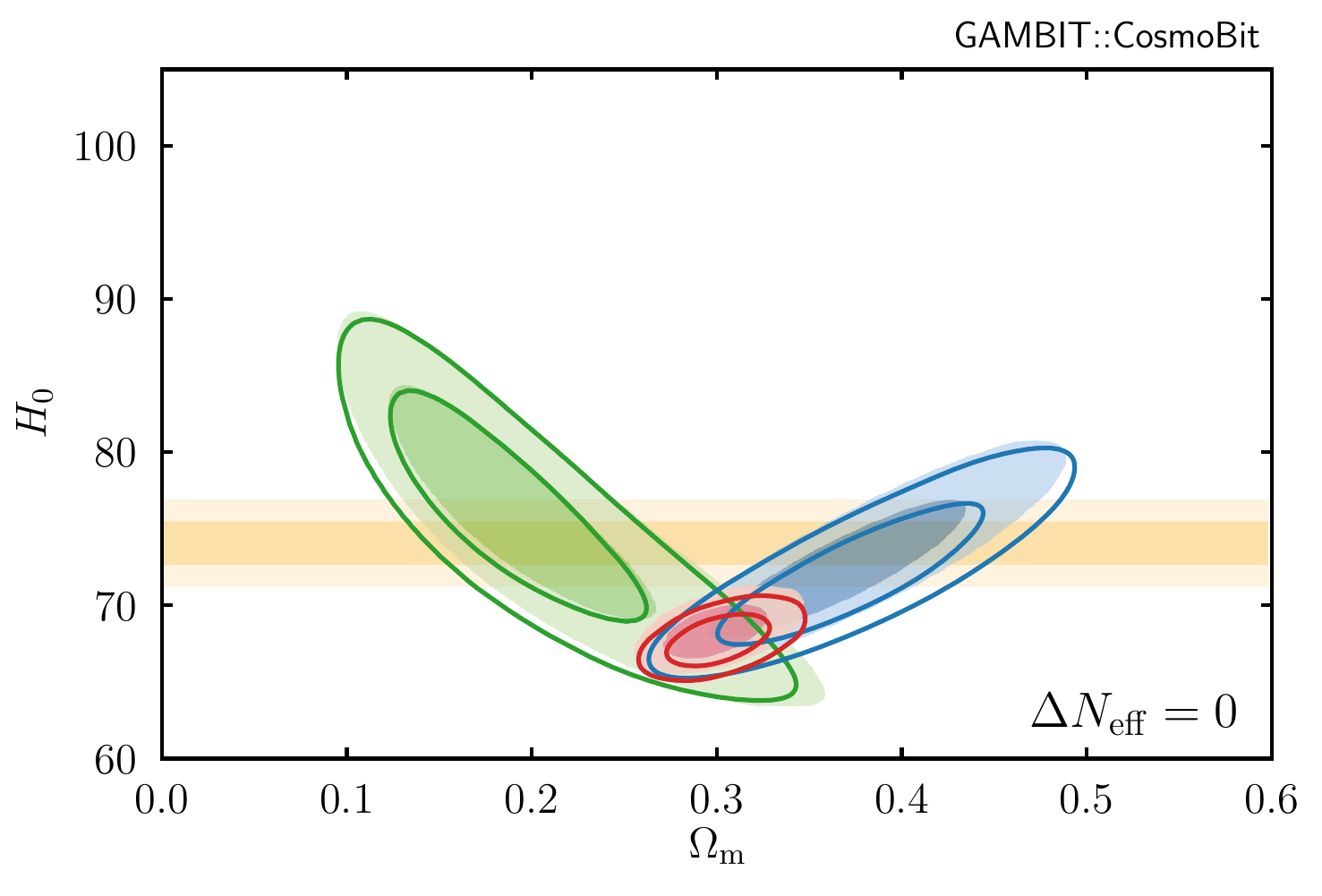}
    \includegraphics[width=0.495\textwidth]{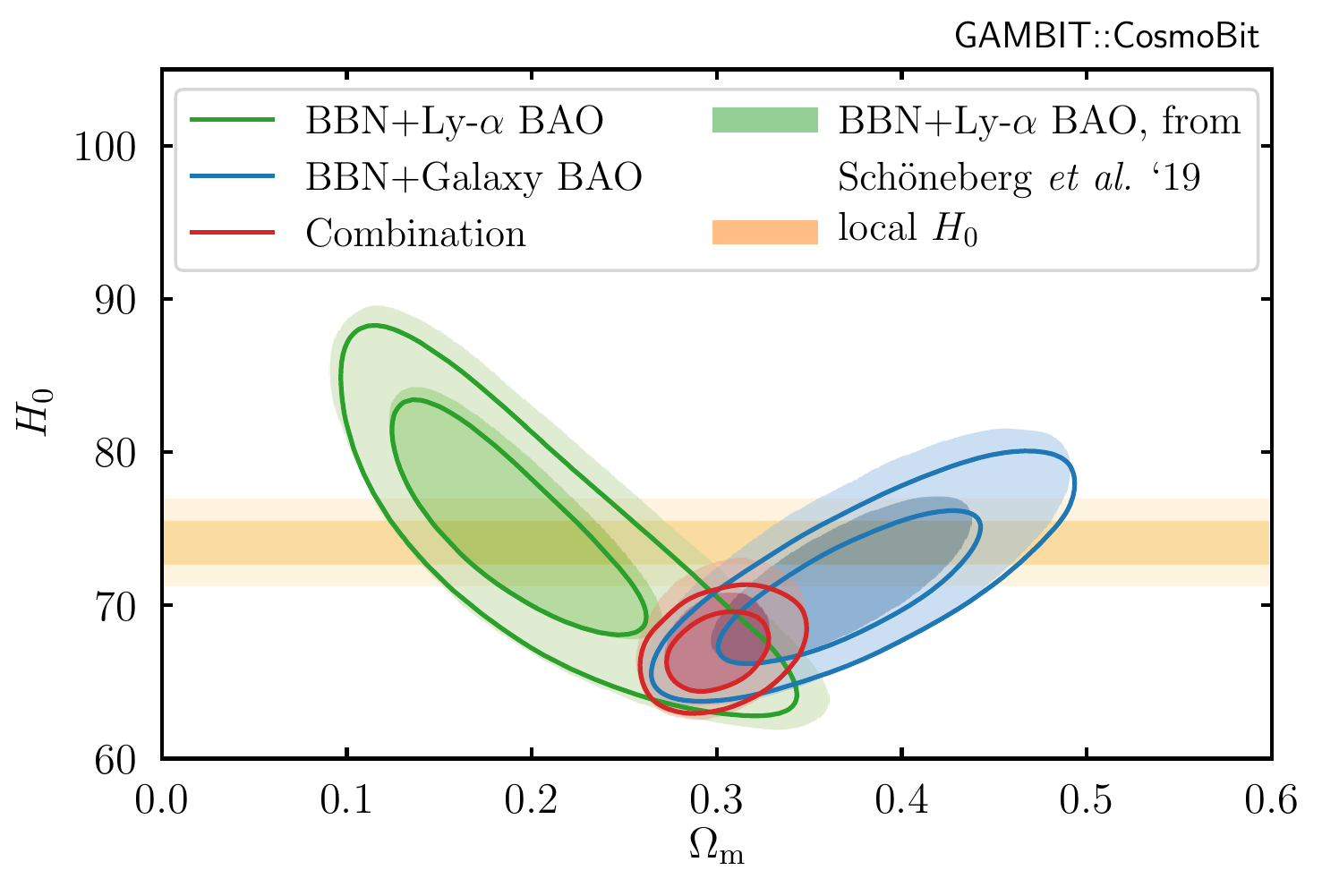}
    \caption{Determination of the present-day Hubble rate $H_0$ from BAO and BBN data for standard $\lcdm$ (\textit{left}) and $\lcdm$ extended by the number of ultra-relativistic species, $N_ {\rm eff}$ (\textit{right}).
    The solid contours indicate the 68\% and 95\% credible regions from our scans. The light shaded regions correspond to the results from Ref.~\cite{Schoneberg:2019wmt}. The constraints shown in blue arise from combining deuterium and helium measurements from BBN with BAO scale measurement from galactic data; the green contours show the combination of BBN data and the BAO scale from Lyman-$\alpha$ measurements. The red contours indicate the constraints from the combination of BBN and galactic as well as Lyman-$\alpha$ BAO measurements.
    For comparison, we show the value inferred from local measurements of Type Ia supernovae calibrated to Cepheids~\cite{Riess:2019cxk} as a yellow band, which demonstrates that the $H_0$ tension is present also in these data sets.
    The differences in the results are sourced by different numerical treatments for calculating the relic abundances of light elements from BBN. We use \alterbbn to obtain these predictions, while Ref.~\cite{Schoneberg:2019wmt} use \textsf{PArthENoPE} (see main text for more details).
    \label{fig:bao_bbn_H0} }
  \end{center}
\end{figure}

As an example of the cosmological constraints obtained by combining different probes, and to illustrate the tension between different ways of inferring $H_0$, we reproduce the results from Ref.~\cite{Schoneberg:2019wmt} in Fig.\ \ref{fig:bao_bbn_H0}.  This paper investigated whether the combination of BBN and BAO data leads to a value of the Hubble constant consistent with measurements from SNe~Ia. The BBN likelihood includes measurements of the helium and deuterium abundances. The BAO data are divided into two groups: measurements from galactic sources and measurements through the Lyman-$\alpha$ forest.%
\footnote{The galactic data are a compilation of the 6dF~\cite{2011MNRAS.416.3017B}, SDSS DR7 MGS~\cite{2015MNRAS.449..835R}, SDSS BOSS DR12 galaxies~\cite{Alam:2016hwk}, and SDSS DR14 eBOSS quasar~\cite{Ata:2017dya} measurements. The Lyman-$\alpha$ set contains the SDSS DR14 eBOSS Lyman-$\alpha$ auto correlation~\cite{Agathe:2019vsu} and cross-correlation with quasars~\cite{Blomqvist2019}. For more details refer to Ref.~\cite{Schoneberg:2019wmt}.} %
The resulting constraints on $H_0$ and the matter energy density $\Omega_m$ are shown in Fig.\ \ref{fig:bao_bbn_H0}. In the left panel, $N_\text{eff}$ is fixed to its standard value, whilst it is treated as a free parameter in the right panel.

Our results are in good agreement with Ref.~\cite{Schoneberg:2019wmt}, which serves as validation of our implementation. In particular, we confirm that the value of $H_0$ inferred by the combination of BBN and BAO data is consistent with the value inferred from the CMB, but in tension with that from SNe~Ia. The slight differences between our results and the ones from Ref.~\cite{Schoneberg:2019wmt} are due to differences in the calculation of the predicted primordial element abundances.  Within \cosmobit, we employ \alterbbn \textsf{2.2} while Ref.~\cite{Schoneberg:2019wmt} used \textsf{PArthENoPE 2.0}~\cite{Pisanti:2007hk}. The shift in the contours for standard $\lcdm$ (left panel of  Fig.\ \ref{fig:bao_bbn_H0}) are similar to the ones found in Fig.~3 of Ref.~\cite{Schoneberg:2019wmt}, which compares the constraints obtained with \textsf{PArthENoPE 2.0} to the ones obtained by using the \textsf{PArthENoPE}-Marcucci implementation~\cite{Marcucci:2015yla} and \textsf{PRIMAT}~\cite{Pitrou:2018cgg}. These differences become more pronounced when $\lcdm$ is extended by the number of ultra-relativistic degrees of freedom $N_{\rm eff}$ (right panel of  Fig.\ \ref{fig:bao_bbn_H0}).

\begin{table}
 \centering
 \small
  \begin{tabular}{cp{11.5cm}}
    \toprule
    Parameter name & Definition \\
    \midrule
    & \textbf{Standard Cosmology} -- Sec.\ \ref{sec:lcdm} \\
    $ H_0 $ & Today's expansion rate in the Universe $ \left[\mathrm{km}\,\mathrm{Mpc}^{-1}\,\mathrm{s}^{-1}\right] $.\\
    $ \omega_\text{b} = \Omega_\text{b}\,h^2 $ & Scaled energy fraction in the form of baryonic matter. \\
    $ \omega_\text{cdm} = \Omega_\text{cdm}\,h^2 $ & Scaled energy fraction in the form of cold dark matter.\\
    $ \tau_\text{reio}$ & Optical depth at reionisation.\\
    $ \ln \left(10^{10}\,A_s\right) $ & Amplitude of the scalar primordial power spectrum. (Not used if power spectrum
      is calculated by \multimodecode).\\
    $ n_s $ & Tilt of the scalar primordial power spectrum. (Not used if power spectrum
    is calculated by \multimodecode).\\
    $ \Tcmb$ & Temperature of CMB photons today.\\
    \midrule
    & \textbf{Inflation} -- Sec.\ \ref{sec:inflation} \\
    $ \lambda$ & Inflaton self-coupling (dimensionless). \\
    $ m_\phi $ & Inflaton mass (in units of $\mplred$). \\
    $ f_\phi $ & Axion-like inflaton decay constant in natural inflation (in units of $\mplred$). \\
    \midrule
    & \textbf{Neutrino masses} -- Sec.\ \ref{sec:Nu-masses} \\
    $ m_{\nu_0} $ & Mass of the lightest neutrino (in units of eV).\\
    $ \Delta m_{21}^2 $ & Difference of the squared masses of $ \nu_2 $ and $ \nu_1 $ (in units of eV$^2$).\\
    $ \Delta m_{3l}^2 $ & Difference of the squared masses of $ \nu_3 $ and $ \nu_l $ (in units of eV$^2$).\\
      & Normal hierarchy: $ \nu_l = \nu_1 $; inverted hierarchy: $ \nu_l = \nu_2$. \\
    \midrule
    & \textbf{Extra radiation} -- Sec.\ \ref{sec:non-std-Neff} \\
    $\Delta N_\text{ur,BBN}$ & Number of extra ultra-relativistic degrees of freedom (i.e. dark radiation) at\\
      & the end of BBN.\\
    $\Delta N_\text{ur,CMB}$ & Number of extra ultra-relativistic degrees of freedom (i.e. dark radiation) at\\
      & the time of recombination. \\
    $r_\text{BBN}$ & Temperature ratio $ T_\nu^\text{BSM} / T_\nu^\text{SM} $ at the end of BBN.\\
    $r_\text{CMB}$ & Temperature ratio $ T_\nu^\text{BSM} / T_\nu^\text{SM} $ at the time of recombination.\\
    $\eta_\text{BBN}$ & Baryon-to-photon-ratio at the end of BBN.\\
      & The respective value at recombination is encoded in $ \omega_b $. \\
    \midrule
    & \textbf{Energy injection} -- Sec.\ \ref{sec:energy-injection} \\
    $ m_\chi $ & Mass of the dark matter candidate (in units of GeV).\\
  $ \langle \sigma v \rangle $ & Thermally averaged cross-section for $s$-wave annihilation (in units of cm$^3$s$^{-1}$).\\
  $ \tau $ & Lifetime of the decaying dark matter candidate (in units of s).\\
  $ {B\!R}_{\rm el} $  (${B\!R}_{\rm ph}$) & Branching ratio into $e^+e^-$ ($\gamma\gamma$) in dark matter annihilation or decay.\\
  $ \xi = \rho_\chi / \rho_\text{cdm}$ & Energy fraction of dark matter in form of $ \chi $ (evaluated in the infinite past\\
    & for decaying dark matter). \\
    \bottomrule
  \end{tabular}
  \caption{List of physical model parameters discussed in this paper, corresponding to cosmology-related models and their simplifications shown in Fig.~\ref{fig:model_tree}.  Standard cosmological parameters are defined and discussed in Sec.\ \ref{sec:lcdm}, and all others in Sec.\ \ref{sec:beyond_lcdm}.  Nuisance parameters do not appear in this table, and are instead detailed in Appendix \ref{app:models-nuisance}. See Appendix~\ref{app:models} for a description of the models and parameter names as implemented in \gambit.}
\label{tab:cosmo_params}
\end{table}

%
\section{Extensions of $\Lambda$CDM in CosmoBit}
%
\label{sec:beyond_lcdm}

So far we have focused on the simple $\lcdm$ scenario defined by 6 independent parameters. In this section we consider several well-founded extensions of this picture, and how they affect the cosmological observables discussed above. An overview of all cosmological models implemented in \cosmobit (a so-called family tree of \gambit models) is shown in Fig.~\ref{fig:model_tree}. Table~\ref{tab:cosmo_params} provides a summary of the relevant model parameters.  A detailed list of these models is provided in Appendix~\ref{app:models}.

\subsection{Inflationary cosmology}\label{sec:inflation}
An early period of accelerated expansion is currently the most widely accepted scenario to generate density perturbations (and to successfully address various initial-value problems of Big Bang cosmology) \cite[see e.g.][]{Baumann:2009ds}.
Periods of accelerated expansion correspond to equation-of-state parameters of $w < -1/3$.  Typically, successful models of inflation sustain $w \approx -1$ for a sufficiently long, but finite
amount of time. The simplest way to achieve this is by introducing the inflaton $\phi$, a scalar field minimally coupled to gravity. Its contribution to the action is
\be \label{eq:phitogravity}
S_\phi = \int {\rm{d}}^4x\sqrt{-g}\left[\frac{1}{2}(\partial_\mu\phi)^2 - V(\phi) \right] \, ,
\ee
where the potential $V(\phi)$ dominates over the kinetic energy (and other potential contributions to the stress-energy tensor).
In an expanding FLRW spacetime this is equivalent to demanding that the evolution of $\phi$ is (and remains) slow compared to the expansion of the Universe, as described by the Klein-Gordon and Friedmann equations,
\begin{eqnarray}
    \ddot{\phi}+3H\dot{\phi}+V'(\phi) = 0 \label{eqn:klein_gordon}\\
    H^2 = \frac{1}{3 \mplred^2}\left[ \frac{1}{2}\dot{\phi}^2 + V(\phi) \right] \label{eqn:friedmann},
\end{eqnarray}
where $\mplred$ is the reduced Planck mass, dots denote differentiation with respect to time, and primes partial differentiation with respect to $\phi$.
This requirement is typically formulated in terms of slow-roll parameters $\epsilon_i$, 
where
$\epsilon_1 \equiv -\frac{\dot{H}}{H^2} = \frac{1}{2\mplred^2}\left(\frac{\dd \phi}{\dd N}\right)^2$
and $\epsilon_{i+1} \equiv \dot{\epsilon}_i/(H\epsilon_i)$~\cite{astro-ph/9408015}.
These satisfy $\epsilon_i\ll1$ during inflation, setting an effective energy scale for inflation of $H\simeq \mplred^{-1}\sqrt{8\pi V(\phi)/{3}}\simeq{\rm constant}$.\footnote{This energy scale can be as large as the GUT scale, $10^{16}$\,GeV, depending on the inflationary model, and can be probed directly with the tensor-to-scalar ratio, which satisfies the relation $r\propto V \mplred^{-4}$. Inflation occurring at the GUT scale corresponds to $r\geq0.01$.}
Here $\dd N\equiv H\dd t=\dd\ln a$,
with $N$ the number of $e$-foldings of expansion, and successful inflation requires
$\epsilon_1,\ \epsilon_2\ll1$.
Alternatively, the first and second slow-roll parameters can be introduced directly in terms of
the inflationary potential,
\begin{equation}
\epsilon_\textrm{v}\equiv\frac{\mplred^2}{2}\left(\frac{V'}{V}\right)^2\ \ \textnormal{and} \ \ \eta_\textnormal{v}\equiv \mplred^2\frac{V''}{V}\,,
\end{equation}
where $V'$ and $V''$ denote the first and second functional derivative w.r.t.\ to the inflaton field~$\phi$, respectively.
These parameters can be directly related to $\epsilon_{i}$ parameters in the slow-roll regime.

Slow-roll inflation produces scalar perturbations as in Eq.~(\ref{Psdef}) and, conveniently, allows one to
express the parameters appearing there directly in terms of the slow-roll parameters,
\begin{eqnarray}
n_s-1= 2\eta_\text{v}-6\epsilon_\text{v}\\
A_s=\frac{1}{24\pi^2\epsilon^\star_\text{v}}\frac{V_\star}{\mplred^4}\,,
\end{eqnarray}
where a star~($\star$) implies that the corresponding quantity is evaluated when the scale of a primordial perturbation
(see~Section~\ref{sec:perturbations}), $k\propto aH$ expands beyond the Hubble scale, $k=k_\star$.
In general, however, the slow-roll parameters are functions of $k$.
Therefore, unlike in $\lcdm$, the spectral index may not actually be exactly constant, but may instead `run' slightly,
\begin{equation}
n_s=n_s^\star+\frac12 \left. \frac{\dd n_s}{\dd\!\ln k}\right|_{k_\star} \ln(k/k_\star)+\ldots,
\end{equation}
which in principle can be used to distinguish inflationary models \cite[see e.g.][]{Martin:2013tda}.
Another predictive feature of slow-roll inflation is that it generates not only scalar but also tensor perturbations, of the form
\begin{equation}
\calP_t(k) = \frac{k^3}{2\pi^2}|h_k|^2 = A_t\left(\frac{k}{k_\star}\right)^{n_t}\,,
\end{equation}
with a tensor-to-scalar ratio given by
\begin{equation}
r\equiv \calP_t(k_\star)/\calP_\zeta(k_\star)=16\epsilon_\text{v}^\star\,.
\end{equation}

In general, the power spectra generated during inflation may exhibit a richer structure than
can be conveniently described with a power-series expansion of the spectral index.  For example,
models of inflation such as axion monodromy \cite{Silverstein:2008sg,McAllister:2008hb,Kaloper:2011jz} cannot be described by simple slow-roll
dynamics. However, going beyond the slow-roll approximation implies that it is no longer sufficient to consider only background
evolution. Instead, one needs to fully take into account the effect of general first-order perturbations to the FLRW metric,
\begin{equation}
\textnormal{d}s^2=-(1+2\Phi)\textnormal{d}t^2  -2a^2\partial_iB\textnormal{d}t\textnormal{d}x^i
+a^2\left[(1-2\Psi)\delta_{ij}-2\partial_{i}\partial_{j}E\right]\textnormal{d}x^i\textnormal{d}x^j\,.
\end{equation}
In this expression, we only keep general scalar metric perturbations  $(\Phi,\Psi,B,E)$, thus neglecting
vector perturbations (which only have decaying solutions) and tensor perturbations (describing gravitational waves).
In the spatially flat gauge ($\Phi=\Psi=E=0$) the Klein-Gordon equation at first order then becomes
\begin{equation}
\label{eq:no_slow_roll}
\frac{\textnormal{d}^2\delta\phi_k }{\textnormal{d}N^2}+(3-\eta)\frac{\textnormal{d}\delta\phi_k }{\textnormal{d}N}+\frac{k^2}{a^2H^2}\delta\phi_k
+ \left[\frac{V''}{H^2}+\frac{2}{H^2}\frac{\textnormal{d}\phi}{\textnormal{d}N}V'+(3-\eta) \frac{\textnormal{d}\phi}{\textnormal{d}N}\right]\delta\phi_k=0\ ,
\end{equation}
where we defined $\eta \equiv \frac{1}{2\mplred^2}(d\phi/dN)^2$ for simplicity, 
and $\delta\phi_k$ are the perturbations of the inflaton background $\phi$, expanded in Fourierspace.

Solving this mode equation,
the (gauge-invariant) power spectrum of the curvature perturbations
is obtained as $\mathcal{P}_\mathcal{R}(k) = \mplred^2\,|\delta\phi_k |^2/(\textnormal{d}\phi/\textnormal{d}N)^2|_{N=\npiv}$,
 where we defined the number of $e$-foldings between the mode $k_\star$ becoming equal to the Hubble scale
 and the end of inflation, $\npiv\equiv \ln (a_\text{end}/a_\star)$.
In the absence of isocurvature fluctuations, $\mathcal{P}_\mathcal{R}(k)$ is constant during this time.

In \cosmobit, we evaluate Eq.~(\ref{eq:no_slow_roll}) using the public package \textsf{MultiModeCode}~\cite{Mortonson:2010er,Price:2014xpa},
which imposes self-consistent initial conditions on the
background evolution of the inflaton in Eq.~\eqref{eqn:klein_gordon}.
This makes use of the fact that for most inflationary models
there exists an attractor solution that is usually approximately the same as (but in general distinct from)
the slow-roll solution. \textsf{MultiModeCode} sets $\dot{\phi}_\mathrm{initial}=-{V^\prime(\phi_\mathrm{initial})}/{\sqrt{3V(\phi_\mathrm{initial})}}$ via
the slow-roll approximation, adjusting $\phi_\mathrm{initial}$ until the solution is in the attractor state for the entire
observable window. The perturbations $\delta\phi_{k}$ are set using Bunch-Davies initial conditions~\citep{1978RSPSA.360..117B} at the initial time.

Note that string theory suggests that inflation models with multiple fields may be more natural from a theoretical point of view. String compactifications, for example, often result in large numbers of scalar fields appearing in the low energy effective action~\cite{Grana:2005jc,Douglas:2006es,Denef:2007pq,Denef:2008wq}. While \textsf{MultiModeCode} is capable of handling multiple inflaton fields, examples for such models are not available in \GB at the time of writing. The current backend interface can in principle accommodate such cases by mirroring the implementation of multiple fields in \textsf{MultiModeCode}. 
This possibility, in fact, is one of the main motivations for our choice of interfacing CosmoBit with \textsf{MultiModeCode}  rather than using \textsf{CLASS} directly to evaluate Eq.~(\ref{eq:no_slow_roll}), along with the larger library of in-built inflaton potentials in the case of \textsf{MultiModeCode}.
However, we anticipate that it will be desirable for users to tweak the backend interface itself if they wish to study more complex cases.

\begin{figure}
  \centering
  \includegraphics[width=14cm]{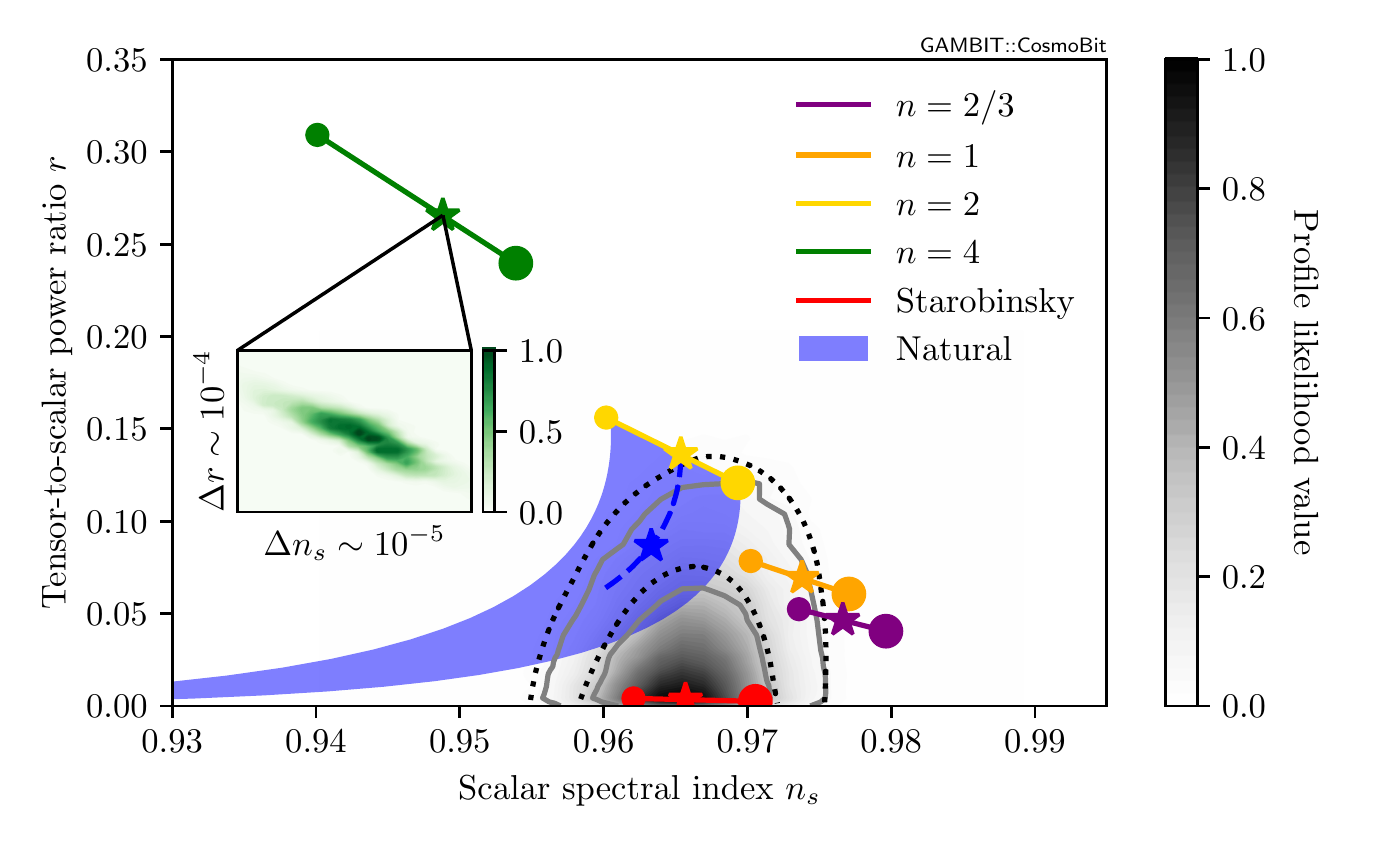}
  \caption{Predictions and constraints on inflation, showing best-fit points for $n_s$ and $r$~(stars) in each of the inflation models implemented in \cosmobit, as well as the slow-roll predictions~(solid lines, blue shaded area) for $N_\star$ between 50~(smaller filled circle) and 65~(larger filled circle). The inset zooms in on the best-fit point for quartic inflation, to demonstrate that we fully scan all nuisance parameters. The dashed line for natural inflation indicates the location of the very narrow 95\% confidence profile likelihood region. Black dotted lines indicate the 68\% and 95\% credible regions from Planck for the corresponding likelihoods~\cite{Planck18inflation}. The grey contours and density map show the profile likelihood with 68\% and 95\% confidence levels from the phenomenological \doublecrosssf{PowerLaw\_ps}{PowerLaw_ps} model.
  \label{fig:inflation_bestfit}}
\end{figure}

Exactly how reheating is assumed to have proceeded following inflation is a significant source of uncertainty when testing inflationary models.
In terms of observables, this is mostly a question of how to relate length scales in the early Universe
to those observed today.  More specifically, this boils down to connecting length scales at the end of inflation to those at the
end of reheating.  We define the end of reheating to be the point at which the Universe has fully thermalised and standard $\lcdm$ cosmology begins.
The matching can be expressed in terms of the equation~\cite{Liddle:1993fq,Liddle:2003as}
\begin{equation}\label{eq:matching_efolds}
\frac{k_\star}{a_0H_0}
=\frac{a_\star}{a_\text{end}}\frac{a_\text{end}}{a_\text{reh}}\frac{a_\text{reh}}{a_\text{eq}}\frac{H_\star}{H_\text{eq}}\frac{a_\text{eq}H_\text{eq}}{a_0H_0}\,,
\end{equation}
where the pivot scale $k_\star \equiv a_\star H_\star$ is defined as usual at `horizon crossing' ($k=aH$), and the various values of
the scale factor $a$ (and Hubble rate $H$) refer, respectively, to the end of inflation and reheating, matter-radiation
equality and today.
In this equation, the quantities evaluated today and at the time of matter-radiation equality are observationally
well constrained, while the parameters $H_\star$ and $\npiv$ are derived quantities that are predicted by the particular inflationary model.
This leaves the ratio $a_\text{end}/a_\text{reh}$ as a quantity that can {\it in principle} be computed from the details
of the reheating process.
More precisely, one can rewrite the matching equation as~\cite{Liddle:2003as,1004.5525,Planck13inflation}
\begin{align}
N_\star \approx 67 &- \ln \left(\frac{k_\star}{a_0H_0}\right) + \frac{1}{4} \ln \left(\frac{V_\star}{\mplred^4}\right) + \frac{1}{4} \ln \left( \frac{V_\star}{\rho_\text{end}}\right) \nonumber \\
&+ \frac{1-3w_\text{int}}{12 \, (1 + w_\text{int})} \, \ln \left(\frac{\rho_\text{reh}}{\rho_\text{end}}\right) - \frac{1}{12} \, \ln (g_\text{reh}) \label{eq:n_star} \\
\approx 71 & -\frac{1}{4} \ln \left(\frac{V_\star}{\mplred^4}\right) + \frac{1}{4} \ln \left( \frac{V_\star}{\rho_\text{end}}\right) \label{eq:n_star_mmc} \, ,
\end{align}
where $\rho_\text{end}$ is the energy density at the end of inflation and $w_\text{int}$ is the average, `intermediate' equation of state between the end of inflation and thermalisation after reheating. At the end of reheating, the energy density is $\rho_\text{reh}$ and there are $g_\text{reh}$ effective bosonic degrees of freedom in the Universe.
In the absence of a concrete model for reheating, $N_\star$ can be used as an additional, effective model parameter that absorbs the uncertainty about the reheating dynamics in a phenomenological way~(e.g.\ in Fig.~\ref{fig:inflation_bestfit}). Currently all inflationary models implemented in \cosmobit assume instant reheating, $a_\text{end}/a_\text{reh}=1$, in which case $\npiv$ is a computable, derived quantity. Note that Eq.~(\ref{eq:n_star}) is only to illustrate how $\npiv$ depends on the inflationary model and reheating scenario, and that -- in what follows -- we use \multimodecode to numerically calculate the value of $\npiv$. Equation~\eqref{eq:n_star_mmc} states the approximation that \multimodecode uses as the \emph{initial guess} for the calculation of $\npiv$ (for instant reheating and $k_\star = 0.05\,\text{Mpc}^{-1}$). Note that \multimodecode ignores the small contribution from the temperature-dependent $\ln(g_\text{reh})/12$ term in Eq.~\eqref{eq:n_star}~\cite{Mortonson:2010er}.

\begin{table}[t]
 \renewcommand{\arraystretch}{1.2}
 \centering
 \begin{tabular}{@{}lcll@{}}
  \toprule
  Inflationary model & Inflaton potential $V(\phi)$ & Parameter & Range \\
  \midrule
  \doublecrosssf{Phenomenological}{PowerLaw_ps}& N/A & $\ln(10^{10}A_s)$ & [2.96, 3.12] \\
  \doublecrosssf{power-law power}{PowerLaw_ps} & & $n_s$ & [0.94, 0.99] \\
  \doublecrosssf{spectrum}{PowerLaw_ps}& & $r$ & [0, 0.2] \\
  \midrule
  \doublecrosssf{$n=2/3$}{Inflation_InstReh_1mono23} & $\frac{3}{2} \, \lambda  \mplred^{10/3}
  \phi^{2/3}$ & $\lambda/10^{-10}$ & [2, 2.6] \\
  \doublecrosssf{$n=1$ (linear)}{Inflation_InstReh_1linear} & $\lambda  \mplred^3 \phi$ & $\lambda/10^{-10}$ & [1.9, 2.2] \\
  \doublecrosssf{$n=2$ (quadratic)}{Inflation_InstReh_1quadratic} & $\frac{1}{2} \, m_\phi^2 \phi^2$ & $m_\phi/(10^{-6} \mplred)$ & [5.8, 6.3] \\
  \doublecrosssf{$n=4$ (quartic)}{Inflation_InstReh_1quartic} & $\frac{1}{4} \, \lambda  \phi^4$ & $\lambda/10^{-13}$ & [1.3, 1.6] \\
  \doublecrosssf{Starobinsky}{Inflation_InstReh_1Starobinsky} & $\lambda^4 \mplred^4\left[1-e^{-\sqrt{2/3}\phi/\mplred}\right]^2$ & $\lambda/10^{-3}$ & [3.18, 3.27] \\
  \doublecrosssf{Natural}{Inflation_InstReh_1natural} & $\lambda^4  \mplred^4\left[1+\cos\left(\phi/f_\phi\right)\right]$ & $\log_{10}(\lambda/10^{-3})$ & [5, 50] \\
  & & $\log_{10}(f_\phi/\mplred)$ & [3.16, 316] \\
  \bottomrule
 \end{tabular}
 \caption{Inflation models presently implemented in \cosmobit, and the ranges of the parameters that we explore in the examples shown in this Section. Where applicable, we quote the associated potential and parameters. For further details on the models, see Appendix~\ref{app:models-inflation}. \label{tab:inflationary_models}}
\end{table}

We now demonstrate how \cosmobit\ can be used to study various aspects of inflation in the well-studied case of instant reheating.
Table~\ref{tab:inflationary_models} provides a quick overview of all inflationary models presently implemented in \cosmobit; for further details, see Appendix~\ref{app:models-inflation}.
We choose these six different inflationary potentials as our benchmark cases: four monomial potentials $V(\phi) \propto \phi^n$ with exponents $n=2/3$, $1$, $2$, and $4$, natural inflation, and Starobinsky inflation~(which is also referred to as $R^2$ inflation, and related to Higgs inflation).
While monomial potentials could be studied in all generality for arbitrary~$n$, it is arguably more interesting to pick individual values of $n$ for which an underlying physical model is known. These models have been studied at length previously, with $n \geq 2$ now strongly disfavoured by observations~\cite{Planck18inflation}.
The original Starobinsky inflation model is particularly relevant because it predicts a rather low tensor-to-scalar ratio, and is therefore highly compatible with all known observations.
Natural inflation is an example of a two-parameter model with a slightly higher degree of complexity than the others we look at (although it should be noted that it is now also in tension with observations~\cite{Planck18inflation}).

We use \diver \cite{ScannerBit} to explore the Planck 2018 lensing, low-$\ell$ TTEE and lightweight (`lite') high-$\ell$ TTTEEE likelihoods for the six inflationary models in combination with the \doublecrosssf{LCDM}{LCDM} cosmological model, setting the arbitrary pivot scale to $k_\star = 0.05\,\text{Mpc}^{-1}$. For comparison, we also scan the phenomenological \doublecrosssf{PowerLaw\_ps}{PowerLaw_ps} with \doublecrosssf{LCDM}{LCDM}. Although we can obtain the full primordial power-spectrum numerically from \multimodecode in each case, and pass it on to \class for predicting the CMB power spectra, we use a simpler finite-difference method to infer only the parameters of the phenomenological power-law approximation to the primordial spectrum. This is sufficient to directly compare with the slow-roll results, which predict the corresponding parameters $n_s$ and $r$. Note, however, that using the full power spectrum would have yielded an equally valid comparison. We also include $A_\text{Planck}$, the nuisance parameters associated with the `lite' Planck likelihoods, in our analysis~(see Appendix~\ref{app:models-nuisance}). Since we perform a frequentist analysis, the $A_\text{Planck}$ prior is re-interpreted as an additional Gaussian nuisance likelihood, with the allowed (necessarily finite) range for $A_\text{Planck}$ chosen to be $\pm 5\sigma$ around the mean. This choice does not influence the outcome of the analysis. We emphasise that, in the general spirit of \GB, we consider and scan the fundamental parameters of the inflationary models, unlike in other work where these are determined by the matching equation. See e.g.\ Refs~\cite{Mortonson:2010er,Planck13inflation} for work following the same rationale.

\begin{table}
  \centering
  \begin{tabular}{@{}lr@{\ }S[table-format=1.2 e-2]S@{}}
    \toprule
    Inflationary model & \multicolumn{2}{c}{Best fit} & \multicolumn{1}{c}{$\Delta \mathrm{ln}(\mathcal{L})$} \\
    \midrule
    \doublecrosssf{$n=2/3$}{Inflation_InstReh_1mono23} & $\lambda=$ & 2.31e-10 & -3.9 \\
    \doublecrosssf{$n=1$ (linear)}{Inflation_InstReh_1linear} & $\lambda=$ & 2.03e-10 & -2.6 \\
    \doublecrosssf{$n=2$ (quadratic)}{Inflation_InstReh_1quadratic} & $m_\phi/\mplred=$ & 6.06e-06 & -3.3 \\
    \doublecrosssf{$n=4$ (quartic)}{Inflation_InstReh_1quartic} & $\lambda=$ & 1.45e-13 & -24.7 \\
    \doublecrosssf{Starobinsky}{Inflation_InstReh_1Starobinsky} & $\lambda=$ & 3.23e-03 & -0.1 \\
    \doublecrosssf{Natural}{Inflation_InstReh_1natural} & $\lambda=$ & 6.97e-3 & -2.0 \\
    & $f_\phi/\mplred=$ & 8.26 & \\
    \bottomrule
  \end{tabular}
  \caption{Best-fit points from scans of inflationary models. We list the difference in log likelihood at the best-fit point with respect to the phenomenological \doublecrosssf{PowerLaw\_ps}{PowerLaw_ps} model, which has an absolute likelihood value of $\mathrm{ln}(\mathcal{L}) = -500.9$.
  \label{tab:inflation_bestfit}}
\end{table}

Table~\ref{tab:inflation_bestfit} lists the best-fit values of the inflationary model parameters. As already noted, instant reheating renders simple single-field inflation models very predictive in terms of~$N_\star$, and the data have a large influence on the preferred values of the model parameters.
As is well-known in the literature, Starobinsky inflation is in better agreement with the data than any of the other models that we examine. This is due to the excellent agreement of predicted and observed values of~$n_s$ and, at the same time, a low value of~$r$ that is compatible with the data. Quartic inflation~($n = 4$) does not posses either of those qualities. While it was already in tension with observations from WMAP~\cite[e.g.][]{wmap3year}, it is now strongly disfavoured by the Planck data.

By using the result of the power-law approximation to their primordial spectra, we can visualise these models in terms of $n_s$ and $r$ (Fig.~\ref{fig:inflation_bestfit}). For each model, we show the best-fit points (stars), as well as solid lines indicating the line that would be traced by the predictions if $N_\star$ were varied from~$50$ to~$65$~(small and large dots, respectively). We calculate these numbers to second order\footnote{The second-order corrections to the slow-roll approximation are important for models with larger values of $r$; the numerical best-fit point for e.g.\ the $n = 4$ model is noticeably off the slow-roll prediction without including this correction.} in the slow-roll approximation~\cite[e.g.][]{astro-ph/9408015}, using the analytical results from Ref.~\cite{Martin:2013tda} for the various inflation models. As the prediction of $N_\star$ is rather specific in the instant-reheating scenario, we do this for illustrative purposes only, in order to demonstrate how the predictions would change in a more general reheating scenario. In such a scenario, the value of $N_\star$ is expected to be smaller than the instant reheating prediction. In this sense, the best-fit points~(stars) can be interpreted as upper limits on the possible values of~$N_\star$. $N_\star = 65$ can therefore be seen to be unrealistically large in all models. The best-fit values for~$N_\star$ range from 56.3 (Starobinsky) to 59.2 ($n = 4$).

To demonstrate that we do indeed fully scan all of the parameters, we also include an example inset in Fig.~\ref{fig:inflation_bestfit}, showing the profile likelihood values around the best-fit point for quartic inflation ($n = 4$). Note that the size of the zoom-in, indicated on the axes, is a few orders magnitude smaller than the size of the best-fit markers.

For natural inflation, the blue dashed line indicates the location of the (narrow) 95\%-confidence region.

The dotted black lines in Fig.~\ref{fig:inflation_bestfit} indicate the 68\% and 95\% credible regions obtained with the same likelihood function in the most recent Planck collaboration papers~\cite{Planck18inflation}. We also performed a similar fit to the general phenomenological power-law model (\doublecrosssf{PowerLaw\_ps}{PowerLaw_ps}), where~$n_s$ and~$r$ are free and independent parameters.  The resulting profile likelihoods are shown in solid grey contours and grey shading in Fig.~\ref{fig:inflation_bestfit}.  These agree well with the Planck results, with some small differences (as expected) due to our use of the `lite' rather than full high-$\ell$ Planck likelihood.

\begin{figure}
  \centering
  \includegraphics[width=0.46340504409369\linewidth]{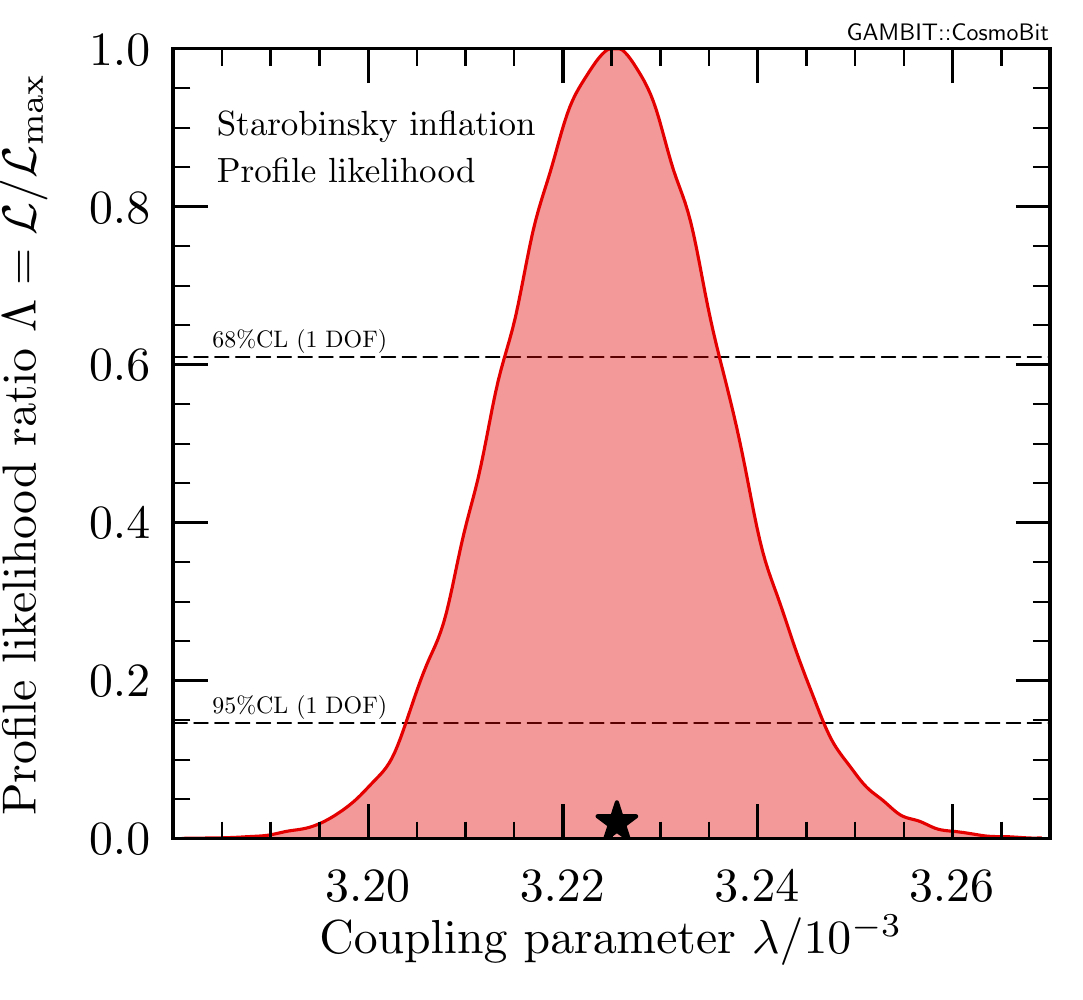}
  \includegraphics[width=0.52659495590631\linewidth]{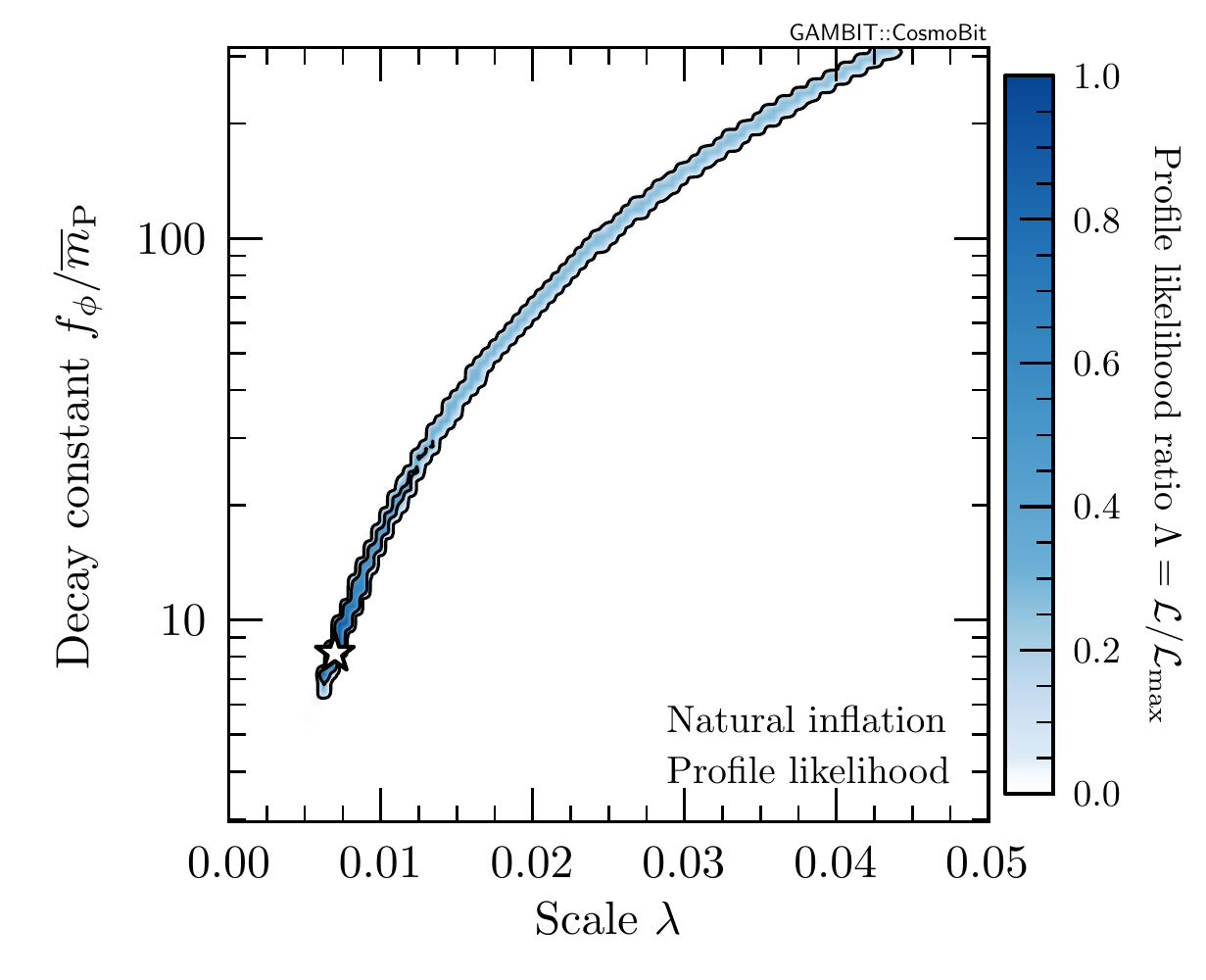}
  \caption{Profile likelihoods of the inflaton potential parameters for Starobinsky (\textit{left}) and  natural inflation (\textit{right}). Stars in both panels indicate the best-fit point, and contours in the right panel indicate 68\% and 95\%~confidence levels. \label{fig:inflation_profile_likelihoods}}
\end{figure}

In Fig.~\ref{fig:inflation_profile_likelihoods} we show examples of the best-fit regions of the inflationary model parameters. The left panel shows the well-constrained coupling parameter for Starobinsky inflation, whereas the right panel shows the two-dimensional profile likelihood for the parameters of natural inflation. The tight degeneracy between $\lambda$ and $f_\phi$ is consistent with the degeneracy seen between $n_s$ and~$r$, and arises from the fact that for $f_\phi \gg \mplred$, $n_s$, $r$ and $\lambda$ can all be determined from $f_\phi$ alone.

\subsection{Neutrino masses}
\label{sec:Nu-masses}

Many of the cosmological observables described in Sec.\ \ref{sec:lcdm-obs}, such as the CMB (Sec.~\ref{sec:CMB}) and the matter power spectrum (Sec.~\ref{sec:late_time}), are sensitive to the total sum of neutrino masses $\sum m_\nu$. Prior to recombination, massive neutrinos are still relativistic and act as radiation, slightly enhancing small scale anisotropies in the CMB. They free-stream with some characteristic thermal velocity close to the speed of light, which also sets a free-streaming length close to the Hubble radius. Sometime after recombination, during matter domination, the neutrinos transition to being non-relativistic, their thermal velocity starts to decay, and the free-streaming length grows more slowly than the scale factor of the Universe.  This creates a characteristic scale $k_{nr}$ in the matter power spectrum given by the free-streaming length at the time of the non-relativistic transition. On scales larger than this ($k < k_{nr}$), there is no effect from neutrino free-streaming and the now non-relativistic neutrinos simply contribute in the same way as cold dark matter. However, on small scales ($k>k_{nr}$), the free-streaming of neutrinos at early times has caused a suppression in the matter power spectrum. This is because the perturbations in the non-relativistic neutrinos are much smaller than for dark matter, and because they did not originally contribute gravitationally to the growth of the baryonic or cold dark matter perturbations. Simply put, non-relativistic massive neutrinos still count towards the total matter budget and expansion rate but, at fixed total matter density, they contribute less to the $k > k_{nr}$ clustering than if the same proportion of the matter density were replaced with cold dark matter.

Hence, the presence of massive neutrinos can be detected in the CMB, as a change in the non-linear matter power spectrum traced by weak lensing and galaxy clustering, as a scale-dependence in the growth rate of structure measured by galaxy clustering in redshift-space, and in the expansion history measured by e.g., BAO and supernovae. Consequently, the simplest $\Lambda$CDM scenario requires non-zero neutrino masses in order to satisfy the observations, but the impact of the individual mass values or their hierarchy is negligible~\cite{Archidiacono:2020dvx}. It is therefore common (e.g.\ in the Planck 2015 and 2018 analyses~\cite{Ade:2015xua, Aghanim:2018eyx}) to use a `baseline' model with only one massive neutrino, with mass $m_{\nu_3} = 0.06$\,eV. However, this approach does not represent a realistic neutrino model as it does not take into account the limits on neutrino mass splittings from oscillation experiments.

Realistic neutrino models are parametrised by the masses of the three flavour-ordered neutrino mass eigenstates $m_{\nu_i}$, $(i=1,2,3)$.  These are related to the flavour eigenstates $\nu_\alpha$ $\alpha=(e,\mu,\tau)$ via the Pontecorvo-Maki-Nakagawa-Sakata (PMNS) matrix $U_{\alpha i}$~\cite{Maki:1962lba,Pontecorvo:1957qd}, as $\nu_\alpha = U_{\alpha i}~ \nu_i$. The PMNS matrix can be written in terms of three mixing angles, $\theta_{12}$,  $\theta_{13}$ and $\theta_{23}$, and one $CP$-violating phase, $\delta_{\text{CP}}$. Non-vanishing values of these mixing angles produce oscillations between the neutrino flavour eigenstates in flight.

The probability of oscillation from one neutrino flavour eigenstate $\nu_\alpha$ to another $\nu_\beta$ is given by
\begin{align}
\notag P_{\alpha \rightarrow \beta } = \delta _{\alpha \beta }&{}-4\sum _{i>j}{\rm {Re}}\left(U_{\alpha i}^{*}U_{\beta i}U_{\alpha j}U_{\beta j}^{*}\right)\sin ^{2}\left({\frac {\Delta m_{ij}^{2}L}{4E}}\right)\\&{}+2\sum _{i>j}{\rm {Im}}\left(U_{\alpha i}^{*}U_{\beta i}U_{\alpha j}U_{\beta j}^{*}\right)\sin \left({\frac {\Delta m_{ij}^{2}L}{2E}}\right),\end{align}
where $\Delta m_{ij}^2  = m^2_i - m^2_j$ is the mass difference among neutrino mass eigenstates, $L$ is the travel distance and $E$ the total energy of the oscillating neutrino. Neutrino oscillations have been measured successfully at various solar (Homestake chlorine \cite{Cleveland:1998nv}, Gallex/GNO \cite{Kaether:2010ag}, SAGE \cite{Abdurashitov:2009tn}, SNO \cite{Aharmim:2011vm}, Super-Kamiokande \cite{Hosaka:2005um,Cravens:2008aa,Abe:2010hy,skiv}, Borexino \cite{Bellini:2011rx,Bellini:2008mr,Bellini:2014uqa}), atmospheric (IceCube/DeepCore \cite{Aartsen:2014yll} and Super-Kamiokande \cite{Abe:2017aap}), reactor (KamLAND \cite{Gando:2013nba}, DoubleChooz \cite{dchooz}, Daya-Bay \cite{An:2016srz,Adey:2018zwh}, Reno \cite{Bak:2018ydk}), and accelerator experiments (MINOS \cite{Adamson:2013whj,Adamson:2013ue}, T2K \cite{t2k,t2k2}, NO$\nu$A \cite{sanchez_mayly_2018_1286758,Acero:2019ksn}).

Simultaneous fits of the results of all oscillation experiments (e.g.~by \textsf{NuFit}~\cite{Esteban:2018azc} and others~\cite{Capozzi:2018ubv,deSalas:2017kay}) strongly constrain the values of the neutrino mass splittings $\Delta m_{ij}^2$, the mixing angles $\theta_{ij}$, and the $CP$-violating phase $\delta_{\text{CP}}$. The absolute value of the lightest neutrino mass $m_{\nu_0}$ remains unconstrained by oscillation data, as (mostly) does the hierarchy of neutrino mass eigenstates. Two hierarchies are possible: the normal hierarchy, where $m_{\nu_1} < m_{\nu_2} < m_{\nu_3}$, and the inverted hierarchy, with $m_{\nu_3} < m_{\nu_1} < m_{\nu_2}$. The neutrino masses can then be re-parametrised in terms of the mass of the lightest neutrino $m_{\nu_0}$ and the mass splittings constrained by oscillation experiments: $\Delta m_{21}^2$ and $\Delta m_{3l}^2$, with $l=1$ for the normal hierarchy and $l=2$ for the inverted hierarchy.

As part of the broader \GB framework, \cosmobit can access the neutrino mass models and likelihoods implemented in \neutrinobit~\cite{RHN}. The neutrino mass parameters $\Delta m_{21}^2$, $\Delta m_{3l}^2$ and $m_{\nu_0}$ are relevant to cosmological likelihoods by way of the sum of the masses $\sum m_\nu$, and thus directly enter scans performed with \cosmobit. The mass splittings are strongly constrained by data from oscillation experiments, so we model them as two-dimensional Gaussian likelihoods following the results of \textsf{NuFit}, whereas the lightest neutrino mass $m_{\nu_0}$ is unconstrained by oscillation data and thus a free parameter. On the other hand, the mixing angles $\theta_{ij}$, responsible for neutrino oscillations, and the $CP$ phase $\delta_\text{CP}$, are of no consequence for the likelihoods presently implemented in \cosmobit.  We therefore simply set these to their central values from the \textsf{NuFit} results in all examples. Section \ref{sec:constraints} and the companion paper \cite{CosmoBit_numass} provide detailed examples of the use of this setup in specific physics applications.

\subsection{Non-standard radiation content}
\label{sec:non-std-Neff}

The radiation content of the Universe is defined as the total energy density in highly relativistic species. In the cosmological standard scenario, it only receives two contributions, from photons (with a present-day temperature of $\Tcmb = 2.7255\,\mathrm{K}$) and from SM neutrinos as long as they are ultra-relativistic.
The radiation content can be modified in three different ways: \emph{i)} allowing variations in the photon temperature, \emph{ii)} allowing variations in the neutrino temperature $T_\nu$ or the neutrino phase-space distribution, and \emph{iii)} allowing extra contributions from additional relativistic species (so-called dark radiation).

The first option, allowing for variations in the photon temperature, is constrained by measurements of the CMB temperature. Within \cosmobit, the user can either fix $\Tcmb$ to a certain value, vary it freely or use a Gaussian likelihood from the COBE/FIRAS measurement as a constraint.

The second option, a non-standard value of the neutrino temperature today, is not constrained by any direct measurement to date, and can be varied with much more freedom than the CMB temperature. Nevertheless, changing the energy density of neutrinos affects the Hubble rate during radiation domination, which has a number of important effects on early Universe cosmology. Firstly, an increased Hubble rate means that the various reactions relevant for BBN fall out of equilibrium at higher redshift. Therefore, a smaller number of neutrons are available for the formation of helium. Hence, the helium abundance decreases with increasing $T_\nu$. Likewise, the point of matter-radiation equality is shifted to smaller redshift, which enhances the early ISW effect. A more subtle effect of changing the neutrino temperature is that it changes the redshift at which neutrinos become non-relativistic and stop free-streaming. Hence, the small scales of the matter power spectrum are also affected by modifications of $T_\nu$.

Modifications of the neutrino temperature arise in many extensions of standard cosmology. For example, any particle species that decays or annihilates after neutrino decoupling can transfer a fraction of its energy into the electron-photon plasma and another fraction into neutrinos. If most of the energy is injected into the electron-photon plasma~-- as in the case of electron-positron annihilations~-- the result is an effective decrease in the neutrino temperature (relative to photons), and vice versa. If, on the other hand, some of the energy is injected into neutrinos, it induces non-thermal distortions. This interesting case however requires a more general parametrisation of the neutrino phase space distribution (see e.g. Refs.~\cite{Cuoco:2005qr,Oldengott:2019lke}) and is not considered further in the present work.

To study the impact of changing the neutrino temperature, we define the ratio
\begin{equation} \label{eq:def-r-nuT}
 r_\nu = \frac{T_\nu}{T_\nu^\lcdm} \; ,
\end{equation}
where $T_\nu^\lcdm$ is the neutrino temperature in $\lcdm$, such that $r_\nu = 1$ in standard cosmology. The simplest assumption is that $r_\nu$ is constant throughout the observable history of the Universe (i.e.\ for $T_\gamma \lesssim 1\,\mathrm{MeV}$). This assumption means that any changes relative to the standard scenario happen in the short time window between neutrino decoupling and the beginning of BBN.\footnote{Note that the value of $r_\nu$ is not affected by electron-positron annihilation, which simply reduces the neutrino temperature (relative to the photon temperature) by a factor $(4/11)^{1/3}$.} It is also conceivable that the photon-to-neutrino temperature ratio changes after BBN, for example due to the decays of a very long-lived exotic particle species into electromagnetic energy. These decays inject energy into the photon bath, increasing the photon temperature with respect to the neutrino temperature. As the former is well constrained by measurements of the CMB temperature, the observable effect of such decays would manifest as a neutrino temperature smaller than expected from SM calculations.
In such a case one generally expects also the baryon-to-photon ratio $\eta_\mathrm{b} = n_b / n_\gamma$ to change between the end of BBN and the beginning of recombination.

In order to study scenarios with non-standard neutrino temperatures in \cosmobit, we introduce three additional model parameters: $r_{\text{CMB}}$, $r_{\text{BBN}}$ and $\eta_{b,\text{BBN}}$.  The values of $r_{\text{CMB}}$ and $r_{\text{BBN}}$ parametrise modification of the neutrino temperature following Eq.~(\ref{eq:def-r-nuT}) at the beginning of recombination and at the end of BBN, respectively. The parameter $\eta_{b,\text{BBN}}$ indicates the value of the baryon-to-photon ratio at the end of BBN. The baryon-to-photon ratio at recombination is fixed by the $\Lambda$CDM parameters $\Omega_b$ and $\Omega_\gamma$. Of course, in realistic models these quantities are not fundamental but derived from other model parameters, such as the abundance and decay rate of an exotic particle species. The \gambit model hierarchy makes it straight-forward to embed the model currently implemented in \cosmobit into a more complete model, such that the underlying parameters are mapped onto the effective parameters introduced above. We leave more detailed studies of such models to future work~\cite{CosmoALP}.

We emphasise that the current implementation in \cosmobit only covers cases in which the neutrino temperature ratio $r_\nu$ remains constant  during BBN ($T \gtrsim 10\,\mathrm{keV}$) and after the beginning of recombination ($T \lesssim 100\,\mathrm{eV}$). For the case that the neutrino temperature is modified during or after recombination through the injection of electromagnetic energy we will discuss relevant complementary constraints in Sec.\ \ref{sec:energy-injection}. The case of exotic particles decaying or annihilating during BBN is a very interesting and active research topic~\cite{Hufnagel:2017dgo, Hufnagel:2018bjp, Escudero:2018mvt, Depta:2019lbe, Sabti:2019mhn}.  Such a scenario, however, is beyond the scope of the first release of \cosmobit.

Finally, the third way to alter the radiation content of the Universe is to introduce additional ultra-relativistic species. These may be e.g.\ sterile neutrinos or dark radiation with a non-negligible energy density. It is common to parametrise these additional species as a non-standard contribution to $N_\text{eff}$, which we denote $\Delta N_\text{ur}$,
\begin{equation}
 \Delta N_\text{ur} (z) = \frac{\rho_\text{dr} (z)}{\tfrac{7}{8} \large( \tfrac{4}{11}\large)^{4/3} \rho_\gamma} \, ,
\end{equation}
where $\rho_\text{dr}(z)$ denotes the redshift-dependent contribution from the additional source of radiation.
Note that the presence of non-standard ultra-relativistic species always increases $\rho_r$, i.e.\ it leads to $\Delta N_\text{ur} \geq 0$.

In general, the time evolution of $\rho_\text{dr}(z)$ is model-dependent. For example, the energy density in dark radiation decreases whenever the temperature drops below the mass of one of the particles in the dark sector~\cite{Blennow:2012de}. Conversely, an increase of $\rho_\text{dr}$ can result from the annihilation or decay of other particle species, as in the case of the conversion of dark matter into dark radiation~\cite{Bringmann:2018jpr}. To capture these effects, we introduce two independent parameters, $\Delta N_\text{ur,BBN}$ and $\Delta N_\text{ur,CMB}$, corresponding to $\Delta N_\text{ur}$ at the end of BBN and the beginning of recombination, respectively. With this parametrisation, we do not allow for $\Delta N_\text{ur}$ to change after recombination and we assume that any particles that contribute to $\rho_\text{dr}$ during recombination are still ultra-relativistic in the present Universe. Further, we assume that dark radiation does not interact with any of the other particle species and has no self-interactions (i.e.\ it is free-streaming). Hence, the only effect of the additional contribution to $\rho_r$ is to increase the Hubble rate during radiation domination. In these scenarios, the presence of additional ultra-relativistic particles is largely equivalent to an increase in the neutrino temperature.

Modified neutrino temperatures and additional ultra-relativistic species both alter the effective number of ultra-relativistic species $N_\text{eff}$, giving
\begin{equation}
 N_\text{eff} = r_\nu^4 \, N_\nu + \Delta N_\text{ur}.
\end{equation}
Here the canonical $\lcdm$ value is $N_\nu = 3.045$. Details of the models by which we implement these modifications of the radiation content of the Universe in \cosmobit can be found in Appendix~\ref{app:models-dNur}.

\begin{figure}
\begin{center}
\includegraphics[width=0.56\textwidth]{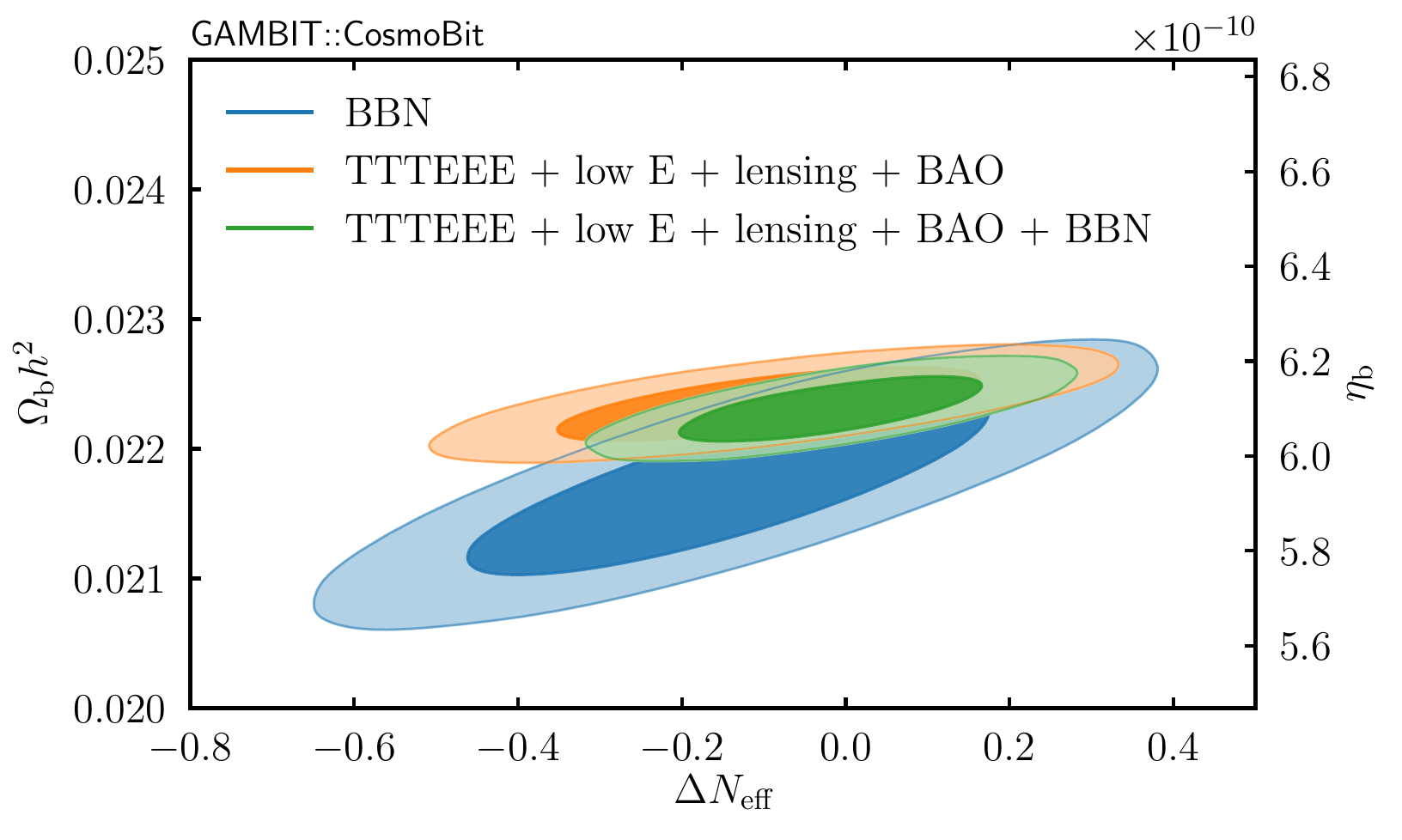}
\includegraphics[width=0.40\textwidth]{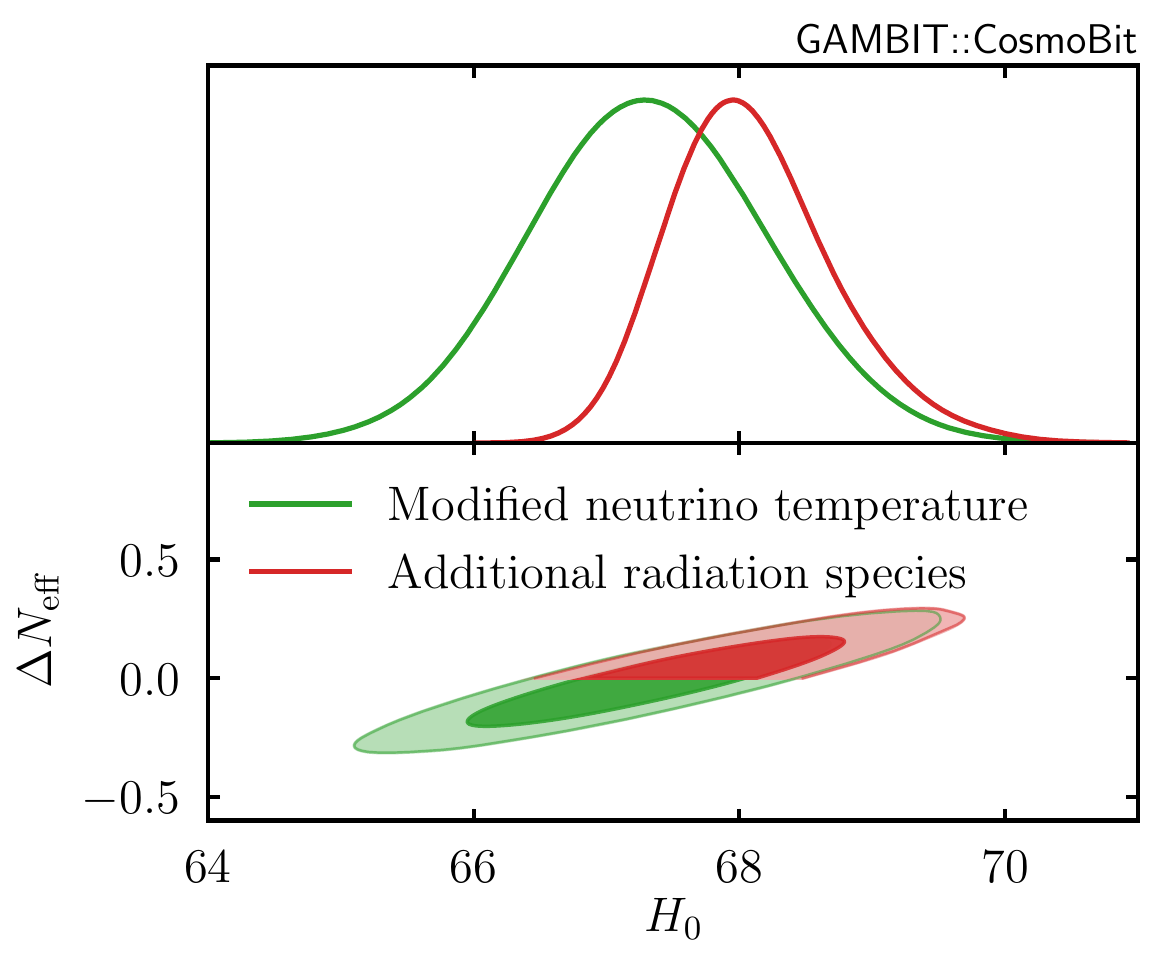}
\caption{\textit{Left}: constraints on $N_\text{eff}$ and $ \Omega_b h^2$ (equivalently $\eta_\mathrm{b}$) from BBN and the CMB alone, and from their combination.  Here we assume that $N_\text{eff}$, $ \Omega_b h^2$ and $\eta_\mathrm{b}$ are constant between the end of BBN and the time of recombination. \textit{Right}: constraints on $N_\text{eff}$ and $H_0$ from the combination of BBN and CMB in the case that the neutrino temperature is varied (allowing $\Delta N_\text{eff} < 0$) or that it is kept fixed and additional ultra-relativistic species are introduced instead (imposing $\Delta N_\text{eff} > 0$). The contours indicate the 68\% and 95\% credible regions.
\label{fig:BBN}}
\end{center}
\end{figure}

In Fig.~\ref{fig:BBN}, we show some examples of the effects of such modifications of the radiation content.  In the left panel, we extend $\lcdm$ by the neutrino temperature ratio $r_\nu$, allowing $\Delta N_\text{eff} \equiv N_\text{eff} - N_\nu$ to take both positive and negative values.  We plot the resulting constraints on $\Delta N_\text{eff}$ and $\Omega_b h^2$ from BBN and from the CMB, as well as the combined constraint obtained when assuming that $N_\text{eff}$ and $\eta_\mathrm{b}$ do not change between the two observations. Our results are in agreement with those obtained by the Planck collaboration~\cite{Aghanim:2019ame}. If we instead keep $r_\nu$ fixed and vary only $\Delta N_\text{ur}$, we impose the constraint $\Delta N_\text{eff} > 0$. As shown in the right panel of Fig.~\ref{fig:BBN}, doing so leads to a visible shift in the posterior probabilities for other cosmological parameters. 
In particular, the prior constraint $\Delta N_\text{eff} > 0$ leads to a shift of the most probable value of $H_0$ to larger values. While it is tempting to interpret this result in the context of the $H_0$ tension (Sec.~\ref{sec:tensions}) it is purely caused by the change of priors, not because higher $H_0$ values accommodate the data better in this scenario. 
We will revisit this effect in a different context in Sec.~\ref{sec:constraints}.

\subsection{Exotic forms of energy injection}
\label{sec:energy-injection}

Any form of energy injection during recombination, or during the ``dark ages'' between recombination and reionisation, affects the thermal history of the primordial plasma. In particular, the evolution of the free electron fraction $ x_e \equiv n_e/n_H $ and the temperature of the intergalactic medium $ T_{\rm M }$ are modified by the injection of energy. These modifications potentially alter the optical depth $ \tau (z) $, leading to distinguishable features in the CMB power spectra, distortions of the CMB blackbody spectrum and observables related to the 21\,cm line~\cite{Furlanetto:2006jb}.

In the three-level atom approximation~\cite{Peebles:1968ja,Zeldovich:1969en} the evolution of $ x_e $
and  $ T_{\rm M} $ are given by
\begin{eqnarray}
  \frac{{\rm d}x_e(z)}{{\rm d}z} &=& \frac{1}{(1+z) H(z)} \left[ R(z) - I(z) -I_\chi (z)\right]\,,
  \label{eq:recombination}\\
  \frac{{\rm d}T_{\rm M}(z)}{{\rm d}z} &=& \frac{1}{1+z} \left[2\,T_{\rm M} + \gamma(T_{\rm M} - T_{\rm CMB})\right] - K_h\,,
    \label{eq:igm_temperature}
\end{eqnarray}
where the functions $ I(z) $ and $ R(z) $ are the ionisation and recombination rates, $K_h$ is the heating term and the factor $ \gamma $ refers to the dimensionless Compton scattering rate (which directly depends on $x_e$).
In the presence of non-standard electromagnetic energy injection, Eqs.~\eqref{eq:recombination} and~\eqref{eq:igm_temperature} are modified by additional ionisation and heating terms (see Ref.~\cite{Poulin:2016anj} for a more detailed discussion).
The dominant ionisation contributions are from direct ionisation,
\begin{equation}
  I_{\chi,\,{\rm ion.}} = -\frac{1}{n_H E_i} \left.\frac{{\rm d}^2 E}{{\rm d}V \, {\rm d}t}\right|_{\rm dep,ion}\,,
\end{equation}
and from excitation into the excited state and subsequent ionisation,
\begin{equation}
  I_{\chi,\,{\rm exc.}} = -\frac{1-\mathcal{C}}{n_H E_\alpha} \left.\frac{{\rm d}^2 E}{{\rm d}V \, {\rm d}t}\right|_{\rm dep,exc}\,,
\end{equation}
where $ E_i $ and $ E_\alpha $ are the average ionisation energy per baryon and the energy of the
Lyman-$ \alpha $ transition, respectively, and $ \mathcal{C} $ is the Peebles C factor (i.e. the probability for an excited hydrogen atom to decay before photoionisation can occur) for case-B recombination \cite{Peebles:1968ja}. The additional heating is
\begin{equation}
  K_h =  - \frac{2}{H(z) (1+z) 3 n_H (1 + f_{\rm He} + x_e)} \left.\frac{{\rm d}^2 E}{{\rm d}V \, {\rm d}t}\right|_{\rm dep,heat}\,.
\end{equation}

All new terms are proportional to the rate per unit volume $ \left.\frac{{\rm d}^2 E}{{\rm d}V \, {\rm d}t}\right|_{\rm dep,c} $ at which energy is deposited into the primordial plasma via the  channel $ c \in \left\lbrace {\rm ionisation, excitation, heating} \right\rbrace$. The rate of energy deposition in turn depends on the rate per unit volume of energy injection $ \left.\frac{{\rm d}^2 E}{{\rm d}V \, {\rm d}t}\right|_{\rm inj} $, which can be directly calculated for a given injection mechanism. The initial version of \cosmobit provides injection rate calculations for two common scenarios of energy injection: $s$-wave annihilation of dark matter and decaying dark matter.

For $s$-wave annihilations, the energy injection rate is given by
\begin{equation}
  \left.\frac{{\rm d}^2 E}{{\rm d}V \, {\rm d}t}\right|_{\rm inj,ann.} = \kappa \xi^2(z) \rho_{\rm cdm}^2 (1+z)^6 \frac{\langle\sigma v\rangle}{m_\chi}\,,
\end{equation}
where $ \xi(z) \equiv \rho_\chi(z) / \rho_{\rm cdm} $ is the fractional abundance of the annihilating dark matter candidate $ \chi $ with respect to the abundance of all dark matter, $ \rho_{\rm cdm} $ at redshift $ z $. The factor $ (1+z)^6 $ takes into account the dilution of $\rho_{\rm cdm}^2$ with the expansion of the universe. The total thermally-averaged cross-section for $ \bar\chi\chi $-annihilation is $ \langle\sigma v\rangle $, and $ m_\chi $ is the dark matter mass. If dark matter is self-conjugate $\kappa =1 $; otherwise $\kappa=1/2$.

For decaying dark matter, the energy injection rate is given by
\begin{equation}
  \left.\frac{{\rm d}^2 E}{{\rm d}V \, {\rm d}t}\right|_{\rm inj,dec.} = \xi(z) \rho_{\rm cdm} (1+z)^3\, \tau^{-1}\,.
\end{equation}
Here, the injection rate is driven by the dark matter lifetime $\tau$. For decaying dark matter, the redshift dependence of the fractional dark matter abundance is $ \xi(z) = \xi_0 \, \exp\left(-t(z)/\tau\right) $.  The reference value $ \xi_0 $ represents the initial fractional dark matter abundance in the infinite past ($ t=0 $).
Note that in the case of annihilating dark matter, the annihilation rate is constrained to be so small that we can treat $\xi$ as constant. In the case of decaying dark matter, the fraction of dark matter that decayed during recombination is constrained to be so small that we can treat all effects at the background level, and neglect the impact on perturbations.

The first release of \cosmobit supports only $e^+ e^-$, $\gamma \gamma$ and $\nu\nu$ annihilation or decay final states, and mixtures thereof. This approach is sufficient if the decaying or annihilating dark matter particles have a mass below about $100\,\mathrm{MeV}$ (such that all other final states are kinematically forbidden), but requires extension for more general models. We provide details on how to include additional final states in Appendix~\ref{app:darkages-new-model}. Future versions of CosmoBit will contain an interface with the \gambit process catalogue, so that the appropriate injection spectra are automatically provided by the \darkbit module.

The rate of energy deposition depends almost exclusively on the energy injected in the form of electrons, positrons and photons.\footnote{In general also the production of protons and antiprotons affects the evolution of $x_e$ and $T_M$. The inclusion of these states in the calculation therefore leads to stronger bounds, but requires an increased computational effort~\cite{Weniger:2013hja}. Here we neglect these final states and therefore provide conservative constraints.}
More precisely, the energy injection rate and energy deposition rate are linked through dimensionless efficiency functions $ f_c(z,x_e) $ that depend on redshift and the injection channel $c$:
\begin{equation}
  \left.\frac{{\rm d}^2 E}{{\rm d}V \, {\rm d}t}\right|_{{\rm dep},c} (z,x_e) = f_c(z,x_e) \,\cdot\, \left.\frac{{\rm d}^2 E}{{\rm d}V\,{\rm d}t}\right|_{\rm inj} (z)\,.
\end{equation}
These can be obtained by convolving the primary rate $ \left.\frac{{\rm d}^2 N}{{\rm d}E\,{\rm d}t} \right\vert_{\rm inj}^{(\ell)} (z',E)$ of electrons, positrons or photons with energy $ E $ injected at redshift $ z' $ with transfer functions $ T_c^{(\ell)} (z',z,E,x_e) $ \cite{Slatyer:2015kla}, such that
\begin{equation}
  f_c(z,x_e) = \frac{\sum\limits_{\ell} \int\limits_{z}^\infty \!{\rm d\,ln}(1 + z')  \frac{(1+z')^{\alpha-3}}{H(z')} \int\limits_{0}^{E_{\rm max}} \!{\rm d}E\, T_c^{(\ell)} (z',z,E,x_e) E \left.\frac{{\rm d}^2 N}{{\rm d}E\,{\rm d}t} \right\vert_{\rm inj}^{(\ell)} (z',E)}{\frac{(1+z)^{\alpha-3}}{H(z)} \int\limits_{0}^{E_{\rm max}} \!{\rm d}E\, E \left.\frac{{\rm d}^2 N}{{\rm d}E\,{\rm d}t} \right\vert_{\rm inj}^{\rm tot} (z,E)}\,,
  \label{eq:energy_injection_efficiency_convolution}
\end{equation}
where $ \alpha = 3  $ for decay and $ \alpha = 6 $ for annihilation. For the case of decaying dark matter, the exponential suppression of the abundance is implicitly included in the redshift dependence of the differential injection rate $\mathrm{d}^2N/\mathrm{d}E\mathrm{d}t$. In this form, the transfer function can be understood as the probability that a particle of species $ \ell \in \left\lbrace e^\pm, \gamma \right\rbrace $ injected at a redshift $ z' $ with energy $ E $ will deposit its remaining energy at redshift $ z < z' $.
Eq.\ \eqref{eq:energy_injection_efficiency_convolution} also takes into account the possibility that a particle can freely stream through the intergalactic medium before it deposits the energy, as opposed to the so-called ``on-the-spot'' approximation~\cite{Lopez-Honorez:2013cua,Diamanti:2013bia}.

It is important to note that these transfer functions~-- and especially the underlying information on the
repartition of the available energy into the deposition channels $ c $~-- depend on the free electron
fraction $ x_e $, which in turn implicitly depends on redshift.
The transfer functions as presented in Ref.~\cite{Slatyer:2015kla} are based on the implicit assumption that the evolution of $ x_e $ is identical to the one in the standard scenario of vanishing energy injection,
which in general leads to an underestimation of the impact on the temperature of the IGM \cite{Liu:2019bbm}.
The effect of a modified free electron fraction can be captured
by assuming the efficiency functions factorise as
\begin{equation}
  f_c(z,x_e) = \chi_c(x_e) \,\cdot\, f_{\rm eff} (z) \,,
  \label{eq:energy_injection_factorisation}
\end{equation}
where the (channel-dependent) repartition functions $\chi_c$ only depend on the free-electron fraction,
and $ f_{\rm eff} (z) $ is an $ x_e $-independent, {\it effective} efficiency function for all channels~\cite{Galli:2013dna} (see Ref.~\cite{Slatyer:2015kla} for details on the construction of $ f_{\rm eff} $).

\begin{figure}
  \begin{center}
    \includegraphics[width=0.4921\textwidth]{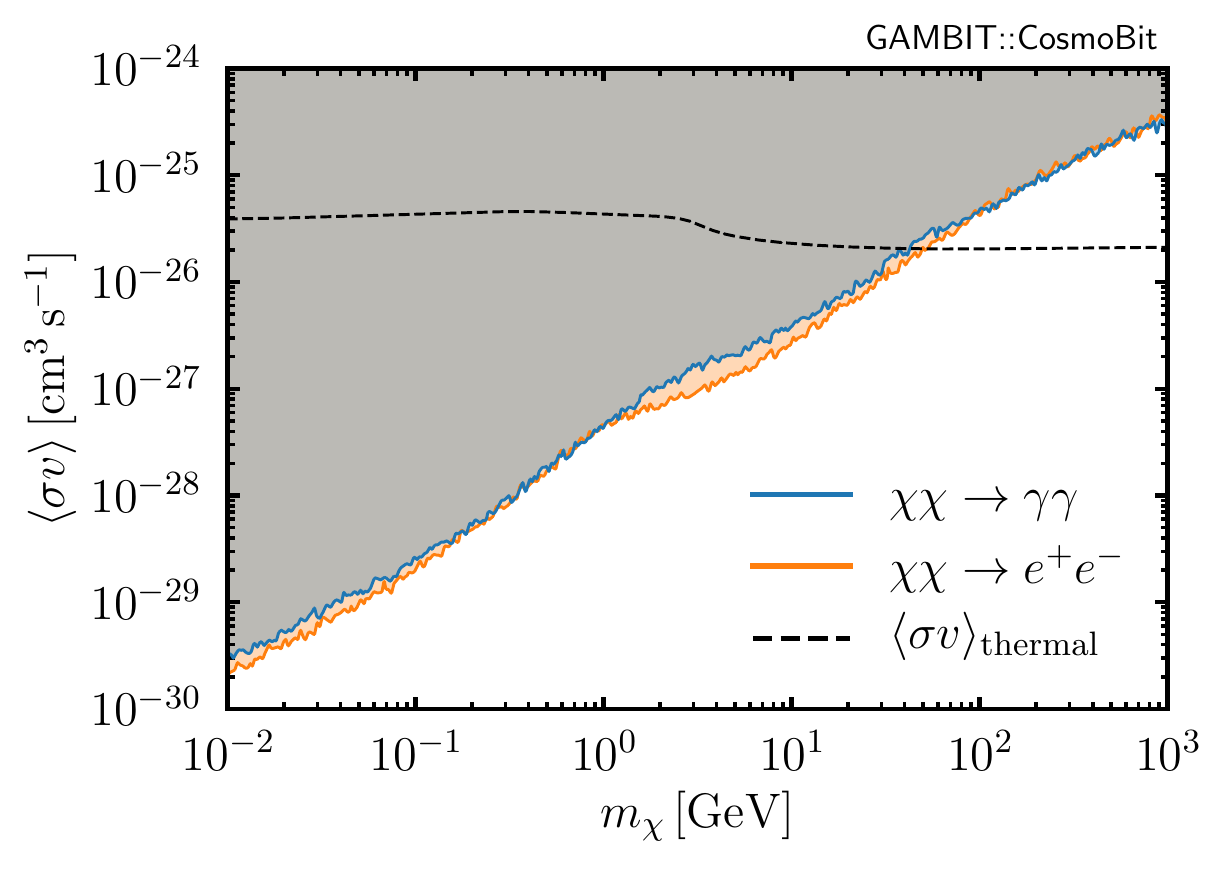}
    \includegraphics[width=0.492\textwidth]{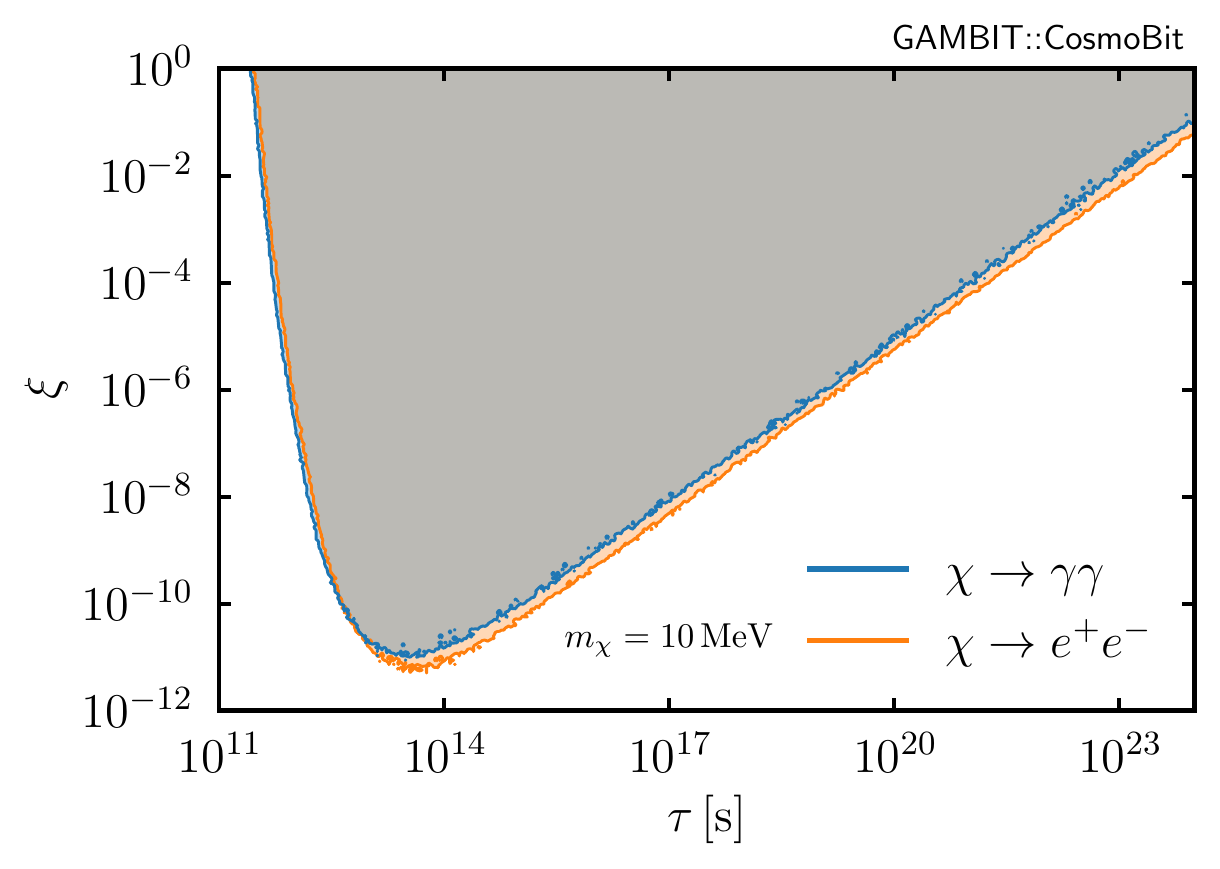}
    \caption{Effects of different forms of exotic energy injection implemented in \cosmobit.
    {\it Left}: $95$\%\,confidence CMB bounds (TT,TE,EE+lowE+lensing) on the velocity-weighted annihilation cross section of s-wave annihilating self-conjugate dark matter $\langle \sigma v \rangle$, as a function of the dark matter mass $m_\chi$. Blue lines show limits on annihilation to photons, and orange on annihilation to $e^+e^-$ pairs.  Shaded areas are excluded. The dashed line shows the value of $\langle \sigma v \rangle$ yielding the observed dark matter abundance from thermal freeze-out~\cite{Bringmann:2020mgx}. {\it Right}: CMB bounds on the decay of a dark matter particle with mass $10$\,MeV, in terms of the initial dark matter fraction $\xi$ and the dark matter lifetime $\tau$.  Colours and shading as in left panel.
    \label{fig:energy_injection}}
  \end{center}
\end{figure}

To convolve the injected particle spectrum with the transfer functions (Eq.\ \ref{eq:energy_injection_efficiency_convolution}), \cosmobit provides an interface to \darkages \cite{Stocker:2018avm}, part of the \exoclass branch of \class. Compared to its initial release, an update to enable the usage of the factorisation scheme (Eq.~\ref{eq:energy_injection_factorisation}) is provided. For implementation details, we refer the reader to Appendix \ref{app:darkages-interface}.

As an illustration, in Fig.~\ref{fig:energy_injection} we show CMB bounds on s-wave annihilating and decaying dark matter, including the Planck 2018 baseline likelihood combination (TT,TE,EE+lowE+lensing). In the left panel, we plot the bound on the thermally-averaged annihilation cross-section as a function of the dark matter mass $m_\chi$, assuming that dark matter is self-conjugate and annihilates exclusively into either photons or electron-positron pairs. The cross section $\langle \sigma v \rangle_\text{thermal}$ that yields the observed dark matter relic abundance (dashed line, calculated following Ref.~\cite{Bringmann:2020mgx}) is excluded for $m_\chi \lesssim 40\,\mathrm{GeV}$ for annihilation into photons. For annihilation into an electron-positron pair, this cross section is excluded for $m_\chi \lesssim 50\,\mathrm{GeV}$. Note that in most realistic models additional annihilation channels need to be included for dark matter masses greater than about $100\,\mathrm{MeV}$, for which CMB constraints tend to be slightly weaker. On the right, we show bounds on the fractional abundance $ \xi $ of dark matter, as a function of its lifetime $ \tau$, assuming a dark matter mass of $ 10\,{\rm MeV} $ and decay into photons or electron-positron pairs.

Note that although we use \multinest for the parameter scan, the limits shown in Fig.~\ref{fig:energy_injection} are obtained in a frequentist approach, i.e.\ they correspond to lines of constant profile likelihood. The reason is that in a Bayesian approach the upper bounds on $\langle \sigma v \rangle$ and $\xi$ would be very sensitive to the chosen priors, which are difficult to motivate from first principles. We have checked explicitly that we can reproduce earlier results derived in Bayesian approaches~\cite{Poulin:2016anj,Stocker:2018avm,Aghanim:2018eyx}, when adopting the same priors. The present setup of \cosmobit
also makes it straightforward to check how the results presented in Fig.~\ref{fig:energy_injection} would
change when directly adopting the channel-dependent efficiency functions $ f_c(z) $ of Ref.~\cite{Galli:2013dna} instead of the default implementation in \cosmobit given by Eq.~(\ref{eq:energy_injection_factorisation}). As expected~\cite{Slatyer:2015kla}, we did not find any significant difference for this particular application.


%
\section{Example: neutrino masses in non-standard cosmologies}
\label{sec:constraints}
%

In this Section, we provide an illustration of the power of \cosmobit by combining several different extensions of $\Lambda$CDM. Specifically, we constrain the sum of neutrino masses in cosmologies with non-standard values of $N_\text{eff}$. As discussed in Sec.\ \ref{sec:non-std-Neff}, these can result from either a modification of the neutrino temperature (parametrised by the rescaling factor $r_\nu$) or from additional ultra-relativistic degrees of freedom $\Delta N_\text{ur}$. When these two effects are considered simultaneously, a degeneracy appears: decreasing the neutrino temperature can be compensated for by introducing additional ultra-relativistic degrees of freedom, leading to a value of $N_\text{eff}$ close to the one in the standard scenario \cite{Nollett:2013pwa, Steigman:2014pfa, Nollett:2014lwa, Wilkinson:2016gsy}. Usually, the observable consequences of neutrino masses and a non-standard energy density in the neutrino sector are studied separately, and are found to be tightly constrained by data. Here, we show that simultaneously including these effects has two interesting consequences 1) the available parameter space in $\sum m_\nu$ is significantly larger than the Planck+BAO ($\Lambda$CDM) constraint $\sum m_\nu \lesssim 0.15$ eV \cite{Aghanim:2018eyx}; and 2) a preference for a \textit{lower} neutrino temperature than predicted by standard cosmology.

The CMB is sensitive to the total energy density in neutrinos, which depends on the number density of neutrinos as well the sum of their masses. A lower neutrino temperature implies a lower equilibrium number density. Therefore, the cosmological bound on the sum of neutrino masses can be relaxed substantially if both the neutrino temperature and the number of additional ultra-relativistic degrees of freedom are allowed to deviate from their standard values. Indeed, as we show below, for sufficiently large values of $\Delta N_\text{ur}$ and sufficiently small neutrino temperatures, cosmological constraints on the mass of the lightest neutrino may be weaker than corresponding laboratory constraints.

To study this scenario in \cosmobit, we simultaneously sample the parameter spaces of the following models:
\begin{itemize}
 \item \emph{Cosmological model:} We vary the six parameters of $\Lambda$CDM, fixing $T_\text{CMB}=2.72548$\,K and the scalar-to-tensor ratio $r$ to zero. In \gambit this corresponds to scanning over the models \doublecrosssf{LCDM}{LCDM} and \doublecrosssf{Minimal\_PowerLaw\_ps}{Minimal_PowerLaw_ps} (see Appendix~\ref{app:models} for detailed model descriptions).

 \item \emph{Neutrino masses:}  We assume the normal neutrino mass hierarchy, and vary the three parameters describing the neutrino masses included in the model \doublecrosssf{StandardModel\_mNudiff}{StandardModel_mNudiff}.

 \item \emph{Non-standard radiation content:} We vary up to four parameters describing possible modifications of $N_\text{eff}$, but in a manner that always provides exactly two additional degrees of freedom. These models are implemented as child models of \doublecrosssf{rBBN\_rCMB\_dNurBBN\_dNurCMB}{rBBN_rCMB_dNurBBN_dNurCMB}. The specific cases we consider are to treat $r_\text{CMB}$ and $\Delta N_\text{ur,CMB}$ as free parameters with two different prescriptions for the two BBN parameters:
  \begin{itemize}
   \item \emph{BBN=CMB}: We fix $r_\text{BBN} = r_\text{CMB}$ and $\Delta N_\text{ur,BBN} = \Delta N_\text{ur,CMB}$, corresponding to the case where any modification of $N_\text{eff}$ occurs before the beginning of BBN.
   \item \emph{Standard BBN}: We fix $r_\text{BBN} = 1$ and $\Delta N_\text{ur,BBN} = 0$, corresponding to the case where any modification of $N_\text{eff}$ occurs after the end of BBN.
  \end{itemize}

 \item \emph{Nuisance parameters:} We vary the nuisance parameters $A_\text{Planck}$ and $M$, corresponding to the absolute calibrations of the Planck likelihoods and the magnitudes of SNe~Ia, respectively. These parameters are included in the \gambit models \doublecrosssf{cosmo\_nuisance\_Planck\_lite}{cosmo_nuisance_Planck_lite} and \doublecrosssf{cosmo\_nuisance\_Pantheon}{cosmo_nuisance_Pantheon}, respectively.
\end{itemize}
Together, this amounts to $6+3+2+2=13$ free parameters.  The priors that we adopt for each parameter are given in Table~\ref{tab:priors}. We include the following likelihoods:
\begin{table}[tbp]
\centering
\begin{tabular}{l l l l}
\toprule
Sector & Param. & Range & Prior \\
\midrule
$\nu$ masses &$m_{\nu_0}$                             & [0, 1]\,eV & flat \\
             &$\Delta m^{2}_{21}$                     & $[5,\ 10] \times 10^{-5}$\,eV$^2$ & flat \\
             &$\Delta m^{2}_{3\ell}$                  & [0.002, 0.003]\,eV$^2$ & flat \\
\midrule
$N_\mathrm{eff}$ & $\Delta N_\mathrm{ur,CMB}$         & [0, 5]     & flat \\
                 & $r_\mathrm{CMB}$                   & [0.2, 1.2] & flat \\
\midrule
Nuisance     &$A_\text{Planck}$                       & [0.9, 1.1] & flat \\
             &$ M $                                   & [-20, -18] & flat \\
\bottomrule
\end{tabular}
\caption{Priors adopted in our scan for the 7 free parameters beyond $\Lambda$CDM. The priors for the 6 $\Lambda$CDM parameters are given in Table~\ref{tab:lcdm_priors}.}
\label{tab:priors}
\end{table}
\begin{itemize}
 \item BBN: $\yhe$~\cite{PDG20} and $D/H$~\cite{Cooke:2017cwo} (see Sec.\ \ref{sec:BBN})
 \item CMB: Planck `lite' high-$\ell$ TTTEEE, low-$\ell$ TT and EE, as well as Planck lensing~\cite{Aghanim:2019ame} (see Sec.\ \ref{sec:CMB})
 \item BAO: 6dF~\cite{2011MNRAS.416.3017B}, SDSS~\cite{2015MNRAS.449..835R}, and BOSS DR12~\cite{Alam:2016hwk} (see Sec.\ \ref{sec:late_time})
 \item SNe~Ia: Pantheon~\cite{Scolnic:2017caz} (see Sec.\ \ref{sec:late_time})
 \item Neutrino oscillations: \textsf{NuFit 4.1}~\cite{Esteban:2018azc} (see Sec.\ \ref{sec:Nu-masses})
\end{itemize}
Note that in order to isolate the constraints on the neutrino mass scale from cosmology we do not include a likelihood for KATRIN~\cite{Aker:2019uuj}, which would slightly disfavour the largest neutrino masses considered in our scans.

\begin{figure}
  \centering
  \includegraphics[width=.85\textwidth]{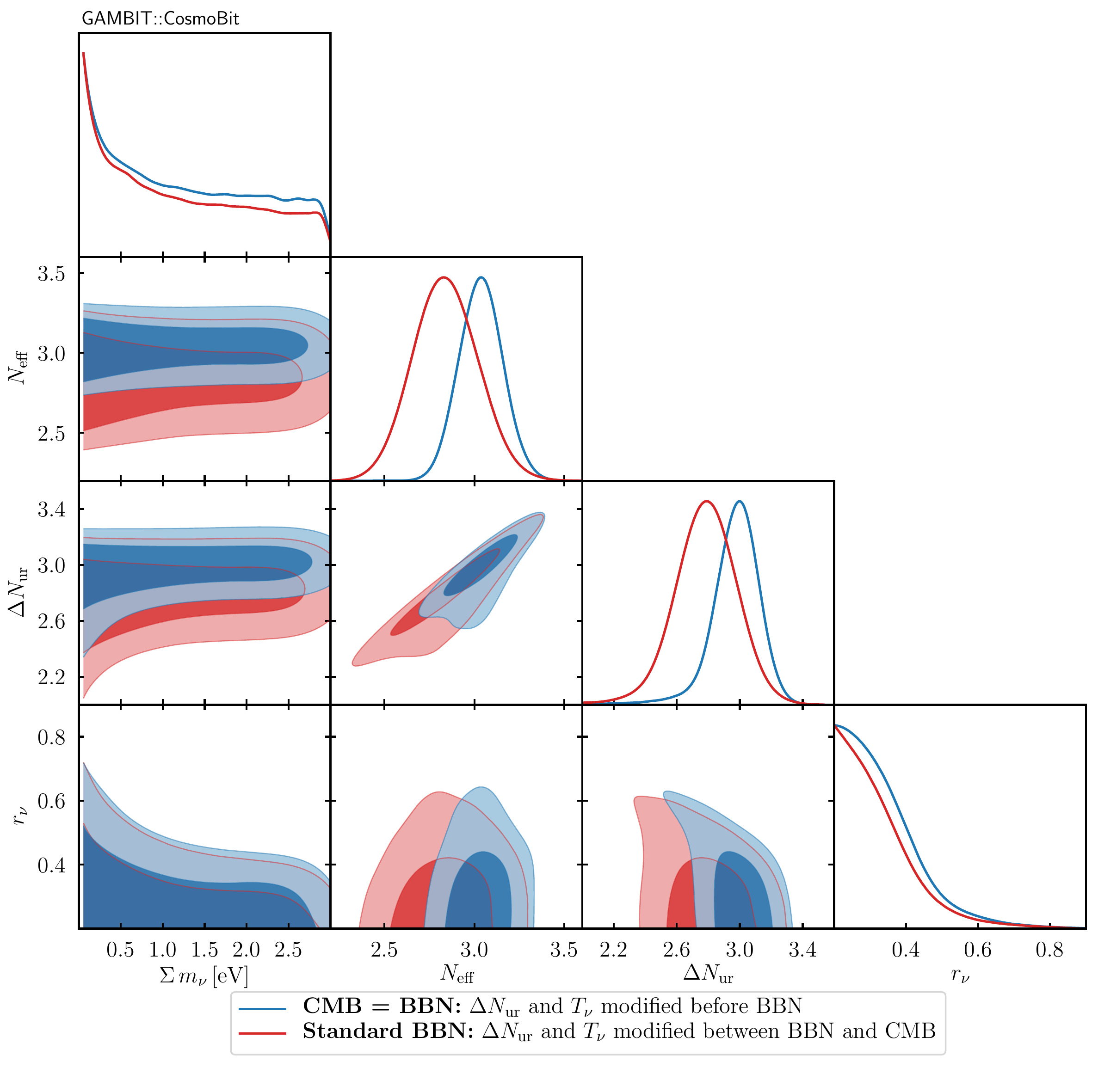}
  \caption{68\% and 95\% credible regions for the sum of the neutrino masses $\sum m_\nu$, the total number of effective relativistic degrees of freedom $N_\text{eff}$, the contribution from exotic ultra-relativistic species $N_\text{ur}$ and the rescaling factor of the neutrino temperature $r_\nu$. Blue lines and contours correspond to the case that all modifications to $N_\text{eff}$ and $r_\nu$ occur before the beginning of BBN, whereas red lines and contours assume that the modifications occur between BBN and CMB. By reducing the neutrino temperature considerably and compensating for the decrease in $N_\text{eff}$ with exotic ultra-relativistic species, the cosmological bound on the sum of neutrino masses can be relaxed considerably.
  \label{fig:dNeff_triangle}}
\end{figure}

The results of our scans are shown in Fig.~\ref{fig:dNeff_triangle}, which shows marginalised posteriors for the free parameters $\Delta N_\text{ur}$ and $r_\nu$, as well as for the derived parameters $\sum m_\nu$ and $N_\text{eff}$. As expected, $N_\text{eff}$ is constrained to be close to 3.045, the standard value. This constraint is tighter when the modification of $N_\text{eff}$ affects both BBN and CMB compared to cases where the synthesis of elements during BBN is not affected by non-standard physics and $N_\text{eff}$ takes non-standard values only after the conclusion of BBN. A similar effect can be seen in Fig.~\ref{fig:BBN}, where the combination of CMB+BAO and BBN gives a tighter constraint on $N_\text{eff}$ than CMB+BAO alone.

To achieve $N_\text{eff} \approx 3$ one can either have a neutrino temperature close to the standard value ($r_\nu \approx 1$) and a small contribution from additional ultra-relativistic degrees of freedom ($\Delta N_\text{ur} \ll 1$), or a reduced neutrino temperature and a sizeable contribution from $\Delta N_\text{ur}$. As the energy density in neutrinos during BBN and recombination is proportional to $r_\nu^4$, even reducing $r_\nu$ slightly leads to a strong suppression of the neutrino contribution to  $N_\text{eff}$. Indeed, for $r_\nu < 0.5$ one obtains $\Delta N_\text{eff} < 0.2$ from neutrinos, such that $N_\text{ur} \approx 3$ is required to be consistent with observations.

From the point of view of $\Delta N_\text{eff}$, the two different configurations ($r_\nu \approx 1$, $\Delta N_\text{ur} \ll 1$ and $r_\nu < 0.5$, $\Delta N_\text{ur} \approx 3$) are indistinguishable, but they have very different implications for the cosmological bound on the sum of neutrino masses. The reason is that after neutrinos become non-relativistic, their energy density is proportional to $r_\nu^3 \sum m_\nu$. Hence, if $r_\nu$ is reduced relative to the standard scenario, the range of neutrino masses consistent with observations increases substantially. Indeed, for late-time probes that are only sensitive to the background evolution of the Universe, such as the BAO scale and the recession velocities of SNe~Ia, there is a complete degeneracy between $r_\nu$ and $\sum m_\nu$, i.e.\ these probes cannot distinguish between scenarios with standard neutrino temperature and small neutrino masses or lower neutrino temperature and larger neutrino masses. The only late-time probe sensitive to this difference included in our scans is the CMB lensing potential, which however does not have a big impact for the parameter ranges considered.\footnote{Additional probes of the matter power spectrum on small scales, like galaxy clustering data, could help to distinguish between the two scenarios.}

Although there is no clear preference in the data for $r_\nu < 1$, we still find that the marginalised posteriors in Fig.~\ref{fig:dNeff_triangle} exhibit a strong preference for the case $r_\nu < 0.5$ and $N_\text{ur} \approx 3$, i.e.\ our scan prefers for the dominant contribution to $N_\text{eff}$ to come from new ultra-relativistic degrees of freedom rather than from SM neutrinos. The reason for this preference is that as $r_\nu$ decreases the bound on the sum of the neutrino masses is relaxed and the available volume of parameter space with high probability increases. This volume then gives a larger contribution to the integral performed when marginalising over other parameters at small $r_\nu$ than at larger values, making small $r_\nu$ preferable to $r_\nu \approx1$ from the Bayesian point of view. In other words, small values of $r_\nu$ are found to be more likely, because they require less tuning in the neutrino masses, even though comparably good fits to the data can be obtained for any value of $r_\nu \lesssim 1$.

\begin{figure}
  \centering
  \includegraphics[width=.85\textwidth]{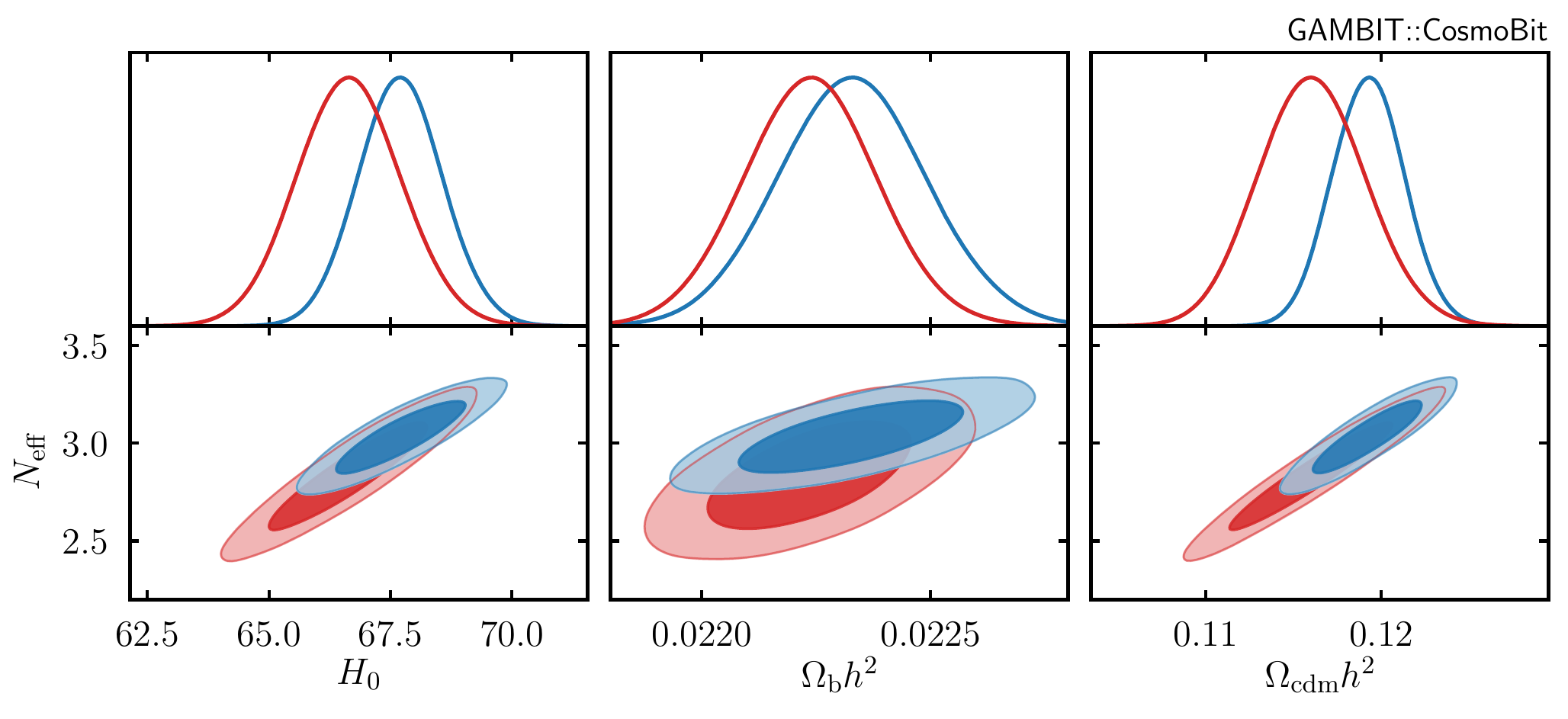}
  \includegraphics[width=.85\textwidth]{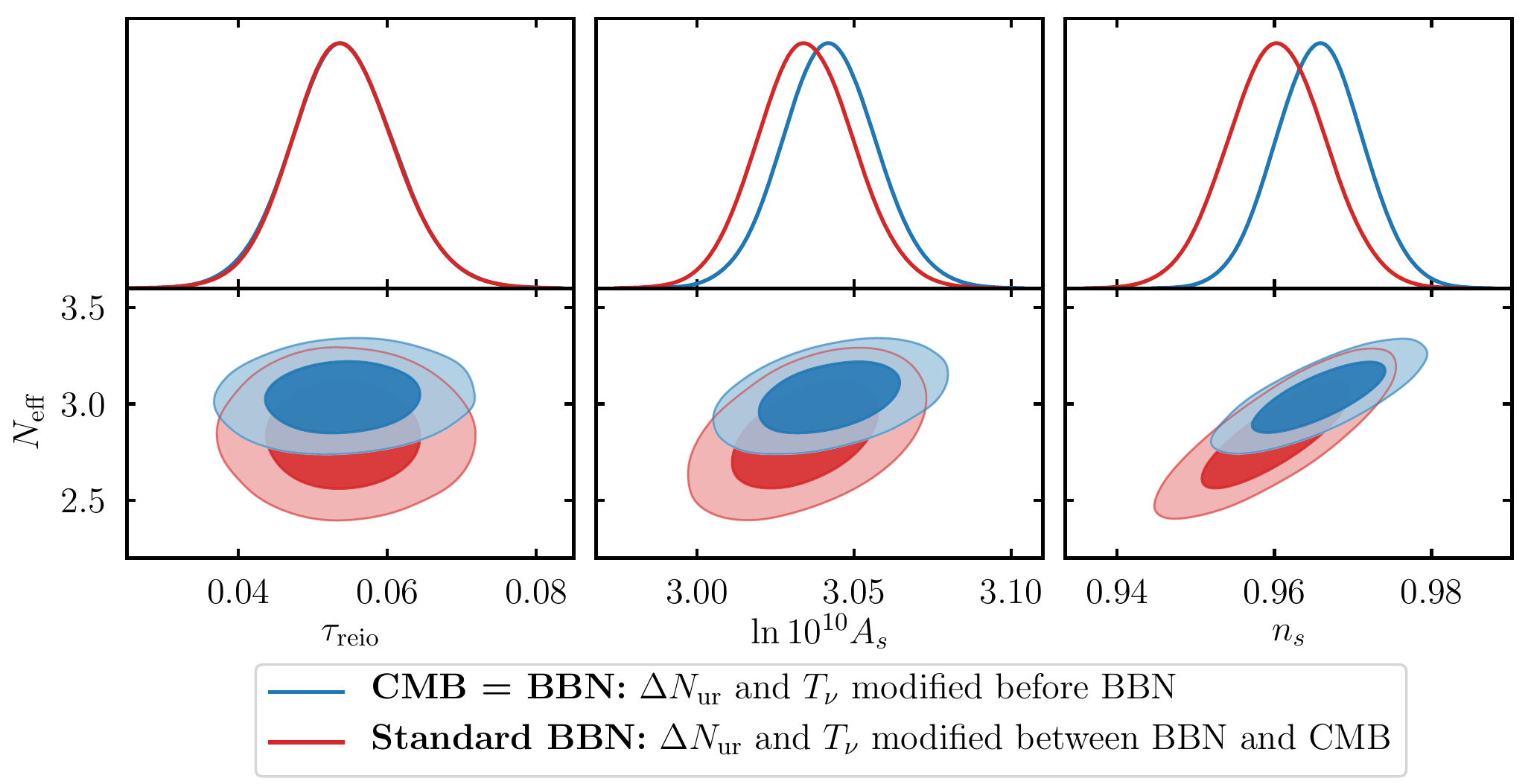}
  \caption{Marginalised posterior probabilities of the six $\Lambda$CDM parameters, as well as their correlations with $N_\text{eff}$, shown in terms of 68\% and 95\% credible regions. Blue lines and contours correspond to the case that all modifications to $N_\text{eff}$ and $r_\nu$ occur before the beginning of BBN, red lines and contours assume that the modifications occur between BBN and CMB. All parameters other than $\tau_\text{reio}$ have strong correlations with $N_\text{eff}$, so allowing $N_\text{eff}$ to vary significantly relaxes the constraints on these parameters compared to the standard $\Lambda$CDM case (see figure~\ref{fig:comparison_planck}).
  \label{fig:dNeff_lcdm}}
\end{figure}

Even though the extent of the preference for large $\Delta N_\text{ur}$ and small $r_\nu$ is dependent upon the time at which  $\Delta N_\text{ur}$ and $T_\nu$ are modified, the corresponding central conclusion is not: if variations in the neutrino temperature and additional ultra-relativistic degrees of freedom are considered simultaneously, the cosmological bound on the sum of neutrino masses disappears. This can be seen most clearly from the 1-dimensional marginalised posterior probability in the top panel of Fig.~\ref{fig:dNeff_triangle}. While data still prefer a neutrino mass close to zero, the probability becomes essentially flat for larger values, so that one cannot quote a sensible upper bound on the sum of the neutrino masses from cosmology if the neutrino temperature is left as a free parameter. This implies that in the scenario considered here, the neutrino mass may saturate the $\sim 1.1\,\mathrm{eV}$ bound from KATRIN \cite{Aker:2019uuj} and lead to a positive detection of the neutrino mass in the near future.

In Fig.~\ref{fig:dNeff_lcdm}, we illustrate the extent to which the resulting deviations of $N_\text{eff}$ lead to correlated changes in the measured values of the parameters of $\Lambda$CDM.  Those parameters most relevant for physics prior to the formation of the CMB are most strongly impacted. Changes to $N_\text{eff}$ have little impact on the measured values of the optical depth.  The impacts are most marked for the scenario where $\Delta N_\text{ur}$ and $r_\nu$ vary between BBN and CMB formation, owing to the fact that $N_\text{eff}$ can be pushed more negative in this scenario than in the case where the variation occurs before BBN. We emphasize again the shifts in the posteriors are driven by the prior ranges and the effect of marginalisation and do not reflect an actual preference in the data (see the text around figure~\ref{fig:BBN} for a related discussion).

Although we have not discussed underlying mechanisms that can give rise to such simultaneous modifications of the neutrino temperature and additional ultra-relativistic species, it is worth noting that the scenario considered here may be similar to the scenarios of exotic energy injection discussed in Sec.\ \ref{sec:energy-injection}. If a decaying dark matter sub-component injects electromagnetic radiation or electrons into the primordial plasma between neutrino decoupling and recombination, the main effect will be an increase in the photon temperature, which is equivalent to a decrease in the neutrino temperature. If the decays produce not only electromagnetic radiation but also dark radiation, they would simultaneously lead to an increase in $\Delta N_\text{ur}$, which could potentially be arranged in such a way that $N_\text{eff}$ remains approximately constant, see e.g.\ Ref.~\cite{Steigman:2013yua}. It is worth noting that a similar effect can be achieved by directly decreasing the neutrino temperature, e.g.\ through  unstable neutrinos that decay into dark radiation on cosmological timescales~\cite{Beacom:2004yd,Serpico:2007pt,Chacko:2019nej,Chacko:2020hmh,Escudero:2020ped}. We leave a more detailed study of such scenarios for future work.

%
\section{Conclusions} \label{sec:conclusion}
%
Cosmology and particle physics are intimately connected. Complementarity of probes from both sectors will become progressively more important for testing theories beyond $\lcdm$ as time goes on. Next-generation CMB experiments such as CMB-S4~\cite{Ade:2018sbj} and the Simons Observatory~\cite{Abazajian:2016yjj} will test the paradigm of inflation, the nature of the neutrino sector, and other light relics. The SKA~\cite{Bull:2018lat} will use the epoch of reionisation to test models of dark energy and gravity, and Euclid~\cite{Laureijs:2011gra,Amendola:2012ys} will employ galaxy clustering and the galaxy power spectrum to probe the sum of neutrino masses and models of dark energy.  The advent of gravitational wave observatories will probe phase transitions in the early Universe and models of modified gravity, and further measurements of the Hubble parameter may well provide evidence of the existence of additional light particle species.  In every case, the results will both serve as important probes of particle theories, and will be strengthened by input from terrestrial particle experiments.

In this paper, we have presented \cosmobit, the cosmology module of the public global-fitting framework \GB. \cosmobit is the first package capable of consistently combining observables and likelihoods from cosmology with those from particle physics. The existing particle physics likelihoods contained within \GB are extensive, including dark matter, collider and flavour physics, the neutrino sector and precision tests of the SM.

\cosmobit ships with a range of models covering this interface between cosmology and particle physics, such as models of inflation, non-standard radiation content, and exotic energy injection from dark matter. Alongside these models \cosmobit also provides two parametrisations of $\lcdm$, and many nuisance models relevant for cosmological experiments and surveys.

\cosmobit provides calculations of primordial power spectra from inflation, BBN abundances in non-standard cosmological scenarios, energy injection spectra from exotic interactions, and anisotropy power spectra for the CMB. The likelihoods contained within \cosmobit include up-to-date measurements of BBN abundances, the latest CMB spectra from Planck, and late-time cosmological observables such as Type Ia supernovae and BAO. These can be consistently combined with relevant BSM likelihoods existing within \GB, such as those for the neutrino sector and dark matter.

The modular nature of \cosmobit means that interfacing it to other cosmological codes is very simple.  It currently features interfaces to \alterbbn, \class, \montepython, \multimodecode, \darkages and the public Planck likelihood code. Planned additional interfaces include the Boltzmann solver \textsf{CAMB}~\cite{Lewis:1999bs} and likelihoods contained within \textsf{CosmoMC}~\cite{CosmoMC,Lewis:2013hha}. The simplicity of interfacing with modified versions of \class also makes it easy to extend to e.g.\ work with modified theories of gravity via \textsf{hi\_class}~\cite{Zumalacarregui:2016pph,Bellini:2019syt}.

We have provided many examples of \cosmobit in action, including vanilla $\lcdm$, studies of various single-field inflationary scenarios, bounds on energy injection from dark matter annihilation, and the effects of varying the neutrino temperature on BBN, the CMB, neutrino masses and cosmological parameter estimation.

\cosmobit fully integrates into the GAMBIT framework and global fits involving cosmological and particle physics models can readily be performed with it.  This opens the possibility of performing combined global fits on various beyond the standard model cosmological and particle physics scenarios to quantify their simultaneous plausibility.  The GAMBIT community plans to use this framework to perform cosmological global fits involving, for example, axions.

Finally, in a companion paper~\cite{CosmoBit_numass}, we perform a global fit of the three-flavour neutrino scenario, using likelihoods from cosmology and terrestrial neutrino experiments. We provide a new upper limit on the mass of the lightest neutrino of 0.037\,eV assuming normal neutrino mass ordering, and 0.042\,eV assuming inverted ordering -- an improvement of nearly 60\% on the previous bound.

Cosmology is at an important historical juncture. As precision cosmological datasets increase in quality and quantity we anticipate that our models of particle physics and the universe will be placed under increasing strain. We believe that \cosmobit will prove invaluable to both communities by providing a unified framework for the theory and inference of particle physics and cosmology.

The \cosmobit source code can be freely downloaded from \href{http://gambit.hepforge.org}{gambit.hepforge.org} as part of the \GB framework.

\acknowledgments

We thank Alexandre Arbey, Thejs Brinckmann, Tamara Davis, Richard Easther, Martina Gerbino, Deanna Hooper, Julien Lesgourgues, Vivian Poulin, Nils Sch\"{o}neberg, Jes\'{u}s Torrado and Sunny Vagnozzi, as well as all members of the \gambit community for discussions. JJR acknowledges support by Katherine Freese through a grant from the Swedish Research Council (Contract No. 638-2013-8993), FK and PSt from the DFG Emmy Noether Grant No.\ KA 4662/1-1, PSc by the Australian Research Council Future Fellowship FT190100814, CB, TEG and MJW by the Australian Research Council Discovery Project DP180102209, CH through the Australian Research Council's Laureate Fellowship funding scheme (project FL180100168), SebH from the Alexander von Humboldt Foundation and the German Federal Ministry of Education and Research, WH by a Gonville \& Caius Research Fellowship and the George Southgate visiting fellowship, and ACV by the Arthur B. McDonald Canadian Astroparticle Physics Research Institute, CFI and MEDJCT.  Research at Perimeter Institute is supported by the Government of Canada through the Department of Innovation, Science, and Economic Development, and by the Province of Ontario through MEDJCT. For access to computing resources, we thank PRACE for access to Marconi at CINECA and Joliot-Curie at CEA, RWTH Aachen University (under project jara0184), the University of Cambridge (CSD3), the University of Adelaide, the Scientific Computing Cluster at GWDG, the joint data centre of the Max Planck Society for the Advancement of Science~(MPG) and the University of {G\"ottingen}.  Plots in this paper made use of \textsf{matplotlib}~\cite{Hunter:2007}, \textsf{GetDist}~\cite{Lewis:2019xzd} and \pippi~\cite{pippi}.

\appendix

%
\section{Quickstart}
\label{app:quickstart}
%

In this appendix we provide a quick-start guide for reproducing the validation results shown in Fig.~\ref{fig:bao_bbn_H0}. This includes configuration and installation of \gambit, how to install all backends relevant for running \cosmobit, and how to run the scan. The \YAML file containing all settings in this example can be found at \term{yaml\_files/CosmoBit\_quickStart.yaml}. This file contains additional comments and explanations of the particular entries. A more detailed explanation of all the features that \cosmobit introduces is given in \term{yaml\_files/CosmoBit\_tutorial.yaml}.

The minimum requirements to build \GB are given in the file \term{README.md} in the \gambit base directory.  These are:
\begin{itemize}\setlength{\itemsep}{-0.2em}
 \item \gpp/\gfortran \textsf{5.1} or \icpc/\ifort \textsf{15.0.2}
 \item \cmake \textsf{2.8.12}
 \item \Python \textsf{2.7} or \Python \textsf{3}
 \item \Python modules: \textsf{yaml}, \textsf{os}, \textsf{re}, \textsf{datetime}, \textsf{sys}, \textsf{getopt}, \textsf{shutil} and \textsf{itertools}
 \item \textsf{git}
 \item \textsf{Boost 1.41}
 \item GNU Scientific Library (\textsf{GSL}) \textsf{2.1}
 \item \textsf{Eigen} \textsf{3.1.0}
 \item \textsf{LAPACK}
 \item \textsf{pkg-config}
 \item 16\,GB of RAM
\end{itemize}
Further optional dependencies are discussed in this file.  Additionally, various backends connected to \cosmobit also require the following \Python packages:
\begin{itemize}\setlength{\itemsep}{-0.2em}
  \item \cython (needed by \classy)
  \item \textsf{scipy} (needed by \montepython and \darkages)
  \item \textsf{numpy 1.12} or greater (needed by \darkages and \classy)
  \item \textsf{dill}, \textsf{future} (needed by \darkages)
  \item \textsf{pandas}, \textsf{numexpr} (needed by \montepython)
\end{itemize}

Running a test scan with \GB can in principle be done in a few seconds with as little as a few GB of RAM and a single core, if appropriately resource-light likelihoods are selected.  The resources required for a production scan depend strongly on the chosen likelihoods and scanner settings. For example, the inflation scans of Sec.\ \ref{sec:inflation} took 1--2 days each on 480 cores (160 MPI processes $\times$ 3 OpenMP threads per process), whilst the main scan of Sec.\ \ref{sec:constraints} took about a day, running without threads on 144 cores.  The example in \term{CosmoBit\_quickStart.yaml} should take on the order of a minute to run through.

To run this example, one must first configure and build \GB. After downloading \GB, run the following from the \GB base directory:
\begin{lstterm}
mkdir build
cd build
# Add all CMake flags in this step, e.g. -Ditch="Flav;Collider" -WITH_MPI=ON etc
cmake ..
# @\metavar{n}@ = number of cores to use for build
make -j@\metavar{n}@ scanners
cmake ..
# @\metavar{n}@ = number of cores to use for build
make -j@\metavar{n}@ gambit
\end{lstterm}
The step \term{make -j}\metavar{n} \term{scanners} will install all available scanners that are implemented in the \scannerbit module \cite{ScannerBit}, including \polychord~\cite{Handley:2015} and \multinest~\cite{Feroz:2007kg,Feroz:2008,1306.2144}.

An additional useful command in the initial \term{cmake} step provides the ability to exclude physics modules from the build (`ditching'). For example, to not include \textsf{FlavBit} nor \textsf{ColliderBit} in the build, simply add \term{-Ditch="Flav;Collider"}. Note that compilation can be achieved with 8\,GB of RAM if sufficient modules are ditched in this manner, leaving e.g.\ only \cosmobit of all the available \GB modules.  Many other options can be passed to \term{cmake}; some commonly-used ones are documented in the file \term{CMAKE_FLAGS.md} in the \gambit base directory.  These include \term{-D WITH\_MPI=ON} to enable parallel computation with several \mpi processes, and \term{-D CMAKE_BUILD_TYPE=Release} to enable high-level optimisation.\footnote{\ldots as well as extreme memory usage ($\gg 16\,$GB) whilst compiling -- beware!}

All backends interfaced with \cosmobit can then be downloaded and built by executing
\begin{lstterm}
make alterbbn classy darkages montepythonlike multimodecode plc plc_data
\end{lstterm}
in the build directory. This will download and install the default version of each backend, as well as the latest Planck data used by \plc.  Specific versions can be built by appending an underscore and the version number, e.g. \term{make classy\_exo\_2.7.2} will build \exoclass \textsf{v2.7.2}.

\begin{figure}
  \centering
   \includegraphics[width=0.74\paperwidth]{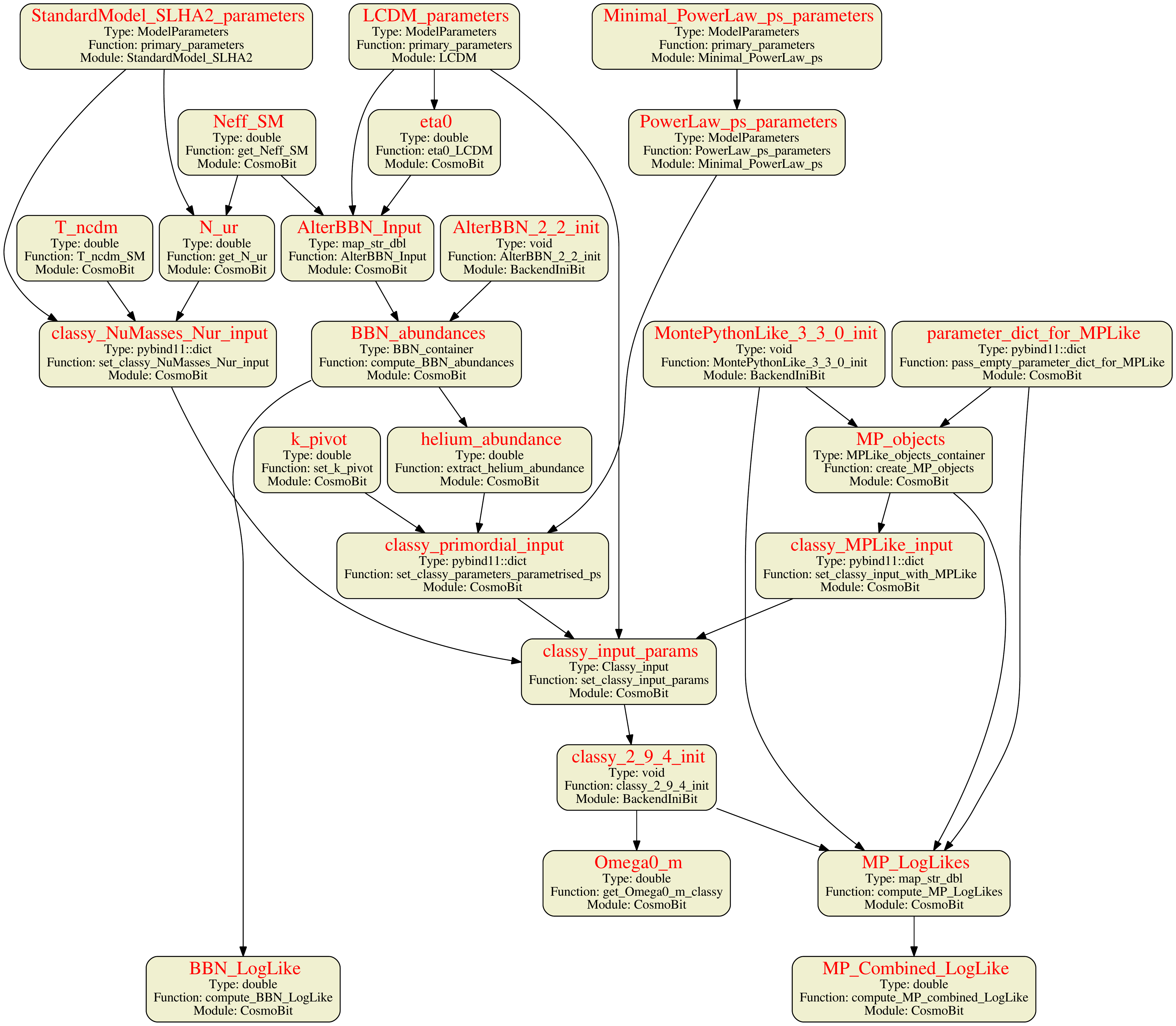}
   \caption{An example dependency tree of the \cosmobit scan used to produce Fig.~\ref{fig:bao_bbn_H0}. This tree can be produced by running \protect\term{CosmoBit_quickStart.yaml}. The models scanned are $\lcdm$ (\doublecrosssf{LCDM}{LCDM}), the minimal power law power spectrum (providing $A_s$ and $n_s$; \doublecrosssf{PowerLaw\_ps}{PowerLaw_ps}), and the SM (\textbf{\textsf{StandardModel\_SLHA2}}), with the Planck `baseline' assumption of a single massive neutrino, $m_{\nu_3}$ = 0.06 eV. The total log-likelihood is constructed from BBN and a selection of BAO likelihoods, provided by \montepython. In this case, the \montepython likelihoods are \yamlvalue{bao\_smallz\_2014} and \yamlvalue{bao\_boss\_dr12} and \yamlvalue{eBOSS\_DR14\_Lya\_combined}, none of which have any associated nuisance models.}
   \label{fig:dep-tree}
\end{figure}

A scan can be started with the command
\begin{lstterm}
./gambit -f yaml_files/CosmoBit_quickStart.yaml
\end{lstterm}
The dependency tree for this scan is shown in Fig.~\ref{fig:dep-tree}.

Finally, when using \gambit the user can obtain information about a given model of name \metavar{model\_name}, and its parameters, by executing \term{gambit} \metavar{model\_name} after a successful build, \eg
\begin{lstterm}
./gambit LCDM
\end{lstterm}
prints information about the model \doublecrosssf{LCDM}{LCDM}, including its parameters and any related models.


\section{Cosmological models in \gambit}
\label{app:models}
%

In this Section, we introduce all relevant models shipped with \cosmobit, including both cosmological models and others needed for cosmological parameter estimation.

\subsection{Standard cosmology -- \textsf{LCDM}} \label{app:models-lcdm}

Standard $\lcdm$ traditionally has 6 free parameters, as described in Sec.\ \ref{sec:lcdm}: the rescaled baryon energy density today, $\omega_b$, the rescaled energy density of cold dark matter today $\omega_c$, the Hubble rate today, $H_0$, the optical depth to reionisation $\taur$, and the tilt and amplitude of the scalar primordial power spectrum, $n_s$ and $A_s$ respectively.

In addition to the standard $\lcdm$ parameters, \cosmobit includes the temperature of the CMB, $\Tcmb$, as model parameter. In the literature, it is common to fix $\Tcmb$ to the mean value inferred from the COBE/FIRAS monopole measurement $\Tcmb = 2.72548 \pm 0.00057$\,K~\cite{Fixsen2009}.  However, with the increased precision of upcoming datasets this treatment can lead to a bias in the other cosmological parameters~\cite{Yoo:2019dyl}.  Our implementation provides the freedom to fix $\Tcmb$, use a Gaussian prior arising form the monopole measurement, or to vary it independently.

The \cosmobit definition of $\lcdm$, however, does not include any parameters describing the spectrum of perturbations. The user must choose a specific model describing the perturbations, either expressed as a purely phenomenological scale-free power spectrum, or derived from a model of inflation. For more information on models describing the power spectrum and inflation see Appendix~\ref{app:models-inflation}.

In the current release, we explicitly assume that the Universe is flat, i.e. vanishing curvature with $\Omega_k = 0$. This assumption is supported to a $1\,\sigma$ accuracy of $0.2 \%$ by recent CMB and BAO data~\cite{Aghanim:2018eyx}. If the user wishes to include $\Omega_k$ as a model parameter, they can follow the instructions given in Appendix~\ref{app:models-new}. Instructions how to pass new model parameters to \class are given in Appendix~\ref{app:class-new-model}. Note that if inflationary models in a non-flat Universe are studied, the calculation of the primordial power spectrum also has to be adopted to hold for $\Omega_k \neq 0$.

When scanning a cosmological model in \GB, one must also select a physical model describing the SM neutrinos; for more details see Appendix~\ref{app:models-NuMass}.

\begin{description}

\item[\textbf{\textsf{LCDM}}:\label{LCDM}] \term{omega\_b, omega\_cdm, H0, tau\_reio, T\_cmb}.

We include four of the six traditional flat $\lcdm$ parameters, as well as the temperature of the CMB as measured today, $\Tcmb$. $H_0$ has units of $\mathrm{km}\,\mathrm{Mpc}^{-1}\,\mathrm{s}^{-1}$.

\item[\textbf{\textsf{LCDM\_theta}}:\label{LCDM_theta}] \term{omega\_b, omega\_cdm, 100theta\_s, tau\_reio, T\_cmb}.

As per \doublecrosssf{LCDM}{LCDM}, but with $\theta^{\star} = r^\star / d^\star_A$, the angular acoustic scale of the CMB multiplied by a factor of 100, replacing $H_0$ as a free parameter.

\end{description}

\subsection{Primordial power spectra \& inflationary models} \label{app:models-inflation}
The $\lcdm$ model does not include any power spectrum parameters; these must be provided by one of the models described in this subsection. The user can scan over a ``traditional'' description of $\lcdm$ with the model \doublecrosssf{PowerLaw\_ps}{PowerLaw_ps}, a purely phenomenological model defining the power spectrum in terms of $n_s$ and $A_s$. Alternatively, the user may scan a physical model of inflation able to provide this information. We describe the parametrised, phenomenological model and all inflationary models included in the first release of \cosmobit in this Section~(see also Table~\ref{tab:inflationary_models} in the main text). The inflationary models are Starobinsky inflation, natural inflation, and various models with a monomial potential,
\begin{equation}
 \label{eq:mono}
   V(\phi) = \lambda \, \mplred^{4-n}\frac{\phi^n}{n} \, .
\end{equation}

All inflationary models below assume instant reheating, which makes $N_\star$ a derived parameter via its relation to instant reheating~(as in Ref.~\cite{1112.0326}). However, we include $N_\star$ as an input parameter for the \doublecrosssf{PowerLaw\_ps}{PowerLaw_ps} model in anticipation of phenomenological models of reheating in future \cosmobit releases.

All inflationary models in \gambit must be used together with a cosmology model, i.e. \doublecrosssf{LCDM}{LCDM} or \doublecrosssf{LCDM\_theta}{LCDM_theta}.

Adding a new inflationary model to \gambit is explained in Appendix~\ref{app:models-new}. The \multimodecode interface to \gambit is explained in Appendix~\ref{app:multimode-new}.

\begin{description}

\item[\textbf{\textsf{PowerLaw\_ps}}:\label{PowerLaw_ps}] \term{ln10A_s}, \term{n_s}, \term{r}, \term{N_pivot}.

This model provides an effective description of the shape of the scalar primordial power spectrum, allowing also for a non-zero scalar to tensor ratio~$r$.
Parameters~$A_s$ and~$n_s$ are identical to the ones traditionally used in $\lcdm$, defined in Eq.~\eqref{Psdef}.
In keeping with Planck conventions, we choose $\ln (10^{10}A_s)$ as the input parameter instead of $A_s$.

$N_\star$ parametrises the uncertainty in reheating.  For the initial release of \cosmobit, it is however included purely for the purposes of easily outputting the value derived when using \multimodecode to translate from one of the true inflationary models to the phenomenological \doublecrosssf{PowerLaw\_ps}{PowerLaw_ps} model. We therefore recommend against scanning over this parameter directly, as its value is neither constrained nor used for any further calculations by any function presently available in \gambit.

\item[\textbf{\textsf{Minimal\_PowerLaw\_ps}}:\label{Minimal_PowerLaw_ps}] \term{ln10A_s}, \term{n_s}.

A simplified description of the spectrum of scalar perturbations; as per \doublecrosssf{PowerLaw\_ps}{PowerLaw_ps}, but with $r = 0$, $N_\star = 55$.

\item[\textbf{\textsf{Inflation\_InstReh\_1mono23}}:\label{Inflation_InstReh_1mono23}] \term{lambda}.

A single-field model, defined by setting $n = 2/3$ in Eq.~(\ref{eq:mono}).
For all monomial models shipped in the first release of \cosmobit ($n=2/3,1,2,4$), the observables $n_s$ and $r$ are provided by the \multimodecode backend.
Solving the equations of motion for the quantum fluctuations up to first order in perturbation theory, one can calculate the power spectrum $P_\mathcal{R}(k)$ as a function of $k$ (c.f.~Sec.\ \ref{sec:inflation}).

\item[\textbf{\textsf{Inflation\_InstReh\_1linear}}:\label{Inflation_InstReh_1linear}] \term{lambda}.

Single-field inflation with a linear potential: $n=1$ in Eq.~(\ref{eq:mono}).

\item[\textbf{\textsf{Inflation\_InstReh\_1quadratic}}:\label{Inflation_InstReh_1quadratic}] \term{m_phi}.

Single-field inflation with a quadratic potential: $n=2$ and $\lambda\mplred^2 \rightarrow m_\phi^2$ in Eq.~(\ref{eq:mono}).

\item[\textbf{\textsf{Inflation\_InstReh\_1quartic}}:\label{Inflation_InstReh_1quartic}] \term{lambda}.

Single-field inflation with a quartic potential: $n=4$ in Eq.~(\ref{eq:mono}).

\item[\textbf{\textsf{Inflation\_InstReh\_1natural}}:\label{Inflation_InstReh_1natural}] \term{lambda}, \term{f_phi}.

Natural inflation, in which the inflaton is a pseudo Nambu-Goldstone boson of a broken global
symmetry~\cite{PhysRevLett.65.3233,adams1993natural}, resulting in a potential of the form
$V(\phi) = \lambda^4 \mplred^4\left[1 + \cos\left(\phi/f_\phi\right)\right]$. \gambit expects the parameter~$f_\phi$ to be in units of the reduced Planck mass~$\mplred$.

\item[\textbf{\textsf{Inflation\_InstReh\_1Starobinsky}}:\label{Inflation_InstReh_1Starobinsky}] \term{lambda}.

Starobinsky inflation results from adding an $R^2$ term to the Einstein-Hilbert action, which
is generically expected to appear due to corrections arising from quantum gravity~\cite{Starobinsky:1980te}.
When transformed to the Einstein frame, this corresponds to a single-field inflation model with potential
\begin{equation}
  V(\phi)=\lambda^4 \, \mplred^4\left(1 - e^{-\sqrt{\frac{2}{3}} \, \frac{\phi}{\mplred}} \right)^2 \, .
\end{equation}

\end{description}

\subsection{Neutrino masses} \label{app:models-NuMass}

Most module functions in \cosmobit require values for the masses of the SM neutrinos.  These are provided by the neutrino mass model.

The `baseline' model of the Planck 2015 \& 2018 analyses~\cite{Ade:2015xua,Aghanim:2018eyx} has only one massive neutrino with $m_{\nu_3} = 0.06$ eV.  This is provided in the \YAML file \term{yaml_files/include/StandardModel_SLHA2_Planckbaseline.yaml}.

\begin{description}

\item[\textbf{\textsf{StandardModel\_mNudiff}}:\label{StandardModel_mNudiff}] \term{mNu\_light, dmNu21, dmNu3l}.

A realistic model of all three SM neutrinos, parametrised by the mass of the lightest neutrino, \term{mNu\_light}, the smallest mass splitting $\Delta m_{21}^2$, \term{dmNu21} and the largest mass splitting $\Delta m_{3l}^2$, \term{dmNu3l}. This parametrisation is agnostic to the mass hierarchy of the neutrinos.  The hierarchy can be chosen by fixing the prior on \term{dmNu3l} appropriately: for a scan only considering the normal (inverted) hierarchy \term{dmNu3l} must be strictly positive (negative), in which case \term{dmNu3l} refers to $\Delta m_{31}^2$ ($\Delta m_{32}^2$).

Note that this model contains additional SM parameters (see~\cite{RHN} for details), including the neutrino mixing angles $\theta_{ij}$ and $\delta_{\text{CP}}$. These parameters are however typically not scanned over in cosmological analyses, and are instead set to their measured central values.  We provide example \YAML files for cosmological scans over both hierarchies:
\begin{itemize}
    \item \term{yaml\_files/include/StandardModel\_mNudiff\_NH\_scan.yaml} for the normal hierarchy
    \item \term{yaml\_files/include/StandardModel\_mNudiff\_IH\_scan.yaml} for the inverted hierarchy
\end{itemize}
which one would import into a \YAML file in the usual way, i.e.
\begin{lstyaml}
Parameters:
  # Scan over neutrino masses in normal hierarchy
  StandardModel_mNudiff: !import include/StandardModel_mNudiff_NH_scan.yaml
\end{lstyaml}

\end{description}

\subsection{Non-standard radiation content} \label{app:models-dNur}

We provide a family of models for a range of effects able to modify $\Delta \Neff$ (see Sec.\ \ref{sec:non-std-Neff} for details). These include the ability to vary the baryon-to-photon ratio $\eta$ before recombination, the temperature ratio between photons and SM neutrinos $r_\nu$, or the number of new ultra-relativistic species $\Delta N_{\rm ur}$.\footnote{Note that $\Delta \Neff = 0$ corresponds to a value of 3.045 in the early Universe at the time of recombination.}

We define $r_\nu$ and $\Delta N_{\rm ur}$ both at the end of BBN and at the beginning of recombination, allowing them to be separately varied at either epoch. Parameters defined at the end of BBN have the subscript `BBN', and those at the beginning of recombination the subscript `CMB'. All of these effects are captured in the general model \doublecrosssf{etaBBN\_rBBN\_rCMB\_dNurBBN\_dNurCMB}{etaBBN_rBBN_rCMB_dNurBBN_dNurCMB}.

When scanning a model in this family tree, $\Delta N_{\rm ur} < 0$ is only permitted if the \YAML option \yaml{allow_negative_delta_N_ur} is set \yaml{true}. We stress that \cosmobit performs no additional checks to ensure the physicality of results obtained in this mode.

\begin{description}

\item[\textbf{\textsf{etaBBN\_rBBN\_rCMB\_dNurBBN\_dNurCMB}}:\label{etaBBN_rBBN_rCMB_dNurBBN_dNurCMB}] \term{dNur\_BBN, dNur\_CMB, r\_BBN, r\_CMB, eta\_BBN}.

The full model allowing modifications to $\Delta \Neff$ via
\begin{itemize}
\item A non-standard baryon-to-photon-ratio at the end of BBN $\eta_\text{BBN}$,
\item Addition of ultra-relativistic degrees of freedom at the end of BBN (start of recombination) $\Delta N_\text{ur,BBN(CMB)}$,
\item Modification of the temperature ratio at the end of BBN (start of recombination) $r_{\nu,\text{BBN(CMB)}} = \left. \frac{T_\nu}{T_\nu^{\Lambda\text{\rm CDM}}} \right|_\text{BBN(CMB)}$.
\end{itemize}

\item[\textbf{\textsf{rBBN\_rCMB\_dNurBBN\_dNurCMB}}:\label{rBBN_rCMB_dNurBBN_dNurCMB}] \term{dNur\_BBN, dNur\_CMB, r\_BBN, r\_CMB}.

As per \doublecrosssf{etaBBN\_rBBN\_rCMB\_dNurBBN\_dNurCMB}{etaBBN_rBBN_rCMB_dNurBBN_dNurCMB}, but with $\eta_\text{BBN} = \eta_0$ as in $\lcdm$, where $\eta_0$ is the value today, inferred from $\omega_b$.

\item[\textbf{\textsf{dNurBBN\_dNurCMB}}:\label{dNurBBN_dNurCMB}] \term{dNur\_BBN, dNur\_CMB}.

As per \doublecrosssf{rBBN\_rCMB\_dNurBBN\_dNurCMB}{rBBN_rCMB_dNurBBN_dNurCMB}, but with no BSM contribution to the temperature of neutrinos, i.e.\ $r_\text{CMB} = r_\text{BBN} = 1$.

\item[\textbf{\textsf{rBBN\_rCMB}}:\label{rBBN_rCMB}] \term{r\_BBN, r\_CMB}.

As per \doublecrosssf{rBBN\_rCMB\_dNurBBN\_dNurCMB}{rBBN_rCMB_dNurBBN_dNurCMB}, but without any extra ultra-relativistic degrees of freedom, whether at the end of BBN or at the start of recombination, i.e.\ $\Delta N_\text{ur,CMB} = \Delta N_\text{ur,BBN} = 0$.

\item[\textbf{\textsf{rCMB}}:\label{rCMB}] \term{r\_CMB}.

As per \doublecrosssf{rBBN\_rCMB}{rBBN_rCMB}, but with the same ratio of neutrino temperatures at the end of BBN and at the start of recombination, i.e.\ $r_\text{CMB} = r_\text{BBN}$.

\item[\textbf{\textsf{dNurCMB}}:\label{dNurCMB}] \term{dNur\_CMB}.

As per \doublecrosssf{dNurBBN\_dNurCMB}{dNurBBN_dNurCMB}, but with the same number of extra ultra-relativistic degrees of freedom at the end of BBN as at the start of recombination, i.e.\ $\Delta N_\text{ur,CMB} = \Delta N_\text{ur,BBN}$.

\item[\textbf{\textsf{etaBBN}}:\label{etaBBN}] \term{eta\_BBN}.

As per \doublecrosssf{etaBBN\_rBBN\_rCMB\_dNurBBN\_dNurCMB}{etaBBN_rBBN_rCMB_dNurBBN_dNurCMB}, but with no contribution to $\Delta \Neff$ from either ultra-relativistic species nor a change of the neutrino temperature, i.e.\ $r_\text{CMB} = r_\text{BBN} = 1$ and $\Delta N_\text{ur,CMB} = \Delta N_\text{ur,BBN} = 0$.

\end{description}

\subsection{Energy injection} \label{app:models-energyinjection}

\cosmobit currently supports two scenarios of energy injection at $30 \le z \le 2000$: $s$-wave annihilation and decay of dark matter.  Annihilation and decay channels explicitly supported are (monochromatic) photons and $e^+e^-$ pairs. The two scenarios are expressed in terms of the dark matter mass $ m_\chi $, its fractional abundance $ \xi $, the total annihilation cross-section $ \langle\sigma v\rangle $ or lifetime $ \tau $, and the branching fractions of annihilation or decays into electrons ${B\!R}_{\rm el}$ and photons ${B\!R}_{\rm ph}$. The complementary branching fraction $1 - {B\!R}_{\rm el} - {B\!R}_{\rm ph}$ is assumed to be in the form of inefficient particle species such as neutrinos.  The electron and photon spectra will then be accordingly normalised.

By providing a new module function, \cosmobit can be extended to final states other than the monochromatic $\gamma\gamma$ and $e^+e^-$ and their mixture.  This new module function has to provide the spectrum of injected electrons and photons (capability \cpp{energy_injection_spectrum}) normalised to a single annihilation/decay event such that inefficient final states or accounted for.  In this case, the general models \doublecrosssf{AnnihilatingDM\_general}{AnnihilatingDM_general} and \doublecrosssf{DecayingDM\_general}{DecayingDM_general} should be used by providing a model relationship of the new model to one of the general models. The implementation of a more realistic scenario via this route, with more complex spectra, is described in \ref{app:darkages-new-model}.

\begin{description}
  \item[\textbf{\textsf{AnnihilatingDM\_general}}:\label{AnnihilatingDM_general}] \term{mass, sigmav}.

  This is the most general scenario of energy injection by $s$-wave annihilating dark matter, and requires an additional function to be written in order to provide the relevant efficiency functions or spectra of injected photons and electrons. The mass is given in units of $ {\rm GeV} $ and the thermal averaged cross-section $ \langle\sigma v\rangle $ is given in units of $ {\rm cm^3\,s}^{-1} $. This scenario implicitly assumes that the particle in question is all of dark matter, such that the fractional abundance $ \xi = 1 $. To consider scenarios where $\xi \ne 1$, the cross-section should be scaled down by $ \xi^2 $.

  \item[\textbf{\textsf{AnnihilatingDM\_mixture}}:\label{AnnihilatingDM_mixture}] \term{mass, sigmav, BR\_el, BR\_ph}.

  This is the basic scenario in which dark matter annihilates into a electron-positron pair with branching fraction (\texttt{BR\_el}) and a photon pair with branching fraction (\texttt{BR\_ph}). The complementary branching fraction ($ 1- $ \texttt{BR\_el} $ - $ \texttt{BR\_ph}) is assumed to be in the form of neutrinos or any other inefficient particle, which is kinematically allowed.  The resulting photons have energies of $ m_\chi$, and the electrons and positrons $m_\chi - m_e$ (i.e.\ no radiative corrections are included).  As for \doublecrosssf{AnnihilatingDM\_general}{AnnihilatingDM_general}, $ \xi = 1 $ is implicit, so treatment of $\xi \ne 1$ requires rescaling of \term{sigmav} by $ \xi^2 $.

  \item[\textbf{\textsf{AnnihilatingDM\_photon}}:\label{AnnihilatingDM_photon}] \term{mass, sigmav}.

  This is a special case of \doublecrosssf{AnnihilatingDM\_mixture}{AnnihilatingDM_mixture}, where dark matter annihilates exclusively into a pair of photons (\texttt{BR\_el} is set to $ 0 $ and \texttt{BR\_ph} is set to $ 1 $).

  \item[\textbf{\textsf{AnnihilatingDM\_electron}}:\label{AnnihilatingDM_electron}] \term{mass, sigmav}.

  This is a special case of \doublecrosssf{AnnihilatingDM\_mixture}{AnnihilatingDM_mixture}, where dark matter annihilates exclusively into an electron-positron pair (\texttt{BR\_el} is set to $ 1 $ and \texttt{BR\_ph} is set to $ 0 $).

  \item[\textbf{\textsf{DecayingDM\_general}}:\label{DecayingDM_general}] \term{mass, lifetime, fraction}.

  This is the most general scenario of energy injection by a decaying subcomponent of dark matter, and requires an additional function to be written in order to provide the relevant efficiency functions or spectra of injected photons and electrons. The mass is given in units of $ {\rm GeV} $ and the lifetime in units of $ {\rm s} $. The fractional abundance $ \xi $ is the value in the infinite past ($ t = 0 $).

  \item[\textbf{\textsf{DecayingDM\_mixture}}:\label{DecayingDM_mixture}] \term{mass, lifetime, fraction, BR\_el, BR\_ph}.

  This is the basic scenario in which dark matter decays into a electron-positron with branching fraction (\texttt{BR\_el}) and a photon pair with branching fraction (\texttt{BR\_ph}). The complementary branching fraction ($ 1- $ \texttt{BR\_el} $ - $ \texttt{BR\_ph}) is assumed to be in the form of neutrinos or any other inefficient particle, which is kinematically allowed. The resulting photons have energies of $ m_\chi/2$, and the electrons and positrons $m_\chi/2 - m_e$ (i.e.\ no radiative corrections are included).

  \item[\textbf{\textsf{DecayingDM\_photon}}:\label{DecayingDM_photon}] \term{mass, lifetime, fraction}.

  This is a special case of \doublecrosssf{DecayingDM\_mixture}{DecayingDM_mixture}, where the dark matter subcomponent decays exclusively into a pair of photons (\texttt{BR\_el} is set to $ 0 $ and \texttt{BR\_ph} is set to $ 1 $).

  \item[\textbf{\textsf{DecayingDM\_electron}}:\label{DecayingDM_electron}] \term{mass, lifetime, fraction}.

  This is a special case of \doublecrosssf{DecayingDM\_mixture}{DecayingDM_mixture}, where the dark matter subcomponent decays exclusively into an electron-positron pair (\texttt{BR\_el} is set to $ 1 $ and \texttt{BR\_ph} is set to $ 0 $).

\end{description}

\subsection{Cosmological nuisance parameters} \label{app:models-nuisance}

Various cosmological likelihoods require the values of additional nuisance parameters.  We implement these as separate models. All nuisance models are located in \term{Models/include/gambit/Models/models/CosmoNuisanceModels.hpp}. To obtain the results presented in Sec.\ \ref{sec:constraints}, we used two cosmological likelihoods with nuisance parameters: the Planck 2018 \cite{Aghanim:2018eyx} likelihood for CMB spectra, and the Pantheon \cite{Scolnic:2017caz} likelihood for measurements of the expansion rate of the Universe using SNe Ia light curves. The corresponding nuisance-parameter models are:
\begin{description}

\item[\textbf{\textsf{cosmo\_nuisance\_Pantheon}}:\label{cosmo_nuisance_Pantheon}] \term{M}.

Nuisance parameter for the Pantheon likelihood~\cite{Scolnic:2017caz}, where $M$ is the absolute magnitude of a Type Ia supernova.

\item[\textbf{\textsf{cosmo\_nuisance\_Planck\_lite}}:\label{cosmo_nuisance_Planck_lite}] \term{A\_planck}.

Nuisance parameter for the condensed (`lite') Planck CMB likelihoods~\cite{Aghanim:2018eyx}, where $A_{\rm Planck}$ is the Planck absolute calibration.
\end{description}
All other nuisance parameter models for likelihoods that are implemented but were not used in the present analysis are listed in Table~\ref{tab:impl-nuisance-models}.

\begin{table}[t]
\small
 \centering
  \begin{tabular}{lcl}
    \toprule
    Model Name & Parameters & Likelihood(s) \\
    \midrule\noalign{\vskip 5pt}
    && \textit{Provided by \textsf{plc}}\\[5pt]
    \textsf{cosmo\_nuisance\_Planck\_TTTEEE}  & 34    & Planck CMB TT-TE-EE\\
    \textsf{cosmo\_nuisance\_Planck\_TT}      & 16    & Planck CMB TT \\
    \textsf{cosmo\_nuisance\_Planck\_lite}    & 1     & Planck `lite' \\
    \midrule\noalign{\vskip 5pt}
    && \textit{Provided by \textsf{MontePython}}\\[5pt]
    \textsf{cosmo\_nuisance\_JLA}             & 4     & {JLA} \\
    \textsf{cosmo\_nuisance\_Pantheon}        & 1     & {JLA\_simple}, {Pantheon} \\
    \textsf{cosmo\_nuisance\_BK14}            & 10    & {BK14} \\
    \textsf{cosmo\_nuisance\_ISW}             & 16    & {ISW} \\
    \textsf{cosmo\_nuisance\_ska1}            & 7     & {ska1\_IM\_band1}, {ska1\_IM\_band2}\\
                                              &       & {ska1\_pk} \\
    \textsf{cosmo\_nuisance\_ska2\_pk}        & 3     & {ska2\_pk} \\
    \textsf{cosmo\_nuisance\_ska\_lensing}    & 1     & {ska1\_lensing}, {ska2\_lensing}\\
    \textsf{cosmo\_nuisance\_euclid\_lensing}                 & 1 & {euclid\_lensing}   \\
    \textsf{cosmo\_nuisance\_euclid\_pk}      & 4     & {euclid\_pk} \\
    \textsf{cosmo\_nuisance\_CFHTLens\_correlation}           & 1 & {CFHTLens\_correlation}\\
    \textsf{cosmo\_nuisance\_kids450\_qe\_likelihood\_public} & 7 & {kids450\_qe\_likelihood\_public} \\
    \textsf{cosmo\_nuisance\_dummy}       & 10    &  \\
    \bottomrule
  \end{tabular}
  \caption{List of all cosmological nuisance parameter models in \gambit, the number of free parameters contained in each, and the name of the likelihood that they influence.  Likelihoods above the line are provided by the \plc backend, while those below are available through \montepython \cite{Audren:2012wb,brinckmann2018montepython}.  Detailed information about the model parameters and their physical meaning can be obtained with the command \protect\term{gambit} \protect\metavar{model\_name}.  The model \doublecrosssf{cosmo\_nuisance\_dummy}{cosmo_nuisance_dummy} is implemented as an example of adding a new nuisance parameter model.  All new functions and files associated with these examples contain comments and instructions for potential developers.
  \label{tab:impl-nuisance-models}}
\end{table}

{
\label{cosmo_nuisance_dummy}
As an example of how to add a new cosmological nuisance parameter model, we include the model \doublecrosssf{cosmo\_nuisance\_dummy}{cosmo_nuisance_dummy}, with 10 nuisance parameters named \term{nuisance\_param\_1}, \term{nuisance\_param\_2}, \ldots  This model can also be used by \montepython likelihoods.  The way the interface between \cosmobit and \montepython is set up, the user does not need to make any modifications to \gambit to use a new \montepython likelihood containing no nuisance parameters (see Appendix~\ref{app:mp-interface} for interface details).  If however the likelihood requires the use of new nuisance parameters, a model containing these parameters must be added to \gambit; instructions for doing so are provided in Appendix~\ref{app:models-new}.  For testing or development purposes, \doublecrosssf{cosmo\_nuisance\_dummy}{cosmo_nuisance_dummy} can simply be used as a temporary nuisance model; the user need only ensure that the names of the nuisance parameters needed by their \montepython likelihood match those in \doublecrosssf{cosmo\_nuisance\_dummy}{cosmo_nuisance_dummy}.
}

\subsection{Neutron lifetime} \label{app:models-neutron}

Lab-based experiments do not agree on the value of the neutron lifetime $\tau_{\rm n}$, with ultra-cold neutron (`bottle') experiments finding values systematically lower than those based on (`beam') observations of decays to protons in flight (see \eg Ref.~\cite{PDG20} for a review).  The neutron lifetime is an important input for BBN, as it determines weak interaction rates, and therefore impacts the neutron-to-proton ratio at freeze-out. Its value therefore affects the helium abundance, which in turn has consequences for recombination.

We allow the user to vary $\tau_{\rm n}$ using the model \doublecrosssf{nuclear\_params\_neutron\_lifetime}{nuclear_params_neutron_lifetime}. If this model is not in use, \cosmobit adopts a default value of $\tau_{\rm n} = 879.4$s, based on the 2020 PDG recommendation \cite{PDG20}.

\begin{description}
  \item[\textbf{\textsf{nuclear\_params\_neutron\_lifetime}}:\label{nuclear_params_neutron_lifetime}] \term{neutron\_lifetime}.

  The \term{neutron\_lifetime} $\tau_{\rm n}$ is given in seconds.

\end{description}

\subsection{Adding a new model} \label{app:models-new}

In the initial release of \cosmobit, we do not include extensions of $\lcdm$ such as curvature or models for dark energy.  However, all capabilities and interfaces are set up in a maximally flexible way, so adding these scenarios requires only a small amount of additional code, with essentially no restructuring or modification of existing code.

To add a new physics model, a model declaration should be added to \term{Models/include/gambit/Models/models/CosmoModels.hpp}. For instructions on adding a new model to \GB, we refer the reader to Chapter 5 of Ref.\ \cite{gambit}. The model parameters will be available within any module function that has \cpp{ALLOW_MODELS(}\metavar{new\_model\_name}\cpp{)} in its rollcall declaration.  When adding a model, adding an informative description of the model and its parameters to \term{config/models.dat} gives the user information about the model when executing the diagnostic \term{gambit} \metavar{new\_model\_name}.

If the new model contains nuisance parameters,
\begin{enumerate}
  \item Add the declaration to \term{Models/include/gambit/Models/models/CosmoNuisanceModels.hpp}.
  \item If the model contains nuisance parameters to be used by a \montepython likelihood, add the model to the \cpp{ALLOW_MODELS} macro for the function \yaml{set\_parameter\_dict\_for\_MPLike} in \term{CosmoBit/include/gambit/CosmoBit/CosmoBit_rollcall.hpp}. This function automatically passes all values of nuisance parameters to \montepython.
\end{enumerate}

If the new model necessitates the passing of extra parameters to \class, this can be achieved following the instructions in Appendix~\ref{app:class-interface}.  An example of how to pass parameters of a new inflationary model to \multimodecode is provided in Appendix~\ref{app:multimode-new}.

An important thing to note when implementing new models is that the names of model parameters should ideally be unique, i.e.\ new parameter names should preferably not be identical to any other model parameter already defined within \gambit.  This avoids parameter clashes in cases where multiple models with common parameter names are being used in the same scan.  Furthermore, \gambit parameter names cannot include characters like \cpp{^\{\}()-:;*+?}\phone\xspace \cite{Gates:1985gk} that could not be used in \Cpp function or variable names.  Hence, if a \montepython likelihood involves a nuisance parameter containing these characters, a translation must be written in the function \yaml{set\_parameter\_dict\_for\_MPLike} in \term{CosmoBit/src/MontePython.cpp}.

%
\section{Backend interfaces}
\label{app:backends}
%

Here, we provide detailed descriptions of the interfaces between \gambit and the backends used by \cosmobit. A further practical introduction to the usage of all interfaces is given in the \cosmobit tutorial \YAML file, \term{yaml\_files/CosmoBit\_tutorial.yaml}.

\subsection{Interface to \alterbbn} \label{app:alterbbn-interface}

\cosmobit makes use of \alterbbn \cite{Arbey:2011nf,Arbey:2018zfh} for computing the abundances of light elements. \alterbbn allows the user to include several non-standard pieces of physics: additional ultra-relativistic species, non-standard neutrino temperatures, a modified expansion rate of the Universe and new sources of entropy injection.

\subsubsection*{Practical usage} \label{app:alterbbn-prac}

By default, \alterbbn is called before every \class run to calculate the primordial helium abundance $\yhe$.  To compute specific primordial abundances, as well as likelihoods based on some of those abundances, one would include the following \YAML entries:
\begin{lstyaml}
ObsLikes:

  # Compute predicted BBN abundances + errors. Those needed for BBN_LogLike are added
  # automatically, but it doesn't hurt to include them here anyway for clarity.
  - purpose:          Observable
    capability:       BBN_abundances
    # Compute the abundances of helium 4, deuterium and lithium 7.
    sub_capabilities:
      - He4
      - D
      - Li7

  # Compute likelihood from BBN, using only helium 4 and deuterium.
  - purpose:          LogLike
    capability:       BBN_LogLike
    sub_capabilities:
      - He4
      - D

Rules:

  # Set path to file containing measured values of light element abundances to use for
  # likelihood calculation (filenames are relative to CosmoBit/data/BBN/ within
  # the GAMBIT base directory)
  - capability: BBN_LogLike
    options:
      DataFile: default.dat
\end{lstyaml}

In this example, the \yaml{sub_capabilities} entry for the likelihood instructs \cosmobit to base the BBN likelihood on the abundances of $^4$He and $D$.  These can also be provided as a list, e.g. \yaml{sub_capabilities:} \yaml{[}\yamlvalue{He4, D}\yaml{]}.  The likelihood will be computed using the observations contained in the \yaml{DataFile} located at \term{gambit/CosmoBit/data/BBN/default.dat}.  This particular \yaml{DataFile} includes helium data from \cite{PDG20} and deuterium data from \cite{Cooke:2017cwo}, and is the one used automatically whenever such a \yaml{Rule} is not included in the user's \YAML file.

The \yaml{BBN\_abundances} entry instructs \GB to save and print the theoretically-predicted abundances (and their uncertainties) computed for each element.  The elements needed for the BBN likelihood are automatically included here.  Additional elements can be given as \yaml{sub_capabilities} and will be printed along with those used in the actual likelihood calculation; in the example above, in addition to $^4$He and $D$, the abundance of $^7$Li is also computed and output by the printer, along with its uncertainty.

Valid BBN \yaml{sub_capabilities} are: \yamlvalue{He4} (equivalently \yamlvalue{Yp}), \yamlvalue{H3}, \yamlvalue{He3}, \yamlvalue{D} (equivalently \yamlvalue{H2}), \yamlvalue{Li6}, \yamlvalue{Li7}, \yamlvalue{Li8}, and \yamlvalue{Be7}. In order to employ any other isotopes than these, the \cpp{abund_map} in \term{backend_types/AlterBBN.cpp} must be extended. We refer the user to the \alterbbn manual~\cite{Arbey:2018zfh} for information about the relevant array indexing and how to achieve this.

\subsubsection*{Theoretical error and covariance matrix calculation} \label{app:alterbbn-err}

In addition to physical input parameters, \alterbbn also takes two input parameters controlling technical details of the calculation: \yaml{failsafe} and \yaml{err}. The parameter \yaml{failsafe} chooses the numerical integration method and corresponding error tolerance. The parameter \yaml{err} specifies if and how the estimation of the theoretical uncertainties of nuclear reactions are calculated. The user can choose to not calculate any errors (\yaml{err: 0}), get uncorrelated error estimates (\yamlvalue{1} or \yamlvalue{2}), or a full covariance matrix (\yamlvalue{3} or \yamlvalue{4}).  For a detailed explanation of all available options, we refer the reader to Appendix B of Ref.~\cite{Arbey:2018zfh}.

We performed several parameter scans comparing the different integration and error calculation modes. For the non-standard scenarios implemented in \cosmobit, we find that the option \yaml{failsafe:7} provides the most stable results, both for the central values and the individual uncertainties of the elemental abundances. Setting \yaml{failsafe:3} significantly reduces computational costs, while still providing acceptable results, in particular if no error estimates are required. Invoking error calculations that include covariances between the different elements by setting \yaml{err} to \yamlvalue{3} or \yamlvalue{4} increase the runtime considerably. We find, however, that the covariance matrix is relatively stable across the parameter space, when normalised to the $1\, \sigma$ uncertainties of the respective elements. Therefore, we also offer an option to override the correlation matrix output by \alterbbn with a fixed matrix specified in the \YAML file, such that in order to obtain the overall covariance, the fixed correlation matrix is rescaled by the individual uncertainties for each isotope computed on a point-to-point basis by \alterbbn. This allows the correlated theoretical uncertainties to be taken into account without additional computational cost.

The syntax for setting these options is
\begin{lstyaml}
Rules:

  # Error calculation precision settings for AlterBBN
  # Defaults: @\ \ \ \ \ \ \ \ \ \ \ \ \ \ \ \ \ \ \ \ \ @failsafe: 7@\ \ \ \ \ \ \ @err: 1
  # For fast testing:@\ \ \ \ \ \ \ \ \ \ \ \ \ \ @failsafe: 1@\ \ \ \ \ \ \ @err: 0
  # For more precise calculations:@\ @failsafe: 3 or 7@\ \ @err: 3 or 4
  - capability: AlterBBN_Input
    options:
      failsafe: 7
      err: 1

  # correlation matrix for BBN error calculation
  - capability: BBN_abundances
    options:
      isotope_basis: [Yp@\yamlvalue{, D, He3, Be7, Li7}@]
      correlation_matrix: [[+1.000e00@\yamlvalue{, +1.524e-2, +2.667e-2, +2.303e-2, +2.429e-2}@]@\yamlvalue{,}@
                              [@\yamlvalue{+1.524e-2, +1.000e00, -8.160e-1, -3.775e-1, -3.653e-1}@]@\yamlvalue{,}@
                              [@\yamlvalue{+2.667e-2, -8.160e-1, +1.000e00, +3.897e-1, +3.780e-1}@]@\yamlvalue{,}@
                              [@\yamlvalue{+2.303e-2, -3.775e-1, +3.897e-1, +1.000e00, +9.974e-1}@]@\yamlvalue{,}@
                              [@\yamlvalue{+2.429e-2, -3.653e-1, +3.780e-1, +9.974e-1, +1.000e00]}@]
\end{lstyaml}
Note here that the \yaml{isotope_basis} option must be provided in order to indicate which rows and columns in the user's \yaml{correlation_matrix} correspond to which isotopes.

For users wishing to use uncertainty estimates that differ from those calculated by \alterbbn, we provide further options to either pass constant relative uncertainties (using option \yaml{relative_errors}) or constant absolute uncertainties (using option \yaml{absolute_errors}). Both options must contain the list of the respective values in the order given by \yaml{isotope_basis}.\footnote{Note that these two options cannot be used simultaneously as they contradict each other. Furthermore, the usage of either of these options implies the option \yaml{err: 0} for \yaml{AlterBBN\_Input}, as any error estimate would be overridden, unless \yaml{isotope_basis} contains only a subset of the isotopes calculated for use in subsequent capabilities. In case \yaml{isotope_basis} contains every isotope that is used and \yaml{err: 0} is not set, the code will raise an error.}

\subsubsection*{Adding new input options} \label{app:alterbbn-new-opt}

The BSM features of \alterbbn employed in \cosmobit are the ability to modify the baryon-to-photon ratio, the number of ultra-relativistic species, and the neutron lifetime.  If the user wishes to implement a new model (Appendix~\ref{app:models-new}) for which further non-standard phenomena during the phase of BBN play a role, they must also pass the corresponding settings to \alterbbn.

Here we describe the interface with \alterbbn used for the \GB model \doublecrosssf{nuclear\_params\_neutron\_lifetime}{nuclear_params_neutron_lifetime}, as an example of how to add support for a new model to \alterbbn. This model has a single parameter \term{neutron\_lifetime}, and is defined in \term{Models/include/gambit/Models/models/nuclear_params.hpp}.

The steps required to interface to \alterbbn are as follows:
\begin{enumerate}
\item Declare that the function \yaml{AlterBBN\_Input} can be used with the model \doublecrosssf{nuclear\_params\_neutron\_lifetime}{nuclear_params_neutron_lifetime}, and provide access to its model parameters.  This can be achieved by adding the lines
\begin{lstcpp}
ALLOW_MODEL_DEPENDENCE(nuclear_params_neutron_lifetime)
MODEL_GROUP(neutron,(nuclear_params_neutron_lifetime))
ALLOW_MODEL_COMBINATION(cosmo,neutron)
\end{lstcpp}
to the declaration of \yaml{AlterBBN\_Input} in \term{CosmoBit\_rollcall.hpp}.  This explicitly declares that \yaml{AlterBBN\_Input} can be used with the \doublecrosssf{nuclear\_params\_neutron\_lifetime}{nuclear_params_neutron_lifetime} model, but only if one of the models in the existing model group \cpp{cosmo} is in use at the same time. The \cpp{cosmo} group is defined already in the declaration of \yaml{AlterBBN\_Input}, and contains \doublecrosssf{LCDM}{LCDM}, \doublecrosssf{LCDM\_theta}{LCDM_theta} and \doublecrosssf{etaBBN\_rBBN\_rCMB\_dNurBBN\_dNurCMB}{etaBBN_rBBN_rCMB_dNurBBN_dNurCMB}; this existing declaration (which does not need to be altered when adding support for a new model) is
\begin{lstcpp}
MODEL_GROUP(cosmo,(LCDM, LCDM_theta, etaBBN_rBBN_rCMB_dNurBBN_dNurCMB))
\end{lstcpp}

\item Add all relevant input parameters that need to be passed to \alterbbn to the \cpp{string}-to-\cpp{double} map \cpp{result} computed by the function \cpp{AlterBBN\_Input} (located in \term{CosmoBit/src/BBN.cpp}). In this example the following \cpp{if} statement is added to the function:
\begin{lstcpp}
void AlterBBN_Input(map_str_dbl &result)
{
  using namespace Pipes::AlterBBN_Input;

  // Neutron lifetime either treated as a free model parameter ..
  if (ModelInUse("nuclear_params_neutron_lifetime"))
  {
    result["neutron_lifetime"] = *Param.at("neutron_lifetime");
  }
  // .. or fixed to PDG 2019 recommendation
  else
  {
    result["neutron_lifetime"] = 879.4;
  }
  ...
}
\end{lstcpp}
Note that the \cpp{ALLOW_MODEL_DEPENDENCE} macro in the header definition of the function allows the user to access the parameters of the \cpp{nuclear_params_neutron_lifetime} model.

\item Pass the input to \alterbbn through the backend convenience function \cpp{fill\_cosmomodel} in the frontend source file (\term{Backends/src/frontends/AlterBBN_2_2.cpp}) by adding the lines
\begin{lstcpp}
  if (AlterBBN_input.count("neutron_lifetime"))
  {
    // Neutron lifetime is known as "life_neutron" in AlterBBN's relicparam type
    input_relicparam->life_neutron = AlterBBN_input["neutron_lifetime"];
  }
\end{lstcpp}
\item Add the relevant \alterbbn input option (in this case \cpp{life_neutron}) to the set of options known to the \alterbbn frontend in \GB. These are stored in string set \cpp{known_relicparam_options} in \term{Backends/src/frontends/AlterBBN_2_2.cpp}.  This string set acts as a safeguard to prevent the user from trying to pass settings from \cosmobit to \alterbbn without actually having implemented the option in the frontend.

\end{enumerate}

\subsubsection*{Technical details} \label{app:alterbbn-tech}

All settings and assumptions relevant for the element abundance calculation in \alterbbn are defined within its \cpp{relicparam} structure.  This structure contains the values of all model parameters needed for the abundance calculation, such as the baryon-to-photon ratio after BBN \cpp{eta0}, the number of ultra-relativistic degrees of freedom \cpp{dNnu}, or the density and temperature of an additional dark component.

To provide access to these variables, the \cpp{relicparam} structure is available for use within \gambit by defining it as a backend type in \term{Backends/include/gambit/Backends/backend\_types/AlterBBN.hpp}.

The communication between \cosmobit and \alterbbn works as follows:
\begin{enumerate}
  \item \textbf{Set \alterbbn input parameters}.

  \yaml{capability}: \yamlvalue{AlterBBN\_Input}, \yaml{function}: \yamlvalue{AlterBBN\_Input}, \yaml{type:} \yamlvalue{map\_str\_dbl}.

  All parameters and settings to be passed to \alterbbn are collected in a \cpp{map_str_dbl} (a \GB\ \cpp{typedef} for \cpp{std::map<std::string,double>}) from parameter name to the respective value. Currently, these are \cpp{eta0}, \cpp{dNnu}, \cpp{Nnu}, the effective number of SM neutrinos, and \cpp{tau_neutron}. The user can also pass values of \yaml{failsafe} and \yaml{err}; the defaults are \yamlvalue{7}, and \yamlvalue{1}, respectively.

  \item \textbf{Get primordial element abundances}.

  \yaml{capability}: \yamlvalue{BBN\_abundances}, \yaml{function}: \yamlvalue{compute\_BBN\_abundances}, \yaml{type:} \yamlvalue{BBN\_Container}.

  Gains access to the \alterbbn input through a dependency on \cpp{AlterBBN\_Input} and stores it within the \cpp{BBN_Container} type (see Table~\ref{tab:type-BBNcontainer}). \cpp{BBN_Container} holds the empty vector \cpp{BBN\_abund} and matrix \cpp{BBN\_covmat} storing the primordial element abundances and their covariances.

  Calls the convenience function \cpp{call_nucl_err}, which creates an instance of \cpp{relicparam}. \cpp{call_nucl_err} calls another convenience function \cpp{fill_cosmomodel}, which fills \cpp{relicparam} with values from \cpp{AlterBBN\_Input} and calls the \alterbbn function \cpp{nucl\_err} to compute the abundances and their covariance.  The abundances and covariances are used to fill the \cpp{BBN_Container} variables \cpp{BBN\_abund} and \cpp{BBN\_covmat} respectively.

  \item \textbf{Compute BBN likelihood}.

  \yaml{capability}: \yamlvalue{BBN\_LogLike}, \yaml{function}: \yamlvalue{compute\_BBN\_LogLike}, \yaml{type:} \yamlvalue{double}.

  Has a dependency on \cpp{BBN\_abundances} computed in step 2; through this dependency we gain access to the abundance calculation results stored in the \cpp{BBN\_Container}.  From these results we calculate the BBN likelihood as described in Sec.\ \ref{sec:BBN}.

\end{enumerate}

When adding a new model to interface with \alterbbn, the routines described in steps 1 and 3 must be altered.

\begin{table}[t]
 \centering
 \small
   \makebox[\linewidth]{
  \begin{tabular}{llp{7.5cm}}
    \toprule
    Type     & Name        & Purpose \\
    \midrule
    \cpp{Class}  & \cpp{BBN\_Container}       &   \\
    \midrule

    \cpp{size_t} & \cpp{NNUC}
              & \cpp{private}, number of element abundances computed by \alterbbn \\

    \cpp{vec<double>}   & \cpp{BBN\_abund}
              & \cpp{private}, vector of length \cpp{NNUC+1} to be filled with element abundances  \\

  \cpp{vec<vec<double>>} & \cpp{BBN\_covmat}
              & \cpp{private}, covariance matrix of element abundances with dimension \cpp{NNUC+1} $\times$ \cpp{NNUC+1}  \\

    \cpp{map_str_int}   & \cpp{abund\_map}
              & \cpp{private}, map from element name to the respective position in \cpp{BBN\_abund} vector, e.g. \cpp{"H2"} $\rightarrow$ \cpp{3} \\
    \midrule
    \cpp{map_str_int}   & \cpp{get\_abund\_map}       ( )
              & \cpp{return} the member \cpp{abund\_map}   \\

    \cpp{double}        & \cpp{get\_BBN\_abund} (\cpp{int} \cpp{i}) & \cpp{return} the $i^{th}$ element of \cpp{BBN\_abund} \\
    \cpp{double}        & \cpp{get\_BBN\_covmat} (\cpp{int} \cpp{i}, \cpp{int} \cpp{j}) & \cpp{return} the $(i,j)^{th}$ element of \cpp{BBN\_covmat} \\
    \bottomrule
  \end{tabular}
  }
  \caption{Members and attributes of the type \protect\cpp{BBN\_Container}. Here \cpp{vec} is shorthand for \cpp{std::vector}, and a \cpp{map_str_int} is a \GB\ \cpp{typedef} for \cpp{std::map<std::string,int>}. The main purpose of these types is to store the result of \alterbbn in a type accessible to module functions in \cosmobit.  The member \protect\cpp{abund\_map} is a hard-coded map from the name of an element to its index in the abundance vector and covariance matrix.  This is to avoid mistakes when accessing the result for a particular element, and to make the code more explicit.}
  \label{tab:type-BBNcontainer}
\end{table}

\subsection{Interface to \class} \label{app:class-interface}

\GB currently includes interfaces to \class \textsf{2.6.3}, \textsf{2.9.3} and \textsf{2.9.4}, as well as \exoclass \textsf{2.7.2}.\footnote{
We used \class \textsf{2.6.3} to produce Figures \ref{fig:BBN_plots}--\ref{fig:BBN}, \ref{fig:dNeff_triangle} and \ref{fig:dNeff_lcdm}, and \exoclass \textsf{2.7.2} for \autoref{fig:energy_injection}.}

Rather than access the \plainC structures and functions of \class through a shared library, we interface \gambit and \class it via the \python wrapper \classy.\footnote{Hence, the \term{make} target to build the interface to \class is called \classy. This is to be explicit about our use of the \python wrapper for the interface.}  The \plainC interface would give \gambit full access to all structures and their members, and the ability to run every \class module individually.%
\footnote{\class is separated into different modules, each responsible for different physics. For example, the \py{background} module solves the background equations, and the spectra are computed in the \py{spectra} module.} %
Although this is not possible through the \python interface, the \python wrapper provides the possibility to easily use cosmological likelihoods implemented in \montepython (Appendix \ref{app:mp-interface}). It also makes the interface version-independent, meaning that if relevant input parameters of a given \class version are known to its \python wrapper, no new code needs to be produced to interface this alternative \class version with \gambit.  This makes it easy to use (privately) developed extensions of \class for non-standard physics scenarios.

At the moment \class is the only Boltzmann solver interfaced with \GB. However, the interface is designed in a way that \class-specific capabilities and functions are clearly separated from capabilities that could in principle be resolved by a different source code. Hence, the design allows for the possibility to implement interfaces to other Boltzmann solvers with minimal effort.

\subsubsection*{Practical usage} \label{app:class-prac}

\GB's dependency resolution process ensures that a dictionary is created containing all input parameters required by \class. These depend on the settings specified in the \YAML file: the chosen models dictate which model parameters \cosmobit passes to \class, and the likelihoods dictate the run-specific settings (e.g.~which spectra need to be computed).  Because \cosmobit sets and passes these values automatically, the user does not need to be aware of any \class-specific syntax, settings nor functions.
(title, first paragraph) Specify energy injection during which epoch

If the user does wish to pass additional settings to \class, they can specify the settings in the \yaml{Rules} sections of the input \YAML file. For example,
\begin{lstyaml}
Rules:
  # Pass additional run options to CLASS
  - capability: classy_input_params
    function: set_classy_input_params
    options:
      classy_dict:
        back_integration_stepsize: 7.e-3
        tol_background_integration: 1.e-2
  \end{lstyaml}
results in a \class run with increased precision for the background integration.  We stress that this should be used for settings related to the \class run, and not scan parameters.  The user can pass any \class setting that is understood by the respective \class version's \python wrapper through this option.  If one specifies conflicting values for a \class setting, \cosmobit will exit with an error identifying any problematic entries. Such a conflict can occur if, for example, the user tries to overwrite a model parameter, or passes an option that conflicts with a setting required by one of the likelihoods in use.

\subsubsection*{Setting \class input for a new model} \label{app:class-new-model}

Consider an example model \textbf{\textsf{DE\_model}}, with parameters \cpp{w0_fld} and \cpp{wa_fld}.  Passing information about the new model from \cosmobit to \class works as follows:
\begin{enumerate}
  \item Add a \cpp{MODEL\_CONDITIONAL\_DEPENDENCY} to the declaration of the function \cpp{set\_classy\_input\_params} in the rollcall header \term{CosmoBit\_rollcall.hpp}, e.g.
  \begin{lstcpp}
MODEL_CONDITIONAL_DEPENDENCY(classy_parameters_DE_model, pybind11::dict, DE_model)
  \end{lstcpp}
  \item Declare the new capability and a corresponding module function in the rollcall header:
  \begin{lstcpp}
#define CAPABILITY classy_parameters_DE_model
START_CAPABILITY
  #define FUNCTION set_classy_parameters_DE_model
   START_FUNCTION(pybind11::dict)
   ALLOW_MODEL(DE_model)
  #undef FUNCTION
#undef CAPABILITY
  \end{lstcpp}
  \item Implement the module function \cpp{set\_classy\_parameters\_DE\_model} containing the model-specific settings in \term{CosmoBit/src/Boltzmann.cpp}:
  \begin{lstcpp}
/// Provide a dictionary containing CLASS settings for the DE_model
void set_classy_parameters_DE_model(pybind11::dict &result)
{
  using namespace Pipes::set_classy_parameters_DE_model;

  // Make sure nothing from previous parameter combination is retained
  result.clear();

  // Set energy density of cosmological constant to 0,
  // and let CLASS infer the energy density of the perfect fluid
  result["Omega_Lambda"] = 0;

  // Set model parameters for parametrised time evolution of w
  result["w0_fld"] = *Param["w0_fld"];
  result["wa_fld"] = *Param["wa_fld"];
}
  \end{lstcpp}
  Note that \cpp{ALLOW_MODEL(DE_model)} in the rollcall header declaration of the module function allows the user to access the values of the parameters of this model via the \cpp{Param} map.

\item Add the additional input parameters to the dictionary that will be passed to \class. This can be achieved by adding the following lines to the function \cpp{set\_classy\_input\_params} in \term{CosmoBit/Boltzmann.cpp}:
\begin{lstcpp}
if (ModelInUse("DE_model"))
{
  // Add options specific to the DE_model to the python dictionary
  // passed to CLASS (consistency checks only executed in first run)
  result.merge_input_dicts(*Dep::classy_parameters_DE_model);
}
\end{lstcpp}

\end{enumerate}

\subsubsection*{Adding a new \class version} \label{app:class-new-version}

We recommend implementing all new versions of \class (even those that have diverged from the mainline public \class repository) as different versions of the \classy backend.  For example, \exoclass \textsf{2.7.2} is implemented as version \term{exo\_2.7.2}, and \textsf{my\_CLASS} version 4.4.0 should be implemented as version \term{my\_4.4.0}. To add such a version of \class as a backend:
\begin{enumerate}

\item Add the default location of version \term{my\_4.4.0} to \term{config/backend_locations.yaml.default}.

\item Add the download and installation commands for version \term{my\_4.4.0} to \term{cmake/backends.cmake}. Copy the code for another version of \classy and replace the version name, download link, and \term{md5} checksum.\footnote{Make sure that only one version of \classy is tagged as the default, using the \cmake command \cpp{set\_as\_default\_version("backend"} \cpp{\$\{name\}} \cpp{\$\{ver\})}.} To use features requiring \class patches, one needs to provide the patch file in the location \term{Backends/patches/classy/my\_4.4.0/classy\_my\_4.4.0}. If no patches need to be applied, one should remove all build commands related to patches.

\item Create a header and source file for the frontend interface to the new \class version. Place the header in \term{Backends/include/gambit/Backends/frontends/} and the source file in \term{Backends/src/frontends/}. Simply copy and rename the corresponding file for the default \classy version and replace the version name. In both header and source file, only the version numbers need to be modified. In the header file, this means modifying the \cpp{#define VERSION} and \cpp{#define SAFE_VERSION} macros, and in the source file it means modifying the name of the header included with \cpp{#include}.

\item Re-run \term{cmake ..; make -j}\metavar{n} \term{gambit} in the \GB build directory.

\end{enumerate}
After running \cmake again, it should be possible to build the new \class version by typing \term{make class_my_4.4.0} in the build directory. All structure members, input parameters and methods that are known to the \Python wrapper of the new version will be automatically available to set and use from within \gambit.

\subsubsection*{Patches applied to \class} \label{app:class-patches}

In principle it is not necessary to patch \class for a basic $\lcdm$ scan with \gambit.  There are however two scenarios in which we need to apply patches:\begin{itemize}
\item[i)] when using an external code (e.g.\ \multimodecode; see Appendix \ref{app:multimode-interface}) to pass primordial power spectra to \class
\item[ii)] when considering additional energy injection into the primordial gas around the time of CMB formation, requiring the passing of annihilation/decay coefficients to \class.
\end{itemize}
Both of these scenarios require the passing of arrays to \class. To accomplish this, we patch \class to accept pointers as inputs.\footnote{We achieve this by converting the memory address to the array's first element to a string, which is converted back to a pointer within \class. Detailed comments on the implementation are given in the patch files, e.g.\ \term{Backends/patches/classy/2.6.3/classy_2.6.3.diff}.} The patches that we apply to modify existing input options are:

\begin{itemize}

   \item \textbf{Additional primordial spectral type \py{"pointer\_to\_Pk"}}: In the standard version of \class, the user can choose different options for the computation of the primordial scalar and tensor power spectra via the input parameter \py{"P\_k\_ini type"}. We add the additional possibility \py{"pointer\_to\_Pk"} to the allowed values of \py{"P\_k\_ini type"}.  Setting \py{"P\_k\_ini type"} to \py{"pointer\_to\_Pk"} passes arrays from \cosmobit to \class, and skips the computation of the primordial power spectra within \class.\footnote{In the function \cpp{primordial\_init} in the \class source file \term{source/primordial.c}.}  This is used whenever one of the specific inflation models (Appendix \ref{app:models-inflation}) is used in conjunction with \multimodecode (i.e.\ when neither explicitly scanning \doublecrosssf{PowerLaw\_ps}{PowerLaw_ps} nor \doublecrosssf{Minimal\_PowerLaw\_ps}{Minimal_PowerLaw_ps}).

   This allows \cosmobit to pass arrays of the wavenumber $k$ (\cpp{k\_array}), the amplitude of scalar modes (\cpp{pks\_array}), and the amplitude of tensor modes (\cpp{pkt\_array}) to \class, as computed anywhere else during a \GB run. The length of the array is passed via the parameter \cpp{lnk\_size}.

  \item \textbf{Additional energy injection coefficient types \term{pointer\_to\_fz\_channel} and \term{pointer\_to\_fz\_eff}}: \exoclass provides different options for the calculation of the efficiency functions of energy injection by decay or annihilation of exotic components during the formation of the CMB.  The user can choose the calculation mode with the input parameter \term{energy\_deposition\_function}.  We add two additional options for this parameter: \term{pointer\_to\_fz\_channel} and \term{pointer\_to\_fz\_eff}. If a decaying or annihilating dark matter model depositing energy at CMB times is scanned, \cosmobit automatically passes one of these options to \exoclass.  Which option \cosmobit chooses depends on the operation mode of \darkages (Appendix \ref{app:darkages-interface}). In each case, the relevant energy efficiency functions are passed to \exoclass and stored internally.

  This patch only applies to \exoclass and is needed whenever a decaying or annihilating dark matter model is scanned (\doublecrosssf{AnnihilatingDM\_general}{AnnihilatingDM_general} or \doublecrosssf{DecayingDM\_general}{DecayingDM_general}) in combination with \doublecrosssf{LCDM}{LCDM}.

\end{itemize}

\begin{table}[t]
 \centering
 \small
   \makebox[\linewidth]{
  \begin{tabular}{llp{8cm}}
    \toprule
    Type     & Name of member variable/function   & Purpose \\
    \midrule
    \cpp{pydict} & \classinput
              & \cpp{private} dictionary filled with all inputs to be passed to \class \\
    \midrule
    \cpp{pydict} & \cpp{get\_input\_dict} ( )
              & \cpp{return} \classinput \\

    \cpp{void}   & \cpp{clear}        ( )
              & clear all entries from \classinput \\

    \cpp{bool}   & \cpp{has\_key}       (\cpp{str} \cpp{key})
            & check if \cpp{key} is already contained in \classinput\\

    \cpp{void}   & \cpp{add\_entry}       (\cpp{str} \cpp{key}, \metavar{val\_type} \cpp{val})
            & add \cpp{key}, \cpp{val} pair to \classinput, for \metavar{val\_type} \cpp{str}, \cpp{double}, \cpp{int} or \cpp{vector<double>}\\

    \cpp{void}   & \cpp{merge\_input\_dicts} (\cpp{pydict} \cpp{extra\_dict})
              & merge entries from \cpp{extra\_dict} into \classinput; includes \class-specific rules on how to treat entries contained in both dictionaries \\
    \bottomrule
  \end{tabular}
  }
  \caption{Member variables and functions of the class \cpp{Classy_input}. The type \cpp{str} is a \GB\ \cpp{typedef} of \cpp{std::string}, and \cpp{pydict} is shorthand for \cpp{pybind11::dict}. The main purpose of the class is to fill the member \protect\classinput with all settings and parameters that must be passed to \class.}
  \label{tab:type-classyInput}
\end{table}

\subsubsection*{Technical details} \label{app:class-tech}

The \classy wrapper provides a \python class \textit{named} \py{Class}.  It is common to instantiate the object as \py{cosmo} in \class examples. We refer to this object as \py{cosmo} for clarity in the following section. The object \py{cosmo} provides all necessary functions to run \class. For communication between \Cpp and \python within \GB, we use \pybind,\footnote{\url{https://github.com/pybind/pybind11}} which for all intents and purposes, makes it possible to work directly with \python objects and execute \python code from within \Cpp.

The first important part of the \class interface to \gambit is that \cosmobit must create a \class input dictionary reflecting all model- and likelihood-dependent run choices.  To achieve this, we implement several different model-specific module functions within \cosmobit.  These functions gather the relevant input parameters for different models and store them in \python dictionaries.  The dictionaries map the parameter names in \class syntax to the corresponding input values. \cosmobit collects these dictionaries into the final input dictionary for \class, \cpp{input\_dict}, within the function \cpp{set\_classy\_input\_params}. This \classinput is returned as a member variable of an instance of the class \cpp{Classy\_input}, which provides some additional helper methods. This type is detailed in Table~\ref{tab:type-classyInput}.

The second important part of the interface is the \classy frontend. Here, the \cpp{static} object \py{cosmo} provides all necessary functions to steer \class:  setting the input parameters (\py{cosmo.set(input_dict)}), running \class (\py{cosmo.compute()}), cleaning the structures (\py{cosmo.struct_cleanup()}), and requesting computed quantities (e.g.\ \py{cosmo.lensed_cl()} for the lensed CMB spectra).  We implemented backend convenience functions to return relevant \class outputs. Besides the CMB spectra, these convenience functions return, e.g. the angular and luminosity distances to a given redshift, or the linear growth rate.

The individual steps executed in \cosmobit related to a \class call are:

\begin{enumerate}

\item {\bf Set input related to neutrino parameters and ultra-relativistic species}.

\yaml{capability}: \yamlvalue{classy\_NuMasses\_Nur\_input}, \yaml{function}: \yamlvalue{set\_classy\_NuMasses\_Nur\_input}, \yaml{type:} \yamlvalue{pybind11::dict}.

Creates and returns a dictionary containing the \class inputs for: the number of ultra-relativistic species today (\py{"N\_ur"}), the three SM neutrino masses (\py{"m\_ncdm"}), and the neutrino-to-photon temperature ratio (\py{"T\_ncdm"}). This function explicitly assumes that there are three massive neutrino eigenstates, and no other non-cold dark matter components. The values for \py{"N\_ur"} and \py{"T\_ncdm"} are obtained from the capabilites \cpp{N\_ur} and \cpp{T\_ncdm}, respectively.

\item {\bf Set primordial input parameters}.

\yaml{capability}: \yamlvalue{classy\_primordial\_input}, \yaml{type:} \yamlvalue{pybind11::dict}.

Pass any relevant primordial quantities from \cosmobit to \class. The primordial helium abundance (\py{"YHe"}) is set through a dependency on the capability \yaml{helium\_abundance}.
There are two separate functions fulfilling the capability \yaml{classy\_primordial\_input}, based on the form of the primordial power spectra:

\begin{enumerate}

  \item \yaml{function}: \yamlvalue{set\_classy\_parameters\_parametrised\_ps}.

  \cosmobit automatically activates this function if the model \doublecrosssf{PowerLaw\_ps}{PowerLaw_ps} (or its child \doublecrosssf{Minimal\_PowerLaw\_ps}{Minimal_PowerLaw_ps}) is in use. The pivot scale (\py{"k\_pivot"}), the amplitude (\py{"ln10^\{10\}A\_s"}) and the tilt (\py{"n\_s"}) of the primordial scalar power spectrum are added to the \python dictionary. If the scalar-to-tensor ratio is not zero, the value of \py{"r"} is included in the dictionary and \py{"modes"} is set to \py{"s,t"}, instructing \class to compute scalar and tensor modes.

  \item \yaml{function}: \yamlvalue{set\_classy\_parameters\_primordial\_ps}.

  \cosmobit automatically activates this function if an inflation model (\textbf{\textsf{Inflation\_\dots}}) is in use. The shape of the primordial power spectra for scalar and tensor perturbations is accessed through a dependency on the capability \cpp{primordial\_power\_spectrum}. We set the \class option \py{"modes"} to \py{"s,t"} to compute scalar and tensor modes, and \py{"P\_k\_ini type"} to \py{"pointer\_to\_Pk"} to enable us to pass pointers to the arrays containing the wavenumber and the primordial scalar and tensor power spectra.

\end{enumerate}

\item {\bf Set input parameters related to energy injection around the time of CMB formation}.

\yaml{capability}: \yamlvalue{classy\_parameters\_EnergyInjection}, \yaml{type:} \yamlvalue{pybind11::dict}.

If any models of energy injection are being scanned, relevant parameters are added to the input dictionary, to be passed to \exoclass. The two functions with capability \yamlvalue{classy\_parameters\_EnergyInjection} are \cpp{set\_classy\_parameters\_EnergyInjection\_AnnihilatingDM} and \cpp{set\_classy\_parameters\_EnergyInjection\_DecayingDM}.  These respectively pass all parameters relevant for exotic energy injection from dark matter annihilation or decay, and are activated if the models \doublecrosssf{AnnihilatingDM\_general}{AnnihilatingDM_general} or
\doublecrosssf{DecayingDM\_general}{DecayingDM_general} (or any of their descendents) are in use.

\item {\bf Set input parameters from Planck likelihoods}.

\yaml{capability}: \yamlvalue{classy\_PlanckLike\_input}, \yaml{function}: \yamlvalue{set\_classy\_PlanckLike\_input}, \yaml{type:} \yamlvalue{pybind11::dict}.

If one of the \plc likelihoods (Appendix \ref{app:plc-interface}) is in use, this function provides the relevant \class input settings for the likelihood calculation. These are the scale up to which the CMB spectra need to be computed (\py{"l\_max\_scalars"}), and the flags indicating to \class that it should include non-linear corrections to the matter power spectrum (\py{"non linear"} = \py{"halofit"}) and the effects of gravitational lensing in its predictions of CMB spectra (\py{"lensing"} = \py{"yes"}). The output spectra are chosen according to which likelihood(s) the user requests (\py{"pCl"} for polarisation, \py{"tCl"} for temperature, \py{"lCl"} for lensing).

\item {\bf Set input parameters from \montepython likelihoods}.

\yaml{capability}: \yamlvalue{classy\_MPLike\_input}, \yaml{type:} \yamlvalue{pybind11::dict}.

If any \montepython likelihoods require any additional output from \class, these are added to the input dictionary in order to ensure that they are computed.
\begin{enumerate}
  \item \yaml{function}: \yamlvalue{set\_classy\_input\_no\_MPLike}.
  Automatically chosen if \montepython is not installed. Returns an empty dictionary.
  \item \yaml{function}: \yamlvalue{set\_classy\_input\_with\_MPLike}.
  If \montepython likelihoods are in use, this adds the required input settings to the returned dictionary. If not, it returns an empty dictionary.
\end{enumerate}

\item {\bf Combine all input parameters for \class run}.

\yaml{capability}: \yamlvalue{classy\_input\_parameters}, \yaml{function}: \yamlvalue{set\_classy\_input\_parameters}, \yaml{type:} \yamlvalue{Classy\_input}.

Sets the cosmological input parameters (\py{"T\_cmb"}, \py{"omega\_b"}, \py{"omega\_cdm"}, \py{"tau\_reio"}, and either \py{"H0"} or \py{"100*theta\_s"}).  This function also combines all \class input dictionaries from the capabilities mentioned above that were activated in a specific run.

To combine all input parameters, we implement a function customised to merge two \class input dictionaries, \cpp{merge\_input\_dicts}\,. If the two dictionaries both contain the same key, the corresponding values can either be concatenated or the lower/larger value chosen. Which rule applies depends on the specific parameter. For example, we concatenate all values given for the \py{"output"} option to ensure that \class calculates all required spectra. For precision parameters that lead to more precise calculations when they take a larger or smaller value, we pass the higher-precision value to \class.  If a key occurs in both dictionaries but no rule to merge the values for this key exists, \cosmobit will exit with an error identifying the problematic entry.  It is left to the user to add a rule for this specific case to the function \cpp{merge\_input\_dicts}\,.

\item {\bf Run \class}.

\cpp{BE\_INI\_FUNCTION}: \cpp{classy\_2_*_*\_init}, \cpp{type: void}.

These functions access the \cpp{classy\_input\_params} dictionary via a dependency, pass it to \class (\py{cosmo.set(input_dict)}), and execute the \class run (\py{cosmo.compute()}). If any errors occur, they catch and propagate them to \gambit.

The \py{compute} step is skipped if all input parameters of the current parameter point are identical to the ones from the previously calculated point.  This avoids running \class to compute results that are still available to access from the \term{cosmo} object.

\end{enumerate}

\subsection{Interface to \darkages} \label{app:darkages-interface}

To derive the efficiency of energy injection at CMB times, we provide an interface to \darkages \textsf{1.2.0}, which is part of the \exoclass branch of \class \cite{Stocker:2018avm}. \darkages is a \python code normally called internally within a run of \exoclass. To allow for more flexibility and modularity, however, we interface \darkages to \gambit as a standalone code, and use the results as external inputs to \exoclass.\footnote{Details of how we pass the results of \darkages to \exoclass can be found in Appendix \ref{app:class-patches}.} The code calculates the efficiency functions $ f_{c/{\rm eff}}(z) $ by means of the convolution of the particle spectrum of injected photon, electrons, and positrons in Eq.~\eqref{eq:energy_injection_efficiency_convolution}. \darkages \textsf{1.2.0} can derive the table of channel-dependent efficiency functions $ f_c(z) $ by using the transfer functions of \cite{Slatyer:2015kla}, or compute the effective efficiency function $ f_{\rm eff}(z) $ needed for employing the factorisation approach of Eq.~\ref{eq:energy_injection_factorisation}.

\subsubsection*{Practical usage} \label{app:darkages-prac}

The interface to \darkages is designed to be very compact, such that all important calculations are automatically performed within the initialisation step of the frontend if \exoclass and one of the annihilating or decaying dark matter models described in Appendix \ref{app:models-energyinjection} is in use.  The default behaviour is to compute the table of energy injection efficiencies $f_c(z)$ separated into the five deposition channels considered in Ref.\ \cite{Slatyer:2015kla}: H ionisation, He ionisation, excitation, heating and low-energy photons. To request that \darkages instead compute the effective efficiency function $ f_{\rm eff}(z) $, the user must set the \cpp{f_eff_mode} option accordingly in the \cpp{Rules} section of the \YAML file.  The full set of available options is:
\begin{lstyaml}
Rules:
  # Options recognised by DarkAges
  - capability: DarkAges_1_2_0_init
    options:
      f_eff_mode: true@\ \ \ \ @# Calculate @\yamlcomment{$f_{\rm eff}(z)$}@ rather than @\yamlcomment{$f_c(z)$}@.@\ \ \ \ \ \ \ \ \ \ \ \ \ \ \ \ \,@ Default: false
      print_table: false@\ \ @# Print the table to stdout. Useful for debugging. @\ \ @Default: false
      z_max: 1.e7@\ \ \ \ \ \ \ \ \ @# Continue efficiency table up to a given redshift. @\ @Default: 1.e7

\end{lstyaml}

\subsubsection*{Technical details} \label{app:darkages-tech}

\begin{table}[t]
  \centering
  \small
  \makebox[\linewidth]{
    \begin{tabular}{llp{10.5cm}}
      \toprule
      Type     & Name        & Purpose \\
      \midrule
      \multicolumn{3}{l}{\cpp{Energy\_injection\_spectrum}} \\
      \midrule
      \cpp{vec<double>} & \cpp{E\_el}
      & kinetic energies of the tabulated injection spectrum of electrons and positrons\\

      \cpp{vec<double>} & \cpp{E\_ph}
      & kinetic energies of the tabulated injection spectrum of photons\\

      \cpp{vec<double>} & \cpp{spec\_el}
      & spectrum $ \frac{{\rm d}N}{{\rm d}E} $ of injected electrons and positrons per interaction\\

      \cpp{vec<double>} & \cpp{spe\_ph}
      & spectrum $ \frac{{\rm d}N}{{\rm d}E} $ of injected photons per interaction\\
      \midrule
      \multicolumn{3}{l}{\cpp{Energy\_injection\_efficiency\_table}} \\
      \midrule
      \cpp{bool}   & \cpp{f\_eff\_mode}
      & flag indicating if table contains $ f_{\rm eff}(z) $ (\cpp{true}) or $ f_c(z) $ (\cpp{false})\\

      \cpp{vec<double>}   & \cpp{f\_eff}
      & effective efficiency $ f_{\rm eff}(z) $ of energy injection\\

      \cpp{vec<double>}   & \cpp{redshift}
      & redshift $ z $ at which the efficiency functions are evaluated\\

      \cpp{vec<double>}   & \cpp{f\_heat}
      & efficiency of IGM heating by energy injection\\

      \cpp{vec<double>}   & \cpp{f\_lya}
      & efficiency of  Ly-$ \alpha $ excitation by  energy injection\\

      \cpp{vec<double>}   & \cpp{f\_hion}
      & efficiency  of hydrogen ionisation by energy injection\\

      \cpp{vec<double>}   & \cpp{f\_heion}
      & efficiency of helium ionisation by energy injection\\

      \cpp{vec<double>}   & \cpp{f\_lowe}
      & ``inefficiency'' of energy injection due to sub-$ 10.2\,{\rm eV} $ energy losses\\

      \bottomrule
    \end{tabular}
  }
  \caption{\protect\cpp{Energy\_injection\_spectrum} and \protect\cpp{Energy\_injection\_efficiency\_table} class member variables. Here \cpp{vec} is shorthand for \cpp{std::vector}.  These types provide structures that conveniently collect inputs and outputs for the \darkages interface. If \protect\cpp{f\_eff\_mode = true}, only the \cpp{f\_eff} member of \protect\cpp{Energy\_injection\_efficiency\_table} is filled.  If \protect\cpp{f\_eff\_mode = false}, the six other members are filled instead.}
  \label{tab:types-darkages}
\end{table}

For convenient input and output, we define two custom datatypes: \cpp{Energy\_injection\_spectrum} and \cpp{Energy\_injection\_efficiency\_table}. The first contains the injected spectra of electron-positron pairs and photons per interaction, and the respective kinetic energies at which these tables are defined. The latter datatype contains the calculated table of energy injection efficiencies and the redshifts $ z $ at which the efficiencies are defined. Depending on the calculation mode, the table contains either the effective efficiency to be used in the factorisation approach, or the efficiencies in the five separate channels. Details of these datatypes can be found in Table~\ref{tab:types-darkages}.

By design, the calculation of the efficiency tables is performed in the initialisation step of the \darkages backend. The backend is only initialised when the following two requirements for a successful calculation of the efficiency tables are met:
\begin{enumerate}
  \item The model considered involves either $s$-wave annihilation of dark matter, or dark matter decay.  This must be indicated by the activation of either the \doublecrosssf{DecayingDM\_general}{DecayingDM_general} or \doublecrosssf{AnnihilatingDM\_general}{AnnihilatingDM_general} model.\footnote{The hybrid scenario involving both annihilating and decaying components is not currently supported.} This means that the actual model being scanned must be translatable to one or the other of these two models, i.e.\ either a direct descendent of it, amongst its direct ``friend'' models, or descended from a friend. Full details on how to define model relationships in \GB can be found in Ref.\ \cite{gambit}.
  \item A function must exist with capability \cpp{energy\_injection\_spectrum} and result type \cpp{Energy\_injection\_spectrum} that is compatible with the model in question.
\end{enumerate}

After initialisation, the \cpp{Energy\_injection\_efficiency\_table} is accessible by the convenience function \cpp{get\_energy\_injection\_efficiency\_table()}.

\subsubsection*{Implementing a new model of energy injection} \label{app:darkages-new-model}

Here we outline the procedure for implementing a new model of energy injection and linking it to the existing code structure. As an example, we choose the scenario of a dark matter candidate producing a pair of neutral pions via $s$-wave annihilation $ \chi\chi \to \pi^0 \pi^0 $. As neutral pions are unstable and primarily decay into a pair of photons, the spectrum of photons is not monochromatic. In fact, the particle spectrum of photons, produced by a boosted neutral pion of energy $ E_{\pi^0} > m_{\pi^0} $ is box shaped and given by
\be
\left. \frac{\mathrm{d}N}{\mathrm{d}E_\gamma} \right\vert_{\pi^0} = \frac{2}{\gamma \beta m_{\pi^0}} \left[\Theta\left(E_\gamma - E_{-}\right) - \Theta\left(E_\gamma - E_{+}\right)\right]
\ee
with $ \gamma = \frac {E_{\pi^0}}{m_{\pi^0}} $, and $ E_{\pm} = \frac{m_{\pi^0}}{2\gamma\,\left(1\mp\beta\right)} $ (c.f \cite{Coogan:2019qpu}). As the annihilation of (cold) dark matter into the pair of neutral pions occurs nearly at rest, we can identify $ E_{\pi^0} = m_\chi $.\footnote{If we were to consider a subcomponent of dark matter decaying into two neutral pions, we would identify $ E_{\pi^0} = m_\chi/2 $. The remainder of this section is analogous for this scenario.} Additionally, we need to scale the spectrum by a factor of two, as we consider the decay of two neutral pions, resulting in the production of four photons in total. Integrating such a model into the existing code structure requires:

\begin{enumerate}
  \item  {\bf Defining a link to the ``flag'' model \doublecrosssf{AnnihilatingDM\_general}{AnnihilatingDM_general}.}

  For simplicity, we parametrise the model by only two parameters: the dark matter mass $ m_\chi $ (in units of GeV), and the thermally-averaged cross-section $ \langle\sigma v\rangle $ (in units of $ {\rm cm^3\,s}^{-1} $).  We then define a new model \textbf{\textsf{AnnihilatingDM\_pi0}} that contains these parameters, and make it a direct child of \doublecrosssf{AnnihilatingDM\_general}{AnnihilatingDM_general}. In this case, the model declaration to be added to \term{Models/include/gambit/Models/models/CosmoEnergyInjection.hpp} is
\begin{lstcpp}
#define MODEL AnnihilatingDM_pi0
#define PARENT AnnihilatingDM_general
  START_MODEL
  DEFINEPARS(mass)      // Mass of dark matter candidate [GeV]
  DEFINEPARS(sigmav)    // Thermally-averaged annihilation cross-section [cm@\cpppragma{$^3$}\,@s@\cpppragma{$^{-1}$}@]
  INTERPRET_AS_PARENT_FUNCTION(AnnihilatingDM_pi0_to_AnnihilatingDM_general)
#undef PARENT
#undef MODEL
\end{lstcpp}
  The model translation function \cpp{AnnihilatingDM\_pi0\_to\_AnnihilatingDM\_general} that should be added to \term{Models/src/models/CosmoEnergyInjection.cpp} is:
\begin{lstcpp}
#define MODEL AnnihilatingDM_pi0
  void MODEL_NAMESPACE::AnnihilatingDM_pi0::
   AnnihilatingDM_pi0_to_AnnihilatingDM_general
   (const ModelParameters &myP, ModelParameters &targetP)
  {
    targetP.setValue("mass", myP.getValue("mass"));
    targetP.setValue("sigmav", myP.getValue("sigmav"));
  }
#undef MODEL
\end{lstcpp}
  Alternatively, a friend relationship can be defined in cases in which the model in question only partially maps to \doublecrosssf{AnnihilatingDM\_general}{AnnihilatingDM_general} or is already defined in another model hierarchy.

  \item {\bf Providing the injected spectrum of electrons, positrons and photons.}

  This is achieved by declaring a new function for the existing \cpp{energy\_injection\_spectrum} capability tailored for the new model in \term{CosmoBit/include/gambit/CosmoBit/CosmoBit_rollcall.hpp}. For the given model, the rollcall declaration of this function is given by
\begin{lstcpp}
#define CAPABILITY energy_injection_spectrum
START_CAPABILITY
  #define FUNCTION energy_injection_spectrum_AnnihilatingDM_pi0
  START_FUNCTION(DarkAges::Energy_injection_spectrum)
  ALLOW_MODEL(AnnihilatingDM_pi0)
  #undef FUNCTION
#undef CAPABILITY
\end{lstcpp}
  and its implementation to be added to \term{CosmoBit/src/models/CMB.cpp} reads:
\begin{lstcpp}
# include "gambit/Utils/numerical_constants.hpp"

/// Compute the energy spectrum for @\cpppragma{$\chi\chi\to\pi^0\pi^0$}@
void energy_injection_spectrum_AnnihilatingDM_pi0(DarkAges::Energy_injection
 _spectrum & spectrum)
{
  using namespace Pipes::energy_injection_spectrum_AnnihilatingDM_pi0;

  // Assume annihilation at rest
  double E_pi0 = *Param["mass"];

  // Get the @\cpppragma{$\pi^0$}@ mass (from numerical_constants.hpp)
  const double m_pi0 = meson_masses.pi0;

  // Clear the spectra from the previous parameter point
  spectrum.E_el.clear();
  spectrum.E_ph.clear();
  spectrum.spec_el.clear();
  spectrum.spec_ph.clear();

  const int resolution = 1000;

  double gamma = E_pi0/m_pi0;
  double beta = pow( 1 - pow(gamma,-2), 0.5);

  const double E_min = m_pi0 / (2 * gamma * (1 + beta));
  const double E_max = m_pi0 / (2 * gamma * (1 - beta));
  const double amplitude = 4 / (gamma * beta * m_pi0);

  // Set the photon spectrum. There are no @\cpppragma{$e^+e^-$}@ produced
  // in this model, so we leave spectrum.spec_el alone.
  spectrum.spec_ph.resize(resolution,amplitude);

  // Set the energy axis. No need to set spectrum.E_el in this example.
  double E = E_min;
  double dE = (E_max - E_min)/(resolution-1);
  spectrum.E_ph.resize(resolution,0.0);
  for (auto& it : spectrum.E_ph)
  {
    it = E;
    E += dE;
  }
}
\end{lstcpp}
\end{enumerate}

\subsection{Interface to \montepython} \label{app:mp-interface}

As \GB has its own dedicated sampling module, \scannerbit, \cosmobit treats \montepython simply as a bank of likelihood functions, which it refers to as ``\montepythonlike''.  This allows one to employ any public or private \montepython likelihood out-of-the-box in a \gambit scan. Each likelihood call within \montepython is to a member function of a \python object. At run time, \cosmobit loads the \montepython likelihoods to be used in a scan by creating instances of each of the relevant objects. This means that if the user implements a new likelihood in \montepython,  there is no need to recompile or add any new code to \gambit.

In the first release of \cosmobit, we provide an interface to \montepython \textsf{3.3.0}, and apply patches to ensure \python \textsf{3} compatibility. Additionally, we add the new likelihood \py{bao\_correlations}. This likelihood includes distance measurements using BAO data from BOSS DR12 and eBOSS DR14, taking into account the cosmology-dependent cross-correlations owing to the volume overlap of the two probes. For details, refer to Ref.~\cite{CosmoBit_numass}, where this likelihood is introduced. All patches and new files are located in \term{Backends/patches/montepythonlike/3.3.0/}.

Note that the current interface to \montepython does not support the creation of files containing spectra computed for a fiducial model. This is the case, for example, for the SKA and Euclid likelihoods. Nevertheless, these likelihoods can still be used if the file containing the fiducial spectrum is provided in advance. This can be achieved by running \montepython as a standalone code, and then copying the resulting fiducial file to the corresponding place in the \montepythonlike installation folder within \gambit.

\subsubsection*{Practical usage} \label{app:mp-usage}

The \montepythonlike interface calls all of the requested \montepython log-likelihood functions, and sums their results to obtain a single contribution to the total likelihood, $\ln\mathcal{L}_{\rm MP}$.  To include a given set of \montepythonlike likelihoods in $\ln\mathcal{L}_{\rm MP}$, the list of likelihoods (as named in \montepython) must be included as \yaml{sub_capabilities} of the function \cpp{MP\_combined\_LogLike}.

An example \YAML entry instructing \GB to include the \montepython likelihoods (`experiments') \py{Pantheon} and \py{bao\_smallz\_2014} in a scan is simply:
\begin{lstyaml}
ObsLikes:
  # Use MontePython likelihoods in scan
  - capability: MP_combined_LogLike
    purpose: LogLike
    # Select default versions of the Pantheon BAO BOSS 2014 likelihoods
    sub_capabilities:
      Pantheon: default
      bao_smallz_2014: default
\end{lstyaml}

The value \yamlvalue{default} for the chosen likelihoods means that the default \term{.data} settings file is used.  These can also be omitted and the sub-capabilities simply given as a list, e.g.\ \yamlvalue{[Pantheon, bao\_smallz\_2014]}, in which case the default data files are implicitly assumed.%
\footnote{The default settings are saved within the folder \term{Backends/installed/montepythonlike/3.1.0/} \term{montepython/likelihoods/}\metavar{likelihood\_name}\term{/}\metavar{likelihood\_name}\term{.data}.}
If the user wishes to use non-standard settings, they can specify a path (relative to the \gambit installation directory) to a custom settings file for the corresponding likelihood.

To print out the contribution of each individual likelihood component, the capability \cpp{MP\_LogLikes} should be added as an \yamlvalue{Observable} to the \YAML file. This capability saves key-value pairs of experiment name and corresponding log-likelihood to the printer for each given \montepythonlike likelihood.

By default, all \yaml{sub_capabilities} of the function \cpp{MP\_combined\_LogLike} are included in the map sent to the printer. It is also possible to request likelihoods from additional experiments via the \yaml{sub_capabilities} of \yaml{MP\_LogLikes}, without adding them to $\ln\mathcal{L}_{\rm MP}$.  This can be useful for testing how the scanned model performs when compared to measurements that are not independent of a dataset used for the parameter scan, or in tension with another dataset.

Continuing on from the above example, one would add the following additional \YAML entry to print the log-likelihood from HST \emph{without} adding it to $\ln\mathcal{L}_{\rm MP}$:
\begin{lstyaml}
ObsLikes:
  ...
  # Print a breakdown of each likelihood component.
  # Pantheon and BAO smallz included by default for the above example.
  - purpose: Observable
    capability: MP_LogLikes
    type: map_str_dbl
    # Also save the output from HST but *do not* add it to lnL_MP
    sub_capabilities:
      hst: default
\end{lstyaml}

If the user request a likelihood in their \YAML file that does not exist in the \montepythonlike installation folder at run-time, \cosmobit will exit with an error and print all currently available likelihoods to the terminal.

The user need not specify which spectra or quantities are required from \class for the likelihood calculations, as this is automatically determined by \cosmobit as long as the following \yaml{Rule} is present:
\begin{lstyaml}
Rules:
  # Add likelihood specific input arguments to CLASS input
  - capability: classy_MPLike_input
    function: set_classy_input_with_MPLike
\end{lstyaml}
This rule makes sure that potential run options for \class required by the chosen \montepython likelihoods are added to the input dictionary for \class. If no \montepython likelihoods are in use, but \montepython is installed, the function \cpp{set\_classy\_input\_with\_MPLike} will simply return an empty dictionary.

It is the user's responsibility to ensure that they have included models in their \YAML file that include all nuisance parameters required by their nominated \montepythonlike likelihoods. One can achieve this by including pre-defined models with the required parameters in the scan.  The naming convention of these pre-defined models is such that the nuisance parameters used by \montepythonlike likelihood \metavar{example\_like} are contained in the corresponding model \mbox{\textbf{\textsf{cosmo\_nuisance\_\metavar{example\_like}}}}.  In general, \gambit automatically checks both if all models being scanned over are used for computing some quantity, and if all nuisance parameters required for computing the requested likelihoods and observables are specified in the \YAML file.  If this is not the case, it exits with an error message indicating the missing parameter and which model must be included.  Likewise, we include a test within the \montepythonlike frontend that no nuisance parameters being scanned are associated with \montepythonlike likelihoods that are not used.

\subsubsection*{Technical details} \label{app:mp-technical}

When run inside \gambit, \montepython is neither allowed to control the parameter sampling nor to internally execute a call to \class.  All information is centrally stored and administrated by \cosmobit.  To achieve this isolated use of the likelihood calculations, we have defined alternative \mplike and \mpdata classes to the ones implemented in \montepython.  When executing the download and build step for {\montepythonlike} within \gambit, we copy these alternative definitions to the installation directory, and apply a patch that makes all likelihoods import and use these alternative definitions instead of the original ones.  Note that a result of this patching procedure is that the specific installation of \montepython in the \gambit backends directory can not be used as a standalone program.

The patch of \montepython for use within \gambit removes all dependencies on \montepython command-line arguments, input and output streams, as well as all calls to \class. These \montepython features are not needed, as all input settings, writing of output files and calls to other backends are managed by \gambit.

\cosmobit executes the following individual steps when \montepython likelihoods are used:
\begin{enumerate}

\item {\bf Set experiment names}.

\yaml{capability}: \yamlvalue{MP\_objects}, \yaml{function}: \yamlvalue{create\_MP\_objects}, \yaml{type: MPLike_objects_container}.

Create a container from the input requested in the \YAML file via the \yaml{sub_capabilities} of \yamlvalue{MP\_Combined\_LogLike} and \yamlvalue{MP\_LogLikes}. This function checks that the requested experiments exist within \montepython, then creates an \yamlvalue{MPLike\_objects\_container}, which contains:

\begin{itemize}

  \item A \python object of type \mpdata, containing all experimental data to be used for likelihood computations

  \item A \cpp{map} from strings to \python objects, with keys given by experiment names and values given by \montepython likelihood objects.

\end{itemize}

This step is only executed once, when the first parameter point is calculated.

\item {\bf Set nuisance parameter values}.

\yaml{capability}: \yamlvalue{parameter\_dict\_for\_MPLike}, \yaml{function}: \yamlvalue{set\_parameter\_dict\_for\_MPLike}, \yaml{type:} \yamlvalue{pybind11::dict}.

The current values of all nuisance parameters must be passed to \montepython when computing likelihoods. This is done by the function \cpp{set\_parameter\_dict\_for\_MPLike}, which has access to all nuisance parameters through the \cpp{ALLOW_MODELS} macro in its rollcall header definition. If none of the active likelihoods require the use of extra nuisance parameters, the function \yamlvalue{pass\_empty\_parameter\_dict\_for\_MPLike} will \emph{automatically} be chosen to satisfy the capability,\footnote{That is, assuming that the dependency resolver option \yaml{prefer_model_specific_functions} has not been changed from its default (\yaml{true}; \cite{gambit}).} simply passing an empty dictionary. If the user wishes to implement a new nuisance parameter model for a \montepython likelihood, they must add that model to the list of allowed models for the function \yamlvalue{set\_parameter\_dict\_for\_MPLike}.

\item {\bf Get \class input settings}.

\yaml{capability}: \yamlvalue{classy\_MPLike\_input}, \yaml{function}: \yamlvalue{set\_classy\_input\_with\_MPLike}, \yaml{type:} \yamlvalue{pybind11::dict}.

Get all arguments for a \class run required by the \montepython likelihoods and add them to the input dictionary that will be passed to \class (see Appendix \ref{app:class-interface} for details of how this input dictionary is filled). These likelihood-dependent settings contain parameters such as specific observables or increased precision settings.  This ensures that \class computes the required output observables, such as the matter power spectrum at a certain redshift.  To obtain the inputs required for \class, we initialise the \mpdata object with the list of all experiment names in use obtained from step 1, and make a call to its method \cpp{cosmo\_arguments}, which returns a \Python dictionary containing all the needed run options.

This step is only executed once, when the first parameter point is calculated.

\item {\bf Calculate \montepython likelihoods}.

\yaml{capability}: \yamlvalue{MP\_LogLikes}, \yaml{function}: \yamlvalue{compute\_MP\_LogLikes}, \yaml{type:} \yamlvalue{map\_str\_dbl}.

After a call to \class, \cosmobit iterates over the requested \montepython likelihoods, querying each of them in turn for their log-likelihood values.  It saves the resulting values to a map, with keys given by the experiment names.

\item {\bf Calculate combined \montepython likelihoods}.

\yaml{capability}: \yamlvalue{MP\_Combined\_LogLike}, \yaml{function}: \yamlvalue{compute\_MP\_combined\_LogLike}, \yaml{type:} \yamlvalue{double}.

Finally, all individual log-likelihood values contained in the map obtained in step 5 are summed to give the combined likelihood contribution from all requested \montepython likelihoods.
\end{enumerate}
%

\subsection{Interface to \multimodecode} \label{app:multimode-interface}

As traditionally described by $\lcdm$, the shape of the primordial spectrum of scalar perturbations is purely phenomenological.  In theories where these density perturbations are sourced by inflation, the fundamental properties that we probe observationally are those of the inflaton potential. In \cosmobit we allow the user to either scan the parameters of a phenomenological description of the primordial scalar and tensor power spectra, or to scan those of a specific inflationary theory. In both cases, one must also specify a cosmological model: either \doublecrosssf{LCDM}{LCDM} or \doublecrosssf{LCDM\_theta}{LCDM_theta}.

To compute primordial power spectra from inflationary theories, \cosmobit provides an interface to the \textsf{Fortran95/2000} inflationary solver \multimodecode~\cite{Price:2014xpa}.  \multimodecode can solve the equations of motion of the background and first-order perturbations for single and multifield inflation models with canonical kinetic terms and minimal coupling to gravity. It is an extension of the single-field inflationary solver \textsf{ModeCode} \cite{Mortonson:2010er}, which is also implemented in \class.

\subsubsection*{Practical usage} \label{app:multimode-prac}

When scanning an inflationary model, the user has the option of returning either the full scalar and tensor power spectra to \cosmobit (to pass to \class), or parameters describing the shape of the power spectra. These are governed by the types \cpp{Primordial_ps} and the \cpp{ModelParameters} of the \doublecrosssf{PowerLaw\_ps}{PowerLaw_ps}~model, respectively. The names of the member variables for \cpp{Primordial_ps} are provided in Table~\ref{tab:powerspectra}, while the \doublecrosssf{PowerLaw\_ps}{PowerLaw_ps}~model is described in Sec.~\ref{app:models-inflation}. Passing the full power spectra can provide physics not captured by the parametrised spectrum, such as running of the scalar spectral index.  This allows the user to explore the full phenomenology of a given inflationary model, and is especially important for models beyond slow-roll. When given the parametrised form of the power spectra, \class will recompute a more approximate version of the full spectra internally anyway, so passing the full spectra is also recommended in order to avoid computing the power spectra twice.

The \cpp{Primordial_ps} class (Table~\ref{tab:powerspectra}) holds vectors of $k$ values, with corresponding values of $P(k)$ for scalar and tensor perturbations.\footnote{For the single-field inflationary models included in this release of \cosmobit, isocurvature modes do not exist. For multifield inflationary models, they can be important, and the interface should be modified in order to pass them to both \cpp{Primordial_ps} and \class.} To elect to use the full power spectra, one would include the following \YAML snippet:
\begin{lstyaml}
Rules:
  # Pass arrays (k, P(k)) to CLASS
  - capability: classy_primordial_input
    function: set_classy_parameters_primordial_ps
\end{lstyaml}
To use the parametrised version of the power spectrum, one would instead include:
\begin{lstyaml}
Rules:
  # Pass A_s, n_s and r to CLASS
  - capability: classy_primordial_input
    function: set_classy_parameters_parametrised_ps
\end{lstyaml}
Whether using the full spectra or their parametrised versions as input to \class, the parametrised power spectra for each point in a scan can be output by including:
\begin{lstyaml}
ObsLikes:
  # Send A_s, n_s, r and N_pivot to the printer
  - purpose:      Observable
    capability:   PowerLaw_ps_parameters
\end{lstyaml}

\subsubsection*{Adding an inflationary model to \gambit} \label{app:multimode-new}

\begin{table}[t!]
 \centering
 \small
  \begin{tabular}{cll}
  \toprule
  \cosmobit type & Member variable and description & \Cpp Type \\
  \midrule
  \cpp{Primordial_ps}   & \cpp{k}, wavenumbers $k$ & \cpp{std::vector<double>} \\
                        & \cpp{P\_s}, scalar power spectrum $P_s(k)$ & \cpp{std::vector<double>} \\
                        & \cpp{P\_t}, tensor power spectrum $P_t(k)$ & \cpp{std::vector<double>} \\
                        & \cpp{N\_pivot}, number of $e$-folds $\npiv$ & \cpp{double} \\
  \bottomrule
  \end{tabular}
  \caption{Table describing the member variables of \cpp{Primordial_ps}. Each variable has a corresponding getter and setter function: \cpp{get_}\metavar{variable} and \cpp{set_}\metavar{variable}. Note that the corresponding structure for the parametrised power spectra is provided by the \doublecrosssf{PowerLaw\_ps}{PowerLaw_ps} model; see Sec.~\ref{app:models-inflation} for details.}
  \label{tab:powerspectra}
\end{table}

When adding a new inflationary potential to \gambit, it must also be defined within \multimodecode. Existing potentials can be found in the \multimodecode file \term{modpk\_potential.f90}, each with a unique (integer) \fortran{case} for the parameter \fortran{potential\_choice}.  If a user wishes to add a \textit{new} potential to \multimodecode, this file must be modified; the procedure for adding a new model is outlined in Appendix B of the \multimodecode manual~\cite{Price:2014xpa}.

\begin{enumerate}

\item When implementing an inflationary model defined in \multimodecode into \GB, this model is included within \GB like any other physics model. The model definition for a \GB model called \textbf{\textsf{new\_inflationary\_model}} with a single parameter $\lambda$ reads
\begin{lstcpp}
#define MODEL new_inflationary_model
  START_MODEL
  DEFINEPARS(lambda)
  INTERPRET_AS_X_FUNCTION(PowerLaw_ps, as_PowerLaw)
  INTERPRET_AS_X_DEPENDENCY(PowerLaw_ps, PowerLaw_ps_parameters, ModelParameters)
#undef MODEL
\end{lstcpp}
in the model file \term{Models/include/gambit/Models/CosmoModels.hpp}. The \cpp{INTERPRET_AS_X_}\metavar{...} entries are needed to allow \GB to map from the inflationary model to the \doublecrosssf{PowerLaw\_ps}{PowerLaw_ps} model. Correspondingly, the following entry must be placed in the definition of \cpp{INFLATION_MODEL_TO_POWER_LAW} in \term{Models/src/models/CosmoModels.cpp},
\begin{lstcpp}
INFLATION_MODEL_TO_POWER_LAW(new_inflationary_model)
\end{lstcpp}

\item Once a new inflationary model is included in \GB, the user must add an entry to the function \cpp{set_multimode_inputs} in \term{CosmoBit/src/Inflation.cpp}. This entry tells \GB which value of \fortran{potential\_choice} to instruct \multimodecode to adopt.

When passing model parameters from \GB to \multimodecode, the parameters passed must correspond directly to \multimodecode's internal parametrisation: \fortran{vparams}, an array containing parameters defining the inflationary potential, \fortran{vparam_rows}, the number of parameters defining the potential \emph{per} inflaton, and \fortran{num_inflaton}, the number of inflatons.

Consider the potential for \textbf{\textsf{new\_inflationary\_model}} implemented as \fortran{case} \fortran{20} in \term{modpk\_potential.f90}. The entry in \cpp{set_multimode_inputs} would read
\begin{lstcpp}
/// Previous if (ModelInUse(...)) statements above
else if (ModelInUse(new_inflationary_model))
{
  // Note: MultiModeCode uses log@\cpppragma{$_{\tt10}$}@ of this parameter
  result.vparams.push_back(log10(*Param[lambda]));
  // This corresponds to the Fortran case 20 in modpk_potential.f90
  result.potential_choice = 20;
  // This corresponds to the number of parameters used defining the potential.
  // In this case it is just one: @\cpppragma{$\lambda$}@.
  result.vparam_rows = 1;
}
\end{lstcpp}

\item Finally, the newly-defined model must be permitted to be used by the functions specific to \multimodecode. This can be achieved with the \cpp{ALLOW_MODEL} macro, by adding
\begin{lstcpp}
ALLOW_MODEL(new_inflationary_model)
\end{lstcpp}
to the module function declaration for \cpp{set_multimode_inputs} in \term{CosmoBit\_rollcall.hpp}.

\end{enumerate}
Note that, by default, models in \cosmobit assume instant reheating. Changing this assumption would require adjustment of the \multimodecode frontend interface, and is planned for future versions of \cosmobit. In such a case, $\npiv$ would no longer be a derived quantity, but an input.

For multifield inflation models (not yet supported by \cosmobit), the initial field conditions may become important. For these scenarios, the user would also need to pass the vectors \cpp{phi_init0} and \cpp{dphi_init0} to \multimodecode, which define the initial conditions for the field values and time derivatives of the inflationary fields. For single-field inflation, the initial conditions can be estimated from the slow-roll approximation, and are already added automatically by \multimodecode to the \cpp{initialphi(phi0)} function in \term{modpk\_potential.f90}.

\subsubsection*{Technical details} \label{app:multimode-tec}

\GB has its own driver function that is patched into \multimodecode when the latter is built. The driver function  calls \multimodecode routines and interfaces with \GB via the object \cpp{gambit_inflation_observables}. This object contains observables associated with the computation of power spectra, such as $A_s$, $n_s$ and $r$ if using a parametrised power spectra, or $k$, $P_s(k)$ and $P_t(k)$ if using the full power spectra.

\begin{table}[t!]
  \centering
  \small
  \makebox[0.8\linewidth]{
    \begin{tabular}{p{4cm} p{1.6cm} p{6.2cm} p{1.5cm}}
    \toprule
    Option & \Cpp Type  & Utility & Default\\
    \midrule

    \cpp{silence_output}  & \cpp{int} & if \cpp{1}, all uncaught error messages from \multimodecode are silenced; if \cpp{0} they are all printed & \cpp{0} \\

    \cpp{dlnk} & \cpp{double} & difference in $k$-space used when calculating observables at~$k_\star$ via finite difference~(units of $\text{Mpc}^{-1})$ & \cpp{0.4} \\

    \cpp{numsteps}$^{*}$ &\cpp{int} & number of $k$-bins when calculating $P(k)$ & \cpp{100} \\

    \cpp{k\_min}$^{*}$ & \cpp{double} & large-scale limit of the wavenumber  when calculating $P(k)$ & {\footnotesize\ttfamily 1e-6} \\

    \cpp{k\_max}$^{*}$   & \cpp{double} & small-scale limit of the wavenumber  when calculating $P(k)$ & {\footnotesize\ttfamily 1e6} \\

    \bottomrule
    \end{tabular}
  }
  \caption{Options for the \multimodecode backend that can be set directly in the \textsf{YAML}~file, as \cpp{runOptions} of the function \cpp{set_multimode_inputs}. Parameters with an asterisk~($^*$) are only used when calculating the full power spectrum (\ie when \cpp{set_classy_parameters_primordial_ps} is specified in the \YAML file). Note that the pivot scale~$k_\star$ can be set using the \cpp{runOptions} for the function \cpp{set_k_pivot} (satisfying capability~\cpp{k_pivot}).}
  \label{tab:if_ps_is_called}
\end{table}

Communication between \cosmobit and \multimodecode proceeds as follows:
\begin{enumerate}

\item {\bf{Initialise inflationary settings}}.

\yaml{capability}: \yamlvalue{multimode\_input\_parameters}, \yaml{function}: \yamlvalue{set\_multimode\_inputs}, \yaml{type}: \yamlvalue{Multimode\_inputs}.

All important settings to be passed to \multimodecode when computing power spectra can be specified in the \YAML file. For a list of all options see Table~\ref{tab:if_ps_is_called}.

This function also sets the \multimodecode internal parameters such \cpp{vparams} and \cpp{potential_choice} from the \GB model parameters.

\item {\bf{Compute power spectra with \multimodecode}}.

There are two options when computing power spectra. The user can either choose to return the full power spectrum, or the parametrised version (see Table~\ref{tab:powerspectra}). The functions returning these types respectively are:

\begin{enumerate}
  \item \yaml{capability}: \yamlvalue{primordial\_power\_spectrum}, \yaml{function}: \yamlvalue{get\_multimode\_primordial\_ps}, \yaml{type}: \yamlvalue{Primordial\_ps}.

  Creates a \cpp{gambit_inflation_observables} object containing the options specified in \cpp{set\_multimode\_inputs}. Calls \multimodecode via the backend convenience function \cpp{multimodecode_primordial_ps} defined in the \GB driver function. This convenience function solves the inflationary equations of motion and fills the \cpp{gambit_inflation_observables} with arrays defining the power spectrum. \cosmobit then probes the \cpp{gambit_inflation_observables} object, and transfers $k$, $P_s(k)$, $P_t(k)$ and $\npiv$ to the \cpp{Primordial_ps} type, ready for use in other calculations in \GB.

  \item \yaml{capability}: \yamlvalue{PowerLaw\_ps\_parameters}, \yaml{function}: \yamlvalue{get\_multimode\_parametrised\_ps}, \yaml{type}: \yamlvalue{ModelParameters} (for model \doublecrosssf{PowerLaw\_ps}{PowerLaw_ps}).

  As above, but calls the convenience function \cpp{multimodecode_parametrised_ps} instead. This fills the \cpp{gambit_inflation_observables} object with $n_s$, $A_s$, $r$, and $\npiv$. These values are returned in the form of a \cpp{ModelParameters} object for \doublecrosssf{PowerLaw\_ps}{PowerLaw_ps}.
\end{enumerate}

From here, one of the power spectrum types has been filled, and will be passed to \classy. For details on how the power spectrum types are handled and passed to \classy, see Appendix~\ref{app:class-tech}.

\end{enumerate}

Finally, note that the remaining \multimodecode parameter choices (excluding those described in Table~\ref{tab:if_ps_is_called}) can be altered by modifying the constructor of \cpp{Multimode_inputs} in \term{CosmoBit_types.cpp}.  These additional parameters become relevant if, for example, the user wishes to interface to a multifield potential.
These options are shown in Table~\ref{tab:options_set_in_cpp}.

\begin{table}[t!]
  \centering
  \small
  \makebox[0.8\linewidth]{
    \begin{tabular}{p{4cm} p{1.6cm} p{6.2cm} p{2.3cm}}
      \toprule
      Option & \Cpp Type  & Utility & Default Value \\
      \midrule

      \cpp{slowroll_inflation_end}  & \cpp{bool}  &  forces inflation to end when the slow-roll parameter is 1 & \cpp{True} \\

      \cpp{use_deltaN_SR} &\cpp{bool} &  determines whether $\delta N$ observables~\cite{Wands:2000dp} are calculated (assuming slow-roll and sum-separable potentials) at the pivot scale & \cpp{False} \\

      \cpp{use_horiz_cross_approx} &\cpp{bool} & determines whether the horizon-crossing-approximation is ignored & \cpp{False} \\

      \cpp{evaluate_modes} & \cpp{bool} & determines whether first-order perturbation equations are solved  &\cpp{True} \\

      \cpp{get_runningofrunning}   & \cpp{bool} & determines whether the derivative of the spectral index with respect to $\ln(k)$ is calculated& \cpp{False} \\

      \bottomrule
    \end{tabular}
  }
  \caption{\multimodecode parameter choices (excluding those described in Table~\ref{tab:if_ps_is_called}) that can be set in the constructor for \cpp{Multimode_inputs} in \protect\term{Backends/src/backend_types/MultiModeCode.cpp}. These parameters exclude those used to determine the precision of the inflationary ODE solutions; these can be modified in the driver function patched into \multimodecode.}
  \label{tab:options_set_in_cpp}
\end{table}

\subsection{Interface to \plc} \label{app:plc-interface}

The \plc backend provides the Planck 2015 and 2018 likelihoods. The relevant \YAML settings for using \plc are:
\begin{lstyaml}
ObsLikes:

  # `Prior' likelihoods

  - purpose:      LogLike
    capability:   Planck_nuisance_prior_loglike

  - purpose:      LogLike
    capability:   Planck_sz_prior_loglike

  # Choose which Planck likelihoods to use, and where to get them from.

  - purpose:      LogLike
    capability:   Planck_lowl_loglike
    function:     function_Planck_lowl_TTEE_2018_loglike

  - purpose:      LogLike
    capability:   Planck_highl_loglike
    function:     function_Planck_highl_TTTEEE_2018_loglike

  - purpose:      LogLike
    capability:   Planck_lensing_loglike
    function:     function_Planck_lensing_2018_loglike
\end{lstyaml}
The different Planck likelihoods, for high- and low-$\ell$ as well as lensing, can be provided by different module functions, depending on which exact combination of data the user wants to include. The different options available within \cosmobit are detailed in Table~\ref{tab:plc_caps}.

For several of the nuisance parameters, associated with the Planck likelihoods, it is advisable to apply specific priors on them \cite{Aghanim:2015xee,Aghanim:2018eyx}. In order to use the plc likelihoods with these priors also in the context of frequentist analyses, they are implemented in \cosmobit as likelihoods such that they need to be included in the \yaml{ObsLikes} section of the \YAML file. The available prior likelihoods are \yaml{Planck_nuisance_prior_loglike} and \yaml{Planck_sz_prior_loglike}. For details, we refer to the original paper and to their implementation in \term{CosmoBit/src/Planck.cpp}.

\begin{table}[t]
  \centering
  \begin{tabular}{ll}
    \toprule
    Capability & Function \\
    \midrule
    \cpp{Planck\_lowl\_loglike}
    & \cpp{function\_Planck\_lowl\_TTEE\_2018\_loglike} \\
    & \cpp{function\_Planck\_lowl\_TT\_2018\_loglike} \\
    & \cpp{function\_Planck\_lowl\_EE\_2018\_loglike} \\
    \arrayrulecolor{gray}\cdashline{2-2}\arrayrulecolor{black}
    & \cpp{function\_Planck\_lowl\_TT\_2015\_loglike} \\
    & \cpp{function\_Planck\_lowl\_TEB\_2015\_loglike} \\
    \midrule
    \cpp{Planck\_highl\_loglike}
    & \cpp{function\_Planck\_highl\_TTTEEE\_2018\_loglike}\\
    & \cpp{function\_Planck\_highl\_TTTEEE\_lite\_2018\_loglike}\\
    & \cpp{function\_Planck\_highl\_TT\_2018\_loglike}\\
    & \cpp{function\_Planck\_highl\_TT\_lite\_2018\_loglike}\\
    \arrayrulecolor{gray}\cdashline{2-2}\arrayrulecolor{black}
    & \cpp{function\_Planck\_highl\_TTTEEE\_2015\_loglike}\\
    & \cpp{function\_Planck\_highl\_TTTEEE\_lite\_2015\_loglike}\\
    & \cpp{function\_Planck\_highl\_TT\_2015\_loglike}\\
    & \cpp{function\_Planck\_highl\_TT\_lite\_2015\_loglike}\\
    \midrule
    \cpp{Planck\_lensing\_loglike}
    & \cpp{function\_Planck\_lensing\_2018\_loglike} \\
    & \cpp{function\_Planck\_lensing\_marged\_2018\_loglike} \\
    \arrayrulecolor{gray}\cdashline{2-2}\arrayrulecolor{black}
    & \cpp{function\_Planck\_lensing\_2015\_loglike} \\
    \bottomrule
  \end{tabular}
  \caption{The capabilities and functions provided by the interface to \plc. The 2018 likelihoods (above the dashed lines) require version \textsf{3.0} of the \protect\term{plc\_data}, whereas the 2015 likelihoods (below) require version \textsf{2.0}.}
  \label{tab:plc_caps}
\end{table}

Running the command \term{make plc} will download and configure \plc \textsf{3.0}, but will not retrieve the likelihood data themselves.  The latest (i.e.\ 2018) likelihood data can be obtained by running \term{make plc_data}, which is equivalent to running \term{make plc_data_3.0}. The 2015 data can be retrieved by running \term{make plc_data_2.0}.  Alternatively, if the user has already downloaded the likelihood data independently of \GB, the file path can just be set with:
\begin{lstyaml}
Rules:
  # Set path to downloaded Planck data
  - capability: plc_3_0_init
    options:
      plc_data_2_path: /path/to/downloaded/Planck2015/data # Data release of 2015 (plc_2.0)
      plc_data_3_path: /path/to/downloaded/Planck2018/data # Data release of 2018 (plc_3.0)
\end{lstyaml}

\section{List of capabilities}
\label{app:capabilities}

We provide a complete list of the capabilities included in the first release of \cosmobit, including the module functions available to use them, dependencies and backend requirements. These include general cosmological quantities (Table~\ref{tab:capabilities_gencosmo}), quantities related to energy injection (Table~\ref{tab:capabilities_ei}), CMB power spectra (Table~\ref{tab:capabilities_cmb}), Planck likelihoods (Tables~\ref{tab:capabilities_planck_I} and \ref{tab:capabilities_planck_II}), inflation (Table~\ref{tab:capabilities_inf}), BBN (Table~\ref{tab:capabilities_bbn}), and the interfaces to \montepython (Table~\ref{tab:capabilities_mp}) and \classy (Table~\ref{tab:capabilities_classy}).



\begin{table*}[tp]
\centering
\scriptsize
   \makebox[\linewidth]{
   \begin{tabular}{p{7.5cm} p{5.0cm} p{4.7cm}}

    \toprule 
   \textbf{Capability} and brief description & \textbf{Function} (\textbf{Return Type}) & \textbf{Dependencies [type] /\newline BE requirement \{type\} }\\ \midrule 

    \cpp{mNu\_tot}\newline   Returns the total mass of neutrinos in eV.
      &\cpp{get\_mNu\_tot} (\cpp{double}) & {\cpp { } } \\
   \midrule 
    
    \cpp{Neff\_SM}\newline   Returns the effective number of neutrino species in the SM. &\cpp{get\_Neff\_SM} (\cpp{double}) & {\cpp { } } \\
   \midrule 
   
    \cpp{Neff}\newline   Returns the effective number of neutrino species. &\cpp{get\_Neff\_classy} (\cpp{double}) & {\cpp {class\_get\_Neff \{double\}} } \\
   \midrule 
    
    \cpp{T\_ncdm}\newline  Returns the neutrino temperature in units of $\Tcmb$.
      &\cpp{T\_ncdm} (\cpp{double}) & {\cpp { } } \\
   \cline{2-3}
	  &\cpp{T\_ncdm\_SM} (\cpp{double}) & {\cpp { } } \\
   \midrule 
   
   \cpp{N\_ur}\newline   Returns the total number of ultra-relativistic species today. &\cpp{get\_N\_ur} (\cpp{double}) & {\cpp {Neff\_SM [double] } } \\
   \midrule 
    
    \cpp{k\_pivot}\newline   Returns the comoving pivot scale, $k_\star$ in $\text{Mpc}^{-1}$. &\cpp{set\_k\_pivot} (\cpp{double}) & {\cpp { } } \\
   \midrule 
    
    \cpp{H0}\newline   Returns  the Hubble rate today, $H_0$, in km/s/Mpc.
      &\cpp{get\_H0\_classy} (\cpp{double}) & {\cpp {class\_get\_H0 \{double\}} } \\
   \midrule 
    
    \cpp{n0\_g}\newline  Returns the number density of photons today, $n_0^\gamma$, in $\text{cm}^{-3}$.
      &\cpp{compute\_n0\_g} (\cpp{double}) & {\cpp { } } \\
   \midrule 

    \cpp{rs\_drag}\newline  Returns the comoving sound horizon at baryon drag.
      &\cpp{get\_rs\_drag\_classy} (\cpp{double}) & {\cpp {class\_get\_rs \{double\}} } \\
   \midrule 
        
    \cpp{S8\_cosmo}\newline  Returns $S_8 = \sigma_8 (\Omega_m/0.3)^{0.5}$, a measure for the amplitude of the matter power spectrum on scale $8/h$ Mpc. &\cpp{get\_S8\_classy} (\cpp{double}) & {\cpp {Omega0\_m [double]} } 
    \newline {\cpp {class\_get\_sigma8 \{double\}} } \\
   \midrule

    \cpp{Omega0\_m}\newline  Returns the total matter content today, $\Omega_{\rm m}$. &\cpp{get\_Omega0\_m\_classy} (\cpp{double}) & {\cpp {class\_get\_Omega0\_m \{double\}} } \\
   \midrule 
    
    \cpp{Omega0\_b}\newline  Returns the total baryon content today, $\Omega_{\rm b}$.  &\cpp{compute\_Omega0\_b} (\cpp{double}) & {\cpp {H0 [double]} } \\
   \midrule 
    
    \cpp{Omega0\_cdm}\newline   Returns the total cold DM content today, $\Omega_{\rm cdm}$. &\cpp{compute\_Omega0\_cdm} (\cpp{double}) & {\cpp {H0 [double]} } \\
   \midrule 
    
    \cpp{Omega0\_r}\newline    Returns the total radiation content today, $\Omega_{\rm r}$.
      &\cpp{get\_Omega0\_r\_classy} (\cpp{double}) & {\cpp {class\_get\_Omega0\_r \{double\}} } \\
   \midrule 
    
    \cpp{Omega0\_g}\newline   Returns the total photon content today, $\Omega_{\gamma}$. &\cpp{compute\_Omega0\_g} (\cpp{double}) & {\cpp {H0 [double]} } \\
   \midrule 
    
    \cpp{Omega0\_ur}\newline   Returns the total content in ultrarelativistic species today. &\cpp{compute\_Omega0\_ur} (\cpp{double}) & {\cpp {Omega0\_g [double]} } \newline
    {\cpp {N\_ur [double]} } \\
    \cline{2-3}&\cpp{get\_Omega0\_ur\_classy} (\cpp{double}) & {\cpp {class\_get\_Omega0\_ur \{double\}} } \\
   \midrule 
    
    \cpp{Omega0\_ncdm}\newline   Returns the total neutrino content today, $\Omega_{\rm ncdm}$. &\cpp{get\_Omega0\_ncdm\_classy} (\cpp{double}) & {\cpp {class\_get\_Omega0\_ncdm\_tot}
      \newline \hphantom{,} \cpp{ \{double\}} } \\
   \midrule 
    
    \cpp{eta0}\newline  Returns the baryon-to-photon ratio today, $\eta_0$.
      &\cpp{eta0\_LCDM} (\cpp{double}) & {\cpp { } } \\
    
   \bottomrule 
    
   \end{tabular}
  }
  \caption{Capabilities in \cosmobit used for general cosmological quantities.}
  \label{tab:capabilities_gencosmo}
\end{table*}

\begin{table*}[tp]
\centering
\scriptsize
   \makebox[\linewidth]{
   \begin{tabular}{p{4.5cm} p{6.4cm} p{4.5cm}}

    \toprule 
   \textbf{Capability} and brief description & \textbf{Function} (\textbf{Return Type}) & \textbf{Dependencies [type] /\newline BE requirement \{type\} }\\ \midrule

    \cpp{energy\_injection\_efficiency}\newline  Gets the energy injection efficiency table from the \darkages frontend.
      &\cpp{energy\_injection\_efficiency\_func} (\cpp{DarkAges::Energy\_injection\_}
      \newline \hphantom{,} \cpp{efficiency\_table}) 
      & {\cpp {get\_energy\_injection\_}
      \newline \hphantom{,} \cpp{efficiency\_table} 
      \newline \cpp{\{DarkAges::Energy\_injection\_}}
      \newline \hphantom{,} \cpp{efficiency\_table\}} \\
   \midrule 
    
    \cpp{energy_injection_spectrum}\newline  Returns the spectrum of electrons/positrons and photons injected by
    the decay or annihilation of DM.
      &\cpp{energy_injection_spectrum_}
      \newline \hphantom{,} \cpp{AnnihilatingDM_mixture} 
      \newline(\cpp{DarkAges::Energy_injection_spectrum}) & \\
    \cline{2-3}
    &\cpp{energy_injection_spectrum_}
      \newline \hphantom{,} \cpp{DecayingDM_mixture} 
      \newline(\cpp{DarkAges::Energy_injection_spectrum}) & \\
   \bottomrule 

   \end{tabular}
  }
  \caption{Capabilities in \cosmobit related to energy injection.}
  \label{tab:capabilities_ei}
\end{table*}

\begin{table*}[tp]
\centering
\scriptsize
   \makebox[\linewidth]{
   \begin{tabular}{p{6.5cm} p{4.8cm} p{4.1cm}} 
    \toprule 
   \textbf{Capability} and brief description & \textbf{Function} (\textbf{Return Type}) & \textbf{Dependencies [type] / \newline BE requirement \{type\} }\\ \midrule 
    \cpp{unlensed\_Cl\_TT}\newline  Returns the unlensed dimensionless CMB temperature-temperature anisotropy spectrum.
      &\cpp{class\_get\_unlensed\_Cl\_TT} (\cpp{std::vector<double>}) & {\cpp {class\_get\_unlensed\_cl}
      \newline \hphantom{,} \cpp{ \{std::vector<double>\}} } \\
   \midrule 
    
    \cpp{unlensed\_Cl\_TE}\newline  Returns the unlensed dimensionless CMB TE cross-correlations.
      &\cpp{class\_get\_unlensed\_Cl\_TE} (\cpp{std::vector<double>}) & {\cpp {class\_get\_unlensed\_cl}
      \newline \hphantom{,} \cpp{ \{std::vector<double>\}} } \\
   \midrule 
    
    \cpp{unlensed\_Cl\_EE}\newline  Returns the unlensed dimensionless CMB E-mode spectrum.
      &\cpp{class\_get\_unlensed\_Cl\_EE} (\cpp{std::vector<double>}) & {\cpp {class\_get\_unlensed\_cl}
      \newline \hphantom{,} \cpp{ \{std::vector<double>\}} } \\
   \midrule 
    
    \cpp{unlensed\_Cl\_BB}\newline  Returns the unlensed dimensionless CMB B-mode spectrum.
      &\cpp{class\_get\_unlensed\_Cl\_BB} (\cpp{std::vector<double>}) & {\cpp {class\_get\_unlensed\_cl}
      \newline \hphantom{,} \cpp{ \{std::vector<double>\}} } \\
   \midrule 
    
    \cpp{unlensed\_Cl\_PhiPhi}\newline  Returns the unlensed dimensionless CMB lensing potential.
      &\cpp{class\_get\_unlensed\_Cl\_PhiPhi} (\cpp{std::vector<double>}) & {\cpp {class\_get\_unlensed\_cl}
      \newline \hphantom{,} \cpp{ \{std::vector<double>\}} } \\
   \midrule 
    
    \cpp{lensed\_Cl\_TT}\newline  Returns the lensed dimensionless CMB temperature-temperature anisotropy spectrum.
      &\cpp{class\_get\_lensed\_Cl\_TT} (\cpp{std::vector<double>}) & {\cpp {class\_get\_lensed\_cl}
      \newline \hphantom{,} \cpp{ \{std::vector<double>\}} } \\
   \midrule 
    
    \cpp{lensed\_Cl\_TE}\newline  Returns the lensed dimensionless CMB TE cross-correlations.
      &\cpp{class\_get\_lensed\_Cl\_TE} (\cpp{std::vector<double>}) & {\cpp {class\_get\_lensed\_cl}
      \newline \hphantom{,} \cpp{ \{std::vector<double>\}} } \\
   \midrule 
    
    \cpp{lensed\_Cl\_EE}\newline  Returns the lensed dimensionless CMB E-mode spectrum.
      &\cpp{class\_get\_lensed\_Cl\_EE} (\cpp{std::vector<double>}) & {\cpp {class\_get\_lensed\_cl}
      \newline \hphantom{,} \cpp{ \{std::vector<double>\}} } \\
   \midrule 
    
    \cpp{lensed\_Cl\_BB}\newline  Returns the lensed dimensionless CMB B-mode spectrum.
      &\cpp{class\_get\_lensed\_Cl\_BB} (\cpp{std::vector<double>}) & {\cpp {class\_get\_lensed\_cl}
      \newline \hphantom{,} \cpp{ \{std::vector<double>\}} } \\
   \midrule 
    
    \cpp{lensed\_Cl\_PhiPhi}\newline  Returns the lensed dimensionless CMB lensing potential.
      &\cpp{class\_get\_lensed\_Cl\_PhiPhi} (\cpp{std::vector<double>}) & {\cpp {class\_get\_lensed\_cl}
      \newline \hphantom{,} \cpp{ \{std::vector<double>\}} } \\
   \bottomrule 
   \end{tabular}
  }
  \caption{Capabilities in \cosmobit related to CMB power spectra.}
  \label{tab:capabilities_cmb}
\end{table*}

 \begin{table*}[tp]
\centering
\scriptsize
   \makebox[\linewidth]{
   \begin{tabular}{p{4.0cm} p{5.5cm} p{6.5cm}}
    \toprule 
   \textbf{Capability} and brief description & \textbf{Function} (\textbf{Return Type}) & \textbf{Dependencies [type] / \newline BE requirement \{type\} }\\ \midrule 

    \cpp{Planck\_lowl\_loglike}\newline   Computes the Planck CMB likelihood for the multipole range 2 $\leq$ $\ell$ $\leq$ 29.
      All module functions described here have the prefix \cpp{function\_Planck\_lowl\_}. & 
    \cpp{..\_TT\_2015\_loglike} (\cpp{double}) & 
         {\cpp {T\_cmb [double]} }
         \newline{\cpp {lensed\_Cl\_TT [std::vector<double>]} }  
         \newline {\cpp {plc\_loglike\_lowl\_TT\_2015 \{double\}} } \\
    \cline{2-3}&\cpp{..\_TEB\_2015\_loglike} (\cpp{double}) & 
         {\cpp {T\_cmb [double]} }
         \newline {\cpp {lensed\_Cl\_TT [std::vector<double>]} }
         \newline {\cpp {lensed\_Cl\_TE [std::vector<double>]} }
         \newline {\cpp {lensed\_Cl\_EE [std::vector<double>]} }
         \newline {\cpp {lensed\_Cl\_BB [std::vector<double>]} }
         \newline {\cpp {plc\_loglike\_lowl\_TEB\_2015 \{double\}} } \\
    \cline{2-3}&\cpp{..\_TTEE\_2018\_loglike} (\cpp{double}) & 
         {\cpp {T\_cmb [double]} }
         \newline{\cpp {lensed\_Cl\_TT [std::vector<double>]} } 
         \newline {\cpp {lensed\_Cl\_EE [std::vector<double>]} }
         \newline {\cpp {plc\_loglike\_lowl\_TT\_2018 \{double\}} }
         \newline {\cpp {plc\_loglike\_lowl\_EE\_2018 \{double\}} } \\
    \cline{2-3}&\cpp{..\_TT\_2018\_loglike} (\cpp{double}) & 
         {\cpp {T\_cmb [double]} } 
         \newline{\cpp {lensed\_Cl\_TT [std::vector<double>]} }
         \newline {\cpp {plc\_loglike\_lowl\_TT\_2018 \{double\}} } \\
    \cline{2-3}&\cpp{..\_EE\_2018\_loglike} (\cpp{double}) & 
         {\cpp {T\_cmb [double]} }
         \newline {\cpp {lensed\_Cl\_EE [std::vector<double>]} }
         \newline {\cpp {plc\_loglike\_lowl\_EE\_2018 \{double\}} } \\
   \midrule 
    
    \cpp{Planck\_lensing\_loglike}\newline   Computes the Planck lensing log-likelihood for a given set of $C_\ell$s. All module functions described here have the prefix \cpp{function\_Planck\_lensing\_}. & 
    \cpp{..\_marged\_2018\_loglike} (\cpp{double}) & {\cpp {lensed\_Cl\_PhiPhi [std::vector<double>]} }
    \newline {\cpp {plc\_loglike\_lensing\_marged\_2018}
      \newline \hphantom{,} \cpp{ \{double\}} } \\
    \cline{2-3}&\cpp{..\_2018\_loglike} (\cpp{double}) & {\cpp {lensed\_Cl\_TT [std::vector<double>]} }
    \newline {\cpp {lensed\_Cl\_TE [std::vector<double>]} }
    \newline {\cpp {lensed\_Cl\_EE [std::vector<double>]} }
    \newline {\cpp {lensed\_Cl\_PhiPhi [std::vector<double>]} }
    \newline {\cpp {T\_cmb [double]} }
    \newline {\cpp {plc\_loglike\_lensing\_2018 \{double\}} }\\
    \cline{2-3}&\cpp{..\_2015\_loglike} (\cpp{double}) & {\cpp {lensed\_Cl\_TT [std::vector<double>]} }
    \newline {\cpp {lensed\_Cl\_TE [std::vector<double>]} } 
    \newline {\cpp {lensed\_Cl\_EE [std::vector<double>]} } 
    \newline {\cpp {lensed\_Cl\_PhiPhi [std::vector<double>]} } 
    \newline {\cpp {T\_cmb [double]} } 
    \newline {\cpp {plc\_loglike\_lensing\_2015 \{double\}} } \\
   \midrule 

    \cpp{Planck\_nuisance\_prior\_}
      \newline \hphantom{,} \cpp{loglike}\newline   Provides Gaussian priors on the nuisance parameters of the Planck likelihoods, as suggested by the Planck collaboration.
      &\cpp{compute\_Planck\_nuisance\_}
      \newline \hphantom{,} \cpp{prior\_loglike} (\cpp{double}) & {\cpp { } } \\
   \midrule 
    
    \cpp{Planck\_sz\_prior\_loglike}\newline   Provides priors on the tSZ and kSZ amplitudes. The correlation is unconstrained by Planck. Uses prior based on SPT and ACT data, cf. Eq. (32) of \cite{Aghanim:2015xee}.
      &\cpp{compute\_Planck\_sz\_prior} (\cpp{double}) & {\cpp { } } \\
   \bottomrule 
   \end{tabular}
  }
  \caption{Capabilities in \cosmobit providing the majority of Planck likelihoods.}
  \label{tab:capabilities_planck_I}
\end{table*}

 \begin{table*}[tp]
\centering
\scriptsize
   \makebox[\linewidth]{
   \begin{tabular}{p{4.0cm} p{5.3cm} p{6.7cm}}
    \toprule 
   \textbf{Capability} and brief description & \textbf{Function} (\textbf{Return Type}) & \textbf{Dependencies [type] / \newline BE requirement \{type\} }\\ \midrule 
  
   \cpp{Planck\_highl\_loglike}\newline   Calculates the Planck CMB likelihood for the multipole range 30 $\leq$ $\ell$ $\leq$ 2508.
      All module functions described here have the prefix \cpp{function\_Planck\_highl\_}. &
   \cpp{..\_TT\_lite\_2015\_loglike} (\cpp{double}) & 
      {\cpp {T\_cmb [double]} } 
      \newline {\cpp {lensed\_Cl\_TT [std::vector<double>]} } 
      \newline {\cpp {plc\_loglike\_highl\_TT\_lite\_2015 \{double\}} } \\
    \cline{2-3}&\cpp{..\_TT\_2015\_loglike} (\cpp{double}) & 
      {\cpp {T\_cmb [double]} } 
      \newline {\cpp {lensed\_Cl\_TT [std::vector<double>]} } 
      \newline {\cpp {plc\_loglike\_highl\_TT\_2015 \{double\}} } \\
    \cline{2-3}&\cpp{..\_TTTEEE\_lite\_2015\_loglike} (\cpp{double}) & 
      {\cpp {T\_cmb [double]} }
      \newline {\cpp {lensed\_Cl\_TT [std::vector<double>]} }
      \newline {\cpp {lensed\_Cl\_TE [std::vector<double>]} }
      \newline {\cpp {lensed\_Cl\_EE [std::vector<double>]} }
    {\cpp {plc\_loglike\_highl\_TTTEEE\_lite\_2015}
      \newline \hphantom{,} \cpp{ \{double\}} } \\
    \cline{2-3}&\cpp{..\_TTTEEE\_lite\_2018\_loglike} (\cpp{double}) & 
      {\cpp {T\_cmb [double]} }
      \newline {\cpp {lensed\_Cl\_TT [std::vector<double>]} }
      \newline {\cpp {lensed\_Cl\_TE [std::vector<double>]} }
      \newline {\cpp {lensed\_Cl\_EE [std::vector<double>]} }
      \newline {\cpp {plc\_loglike\_highl\_TTTEEE\_lite\_2018}
      \newline \hphantom{,} \cpp{ \{double\}} } \\
    \cline{2-3}&\cpp{..\_TT\_lite\_2018\_loglike} (\cpp{double}) & 
      {\cpp {T\_cmb [double]} }
      \newline{\cpp {lensed\_Cl\_TT [std::vector<double>]} }
      \newline {\cpp {plc\_loglike\_highl\_TT\_lite\_2018 \{double\}} } \\
    \cline{2-3}&\cpp{..\_TTTEEE\_2018\_loglike} (\cpp{double}) & 
      {\cpp {T\_cmb [double]} }
      \newline {\cpp {lensed\_Cl\_TT [std::vector<double>]} }
      \newline {\cpp {lensed\_Cl\_TE [std::vector<double>]} }
      \newline {\cpp {lensed\_Cl\_EE [std::vector<double>]} }
      \newline {\cpp {plc\_loglike\_highl\_TTTEEE\_2018 \{double\}} } \\
    \cline{2-3}&\cpp{..\_TT\_2018\_loglike} (\cpp{double}) & 
      {\cpp {T\_cmb [double]} }
      \newline \cpp{lensed\_Cl\_TT [std::vector<double>]}
      \newline {\cpp {plc\_loglike\_highl\_TT\_2018 \{double\}} } \\
    \cline{2-3}&\cpp{..\_TTTEEE\_2015\_loglike} (\cpp{double}) & 
      {\cpp {T\_cmb [double]} }
      \newline {\cpp {lensed\_Cl\_TT [std::vector<double>]} }
      \newline {\cpp {lensed\_Cl\_TE [std::vector<double>]} }
      \newline {\cpp {lensed\_Cl\_EE [std::vector<double>]} }
      \newline {\cpp {plc\_loglike\_highl\_TTTEEE\_2015 \{double\}} } \\
   \bottomrule 
   \end{tabular}
  }
  \caption{Module functions in \cosmobit providing the high-$\ell$ Planck likelihood.}
  \label{tab:capabilities_planck_II}
\end{table*}

\begin{table*}[tp]
\centering
\scriptsize
   \makebox[\linewidth]{
   \begin{tabular}{p{5cm} p{4.8cm} p{5.9cm}}

    \toprule 
   \textbf{Capability} and brief description & \textbf{Function} (\textbf{Return Type}) & \textbf{Dependencies [type] / \newline BE requirement \{type\} }\\ \midrule 

    \cpp{multimode\_input\_parameters}\newline   Creates object containing the generic inputs for calling \multimodecode.
      &\cpp{set\_multimode\_inputs} (\cpp{Multimode\_inputs}) & {\cpp {k\_pivot [double]} } \\
   \midrule 
    
    \cpp{primordial\_power\_spectrum}\newline   Creates object containing the primordial power spectra of scalar, tensor, and isocurvature perturbations, for a given array of $[k]$ values.
      &\cpp{get\_multimode\_primordial\_ps} (\cpp{Primordial\_ps}) & {\cpp {multimode\_input\_parameters}
      \newline \hphantom{,} \cpp{ [Multimode\_inputs]} } 
    \newline {\cpp {multimodecode\_primordial\_ps}
      \newline \hphantom{,} \cpp{ \{gambit\_inflation\_observables\}} } \\
   \midrule 
    
    \cpp{PowerLaw\_ps\_parameters}\newline   Returns the purely phenomenological parametrised scale-free power spectrum.
      See \term{./gambit} \term{PowerLaw\_ps} for details.
      &\cpp{get\_multimode\_parametrised\_ps} (\cpp{ModelParameters}) & {\cpp {multimode\_input\_parameters}
      \newline \hphantom{,} \cpp{ [Multimode\_inputs]} }
    \newline {\cpp {multimodecode\_parametrised\_ps}
      \newline \hphantom{,} \cpp{ \{gambit\_inflation\_observables\}} } \\
   \bottomrule 

   \end{tabular}
  }
  \caption{Capabilities provided by \cosmobit relating to inflation.}
  \label{tab:capabilities_inf}
\end{table*}

\begin{table*}[tp]
\centering
\scriptsize
   \makebox[\linewidth]{
   \begin{tabular}{p{5cm} p{6cm} p{5.5cm}}

    \toprule 
   \textbf{Capability} and brief description & \textbf{Function} (\textbf{Return Type}) & \textbf{Dependencies [type] /\newline BE requirement \{type\} }\\ \midrule 

    \cpp{AlterBBN\_Input}\newline    Sets input parameters for an \alterbbn run.
      &\cpp{AlterBBN\_Input} (\cpp{map\_str\_dbl}) & {\cpp {Neff\_SM [double]} } \newline {\cpp { eta0 [double]} } \\
   \midrule 
    
    \cpp{BBN\_abundances}\newline    Fills \cpp{BBN\_container} with theoretically expected element abundances and covariance matrix from BBN.
      &\cpp{compute\_BBN\_abundances} (\cpp{BBN\_container}) & {\cpp {AlterBBN\_Input [map\_str\_dbl]} } 
    \newline {\cpp {call\_nucl\_err \{int\}} } 
    \newline {\cpp {get\_NNUC \{size\_t\}} }
    \newline {\cpp {get\_abund\_map\_AlterBBN}
      \newline \hphantom{,} \cpp{ \{map\_str\_int\}} } \\
   \midrule 
    
    \cpp{helium\_abundance}\newline   Returns the $\prescript{4}{}{\rm He}$ abundance.
      &\cpp{extract\_helium\_abundance} (\cpp{double}) & {\cpp {BBN\_abundances [BBN\_container]} } \\
   \midrule 
    
    \cpp{BBN\_LogLike}\newline    Computes likelihood for BBN data.
      &\cpp{compute\_BBN\_LogLike} (\cpp{double}) & {\cpp {BBN\_abundances [BBN\_container]} } \\
   \bottomrule 

   \end{tabular}
  }
  \caption{Capabilities provided by \cosmobit relating to Big Bang Nucleosynthesis.}
  \label{tab:capabilities_bbn}
\end{table*}

 \begin{table*}[tp]
\centering
\scriptsize
   \makebox[\linewidth]{
   \begin{tabular}{p{5.3cm} p{6cm} p{5.5cm}}

    \toprule 
   \textbf{Capability} and brief description & \textbf{Function} (\textbf{Return Type}) & \textbf{Dependencies [type] / \newline BE requirement \{type\} }\\ \midrule 

    \cpp{parameter\_dict\_for\_MPLike}\newline   Returns \cpp{pybind11::dict} filled with the \py{mcmc\_parameters} dictionary of \montepython's \py{Data} object, with current values of nuisance parameters.
      &\cpp{pass\_empty\_parameter\_dict\_for\_MPLike} (\cpp{pybind11::dict}) & {\cpp { } } \\
    \cline{2-3}&\cpp{set\_parameter\_dict\_for\_MPLike} (\cpp{pybind11::dict}) & {\cpp { } } \\
   \midrule 
    
    \cpp{MP\_objects}\newline   Creates an object containing the \montepython ~\py{Data} and \py{Likelihood} objects, determining which experiments are in use in the process.
      &\cpp{create\_MP\_objects} (\cpp{MPLike\_objects\_container}) & {\cpp {create\_MP\_data\_object}
      \newline \hphantom{,} \cpp{ \{pybind11::object\}} } 
    \newline {\cpp {get\_MP\_available\_likelihoods}
      \newline \hphantom{,} \cpp{ \{std::vector<str>\}} } 
    \newline {\cpp {create\_MP\_likelihood\_objects}
      \newline \hphantom{,} \cpp{ \{map\_str\_pyobj\}} } 
    \newline {\cpp {parameter\_dict\_for\_MPLike}
      \newline \hphantom{,} \cpp{ [pybind11::dict]} } \\
   \midrule 
    
    \cpp{MP\_LogLikes}\newline    Calls \montepython to get the likelihood of all experiments (tagged with \cpp{Likelihood} and \cpp{Observable}). The results are stored as map from experiment name to LogLike result in the type \cpp{MPLike\_result\_container}.
      &\cpp{compute\_MP\_LogLikes} (\cpp{map\_str\_dbl}) & {\cpp {parameter\_dict\_for\_MPLike}
      \newline \hphantom{,} \cpp{ [pybind11::dict]} } 
    \newline {\cpp {check\_likelihood\_classy\_combi}
      \newline \hphantom{,} \cpp{ \{void\}} } 
    \newline {\cpp {MP\_objects}
      \newline \hphantom{,} \cpp{ [MPLike\_objects\_container]} } 
    \newline {\cpp {get\_classy\_backendDir}
      \newline \hphantom{,} \cpp{ \{std::string\}} } 
    \newline {\cpp {get\_classy\_cosmo\_object}
      \newline \hphantom{,} \cpp{ \{pybind11::object\}} } 
    \newline {\cpp {get\_MP\_loglike \{double\}} } \\
   \midrule 
    
    \cpp{MP\_Combined\_LogLike}\newline    Returns combined LogLike of all \montepython Likelihoods used in the scan.
      &\cpp{compute\_MP\_combined\_LogLike} (\cpp{double}) & {\cpp {MP\_LogLikes [map\_str\_dbl]} } \\
   \midrule 
    
    \cpp{bao\_like\_correlation}\newline   Adds the correlation coefficients of the \term{bao\_correlations} likelihood to the output file. Optional to use in combination with the \term{bao\_correlations} \montepython likelihoods. 
      &\cpp{get\_bao\_like\_correlation} (\cpp{map\_str\_dbl}) & {\cpp {MP\_LogLikes [map\_str\_dbl]} } 
      \newline {\cpp {MP\_objects}
      \newline \hphantom{,} \cpp{ [MPLike\_objects\_container]} } \\
   \bottomrule

   \end{tabular}
  }
  \caption{Capabilities provided by \cosmobit relating to the \montepython interface.}
  \label{tab:capabilities_mp}
\end{table*}

\begin{table*}[tp]
\centering
\scriptsize
   \makebox[\linewidth]{
   \begin{tabular}{p{7.2cm} p{5cm} p{4.5cm}}

    \toprule 
   \textbf{Capability} and brief description & \textbf{Function} (\textbf{Return Type}) & \textbf{Dependencies [Type] /\newline BE requirements \{Type\} }\\ \midrule 
 
    \cpp{classy\_input\_params}\newline  Creates an object of type \cpp{Classy\_input}, containing a \Python dictionary with all relevant input parameters for the \classy run. 

    Here, the parameters $H_0$, $\Tcmb$, $\omega_{\rm b}$, $\tau_{\rm reio}$, and $\omega_{\rm cdm}$ are set. Further, all input parameters from the capabilities \cpp{classy\_NuMasses\_Nur\_input}, \cpp{classy\_MPLike\_input}, \cpp{classy\_PlanckLike\_input}, \cpp{classy\_parameters\_EnergyInjection}, and \cpp{classy\_primordial\_input} are gathered and merged into one dictionary which will be passed to \classy.
      &\cpp{set\_classy\_input\_params} (\cpp{Classy\_input}) & {\cpp {classy\_MPLike\_input}
      \newline \hphantom{,} \cpp{ [pybind11::dict]} } 
    \newline {\cpp {classy\_NuMasses\_Nur\_input}
      \newline \hphantom{,} \cpp{ [pybind11::dict]} } 
    \newline {\cpp {classy\_primordial\_input}
      \newline \hphantom{,} \cpp{ [pybind11::dict]} } \\
   \midrule 
    
    \cpp{classy\_MPLike\_input}\newline  Sets all input parameters for \classy from active \montepython likelihoods and stores them in a \Python dictionary. If no \montepython likelihoods are in use, this will simply pass an empty dictionary.
      &\cpp{set\_classy\_input\_with\_MPLike} (\cpp{pybind11::dict}) & {\cpp {MP\_objects}
      \newline \hphantom{,} \cpp{ [MPLike\_objects\_container]} } \\

    \cline{2-3}&\cpp{set\_classy\_input\_no\_MPLike} (\cpp{pybind11::dict}) & {\cpp { } } \\
   \midrule 
    
    \cpp{classy\_primordial\_input}\newline  Sets all primordial input parameters for \classy. These are the primordial helium abundance, and parameters related to the primordial power spectrum. This can either be just $A_s$, $n_s$ (and $r$) through the function \cpp{set\_classy\_parameters\_parametrised\_ps}, or the full shape of the spectrum calculated from an inflationary model through the function \cpp{set\_classy\_parameters\_primordial\_ps}.
      &\cpp{set\_classy\_parameters\_}
      \newline \hphantom{,} \cpp{primordial\_ps} 
      \newline(\cpp{pybind11::dict}) & {\cpp {primordial\_power\_spectrum}
      \newline \hphantom{,} \cpp{ [Primordial\_ps]} } 
    \newline {\cpp {helium\_abundance [double]} } 
    \newline {\cpp {k\_pivot [double]} } \\
    \cline{2-3}
    &\cpp{set\_classy\_parameters\_}
      \newline \hphantom{,} \cpp{parametrised\_ps} 
      \newline (\cpp{pybind11::dict}) & {\cpp {helium\_abundance [double]} } 
    \newline {\cpp {k\_pivot [double]} } \\
   \midrule 
    
    \cpp{classy\_parameters\_EnergyInjection}\newline  Sets all input parameters for \exoclass relevant for energy injection in the early Universe. These parameters are for example the DM mass and cross section (for annihilating DM models) or the DM lifetime and fraction (decaying DM models). Further, the energy injection coefficients and the injection efficiency functions are passed.
      &\cpp{set\_classy\_parameters\_}
      \newline \hphantom{,} \cpp{EnergyInjection\_DecayingDM} 
      \newline(\cpp{pybind11::dict}) & {\cpp {energy\_injection\_efficiency}
      \newline \hphantom{,} \cpp{ [DarkAges::Energy\_}
      \newline \hphantom{,} \cpp{injection\_efficiency\_table]} } \\
    \cline{2-3}&\cpp{set\_classy\_parameters\_}
      \newline \hphantom{,} \cpp{EnergyInjection\_AnnihilatingDM}
      \newline (\cpp{pybind11::dict}) & {\cpp {energy\_injection\_efficiency}
      \newline \hphantom{,} \cpp{ [DarkAges::Energy\_}
      \newline \hphantom{,} \cpp{injection\_efficiency\_table]} } \\
   \midrule 
    
    \cpp{classy\_PlanckLike\_input}\newline  Sets all input parameters for \classy needed for the calculation of active Planck CMB likelihoods. This sets the non-linear treatment to \py{"halofit"}, the maximum angular scale up to which the CMB spectra are computed, and which spectra need to be computed (temperature, polarisation and/or lensing).
      &\cpp{set\_classy\_PlanckLike\_input} (\cpp{pybind11::dict}) & {\cpp {plc\_required\_Cl \{void\}} } \\
   \midrule 
    
    \cpp{classy\_NuMasses\_Nur\_input}\newline  Sets all neutrino related input parameters for \classy (\term{N\_ur}, \term{N\_ncdm}, \term{T\_ncdm}, \term{m\_ncdm}) and stores them in a \Python dictionary.
      &\cpp{set\_classy\_NuMasses\_Nur\_input} (\cpp{pybind11::dict}) & {\cpp {T\_ncdm [double]} } 
      \newline \cpp {N\_ur [double]} \\
   \bottomrule 
   \end{tabular}
  }
  \caption{Capabilities used to collate parameter values and information used to pass to the Boltzmann solver \classy.}
  \label{tab:capabilities_classy}
\end{table*}

\bibliography{R2}

\end{document}